%% file: article.tex
\documentclass[12pt,a4paper,french,english]{article}
\usepackage[T1]{fontenc}
\usepackage[latin9]{inputenc}
\usepackage{color}
\usepackage{babel}
\addto\extrasfrench{\providecommand{\fg}{\ifdim\lastskip>\z@\unskip\fi~\frqq}}

\usepackage{amsthm}
\usepackage{amsmath}
\usepackage{graphicx}
\usepackage{amssymb}
\usepackage[unicode=true, pdfusetitle,
 bookmarks=true,bookmarksnumbered=false,bookmarksopen=false,
 breaklinks=false,pdfborder={0 0 0},backref=false,colorlinks=false]
 {hyperref}

\makeatletter

\theoremstyle{plain}
\newtheorem{thm}{Theorem}
  \theoremstyle{definition}
  \newtheorem{defn}[thm]{Definition}
  \theoremstyle{plain}
  \newtheorem{prop}[thm]{Proposition}
  \theoremstyle{plain}
  \newtheorem{lem}[thm]{Lemma}
  \theoremstyle{plain}
  \newtheorem{cor}[thm]{Corollary}

\usepackage{a4wide}
\usepackage{color}
\usepackage{graphics}

\makeatother

\begin{document}

\title{Upper bound on the density of Ruelle resonances for Anosov flows.}

\date{24 February 2010}

\author{Frédéric Faure\textit{}%
\thanks{Institut Fourier, UMR 5582, 100 rue des Maths, BP74 38402 St Martin
d'Hères. frederic.faure@ujf-grenoble.fr http://www-fourier.ujf-grenoble.fr/\textasciitilde{}faure%
}, Johannes Sjöstrand%
\thanks{IMB, UMR 5584, Université de Bourgogne 9, Avenue Alain Savary, BP
47870 Dijon cedex. Johannes.Sjostrand@u-bourgogne.fr%
}}
\maketitle
\begin{abstract}
Using a semiclassical approach we show that the spectrum of a smooth
Anosov vector field $V$ on a compact manifold is discrete (in suitable
anisotropic Sobolev spaces) and then we provide an upper bound for
the density of eigenvalues of the operator $\left(-i\right)V$, called
Ruelle resonances, close to the real axis and for large real parts.
\end{abstract}
~
\selectlanguage{french}%
\begin{abstract}
Par une approche semiclassique on montre que le spectre d'un champ
de vecteur d'Anosov $V$ sur une variété compacte est discret (dans
des espaces de Sobolev anisotropes adaptés). On montre ensuite une
majoration de la densité de valeurs propres de l'opérateur $\left(-i\right)V$,
appelées résonances de Ruelle, près de l'axe réel et pour les grandes
parties réelles.
\end{abstract}
\selectlanguage{english}%
\footnote{
operators, and other functional analytic techniques in dynamical systems
81Q20 Semi-classical techniques

}

\newpage

\tableofcontents{}

\newpage

\section{Introduction}

Chaotic behavior of certain dynamical systems is due to hyperbolicity
of the trajectories. This means that the trajectories of two initially
close points will diverge in the future or in the past (or both) \cite{brin-02,katok_hasselblatt}.
As a result the behavior of an individual trajectory appears to be
complicated and unpredictable. However evolution of a cloud of points
seems more simple: it will spread and equidistribute according to
an invariant measure, called an equilibrium measure (or S.R.B. measure).
Also from the physical point of view, a distribution reflects the
unavoidable lack of knowledge about the initial point. Following this
idea, D. Ruelle in the 70' \cite{ruelle_75,ruelle_86}, has shown
that instead of considering individual trajectories, it is much more
natural to consider evolution of densities under a linear operator
called the Ruelle transfer operator or the Perron Frobenius operator.

For dynamical systems with strong chaotic properties, such as uniformly
expanding maps or uniformly hyperbolic maps, Ruelle, Bowen, Fried,
Rugh and others, using symbolic dynamics techniques (Markov partitions),
have shown that the transfer operator has a discrete spectrum of eigenvalues.
This spectral description has an important meaning for the dynamics
since each eigenvector corresponds to an invariant distribution (up
to a time factor). From this spectral characterization of the transfer
operator, one can derive other specific properties of the dynamics
such as decay of time correlation functions, central limit theorem,
mixing, etc. In particular a spectral gap implies exponential decay
of correlations. 

This spectral approach has recently (2002-2005) been improved by M.
Blank, S. Gouëzel, G. Keller, C. Liverani \cite{liverani_02,liverani_04,liverani_05,liverani_butterley_07},
V. Baladi and M. Tsujii \cite{baladi_sobolev_05,baladi_05} (see \cite{baladi_05}
for some historical remarks) and in \cite{fred-roy-sjostrand-07},
through the construction of functional spaces adapted to the dynamics,
independent of every symbolic dynamics. 

The case of flows i.e. dynamical systems with continuous time is more
delicate (see \cite{melbourne_07} for historical remarks). This is
due to the direction of time flow which is neutral (i.e. two nearby
points on the same trajectory will not diverge from one another).
In 1998 Dolgopyat \cite{dolgopyat_98,dolgopyat_02} showed the exponential
decay of correlation functions for certain Anosov flows, using techniques
of oscillatory integrals and symbolic dynamics. In 2004 Liverani \cite{liverani_contact_04}
adapted Dolgopyat's ideas to his functional analytic approach, to
treat the case of contact Anosov flows. In 2005 M. Tsujii \cite{tsujii_05}
obtained an explicit estimate for the spectral gap for the suspension
of an expanding map. In 2008 M. Tsujii \cite{tsujii_08} obtained
an explicit estimate for the spectral gap, in the case of contact
Anosov flows.

\paragraph{Semiclassical approach for transfer operators:}

It also appeared recently \cite{fred-RP-06,fred-roy-sjostrand-07,fred_expanding_09}
that for hyperbolic dynamics on a manifold $X$, the study of transfer
operators is naturally a semiclassical problem in the sense that a
transfer operator can be considered as a {}``Fourier integral operator''
and using standard tools of semiclassical analysis, some of its spectral
properties can be obtained from the study of {}``the associated classical
symplectic dynamics'', namely the initial hyperbolic dynamics on
$X$ lifted to the cotangent space $T^{*}X$ (the phase space).

The simple idea behind this, crudely speaking, is that a transfer
operator transports a {}``wave packet'' (i.e. localized both in
space and in Fourier space) into another wave packet, and this is
exactly the characterization of a Fourier integral operator. A wave
packet is characterized by a point in phase space (its position and
its momentum), hence one is naturally led to study the dynamics in
phase space. Moreover, since every function or distribution can be
decomposed into a linear superposition of wave packets, the dynamics
of wave packets characterizes completely the transfer operators.

Following this approach, in the papers \cite{fred-RP-06,fred-roy-sjostrand-07}
we studied hyperbolic diffeomorphisms. The aim of the present paper
is to show that semiclassical analysis is also well adapted to hyperbolic
systems with a neutral direction since it induces a natural semiclassical
parameter ($\alpha$ in Theorem \ref{thm:Upper-bound-for-resonances}
page \pageref{thm:Upper-bound-for-resonances}), the Fourier component
in the neutral direction. In the paper \cite{fred_expanding_09} one
of us has considered a partially expanding map and showed that a spectral
gap develops in the limit of large oscillations in the neutral direction
(which is a semiclassical limit). In this paper we consider a hyperbolic
flow on a manifold $X$ generated by a vector field $V$. In Section
2 we recall the definition of a hyperbolic flow and some of its properties.
In Section 3 we describe the dynamics induced on the cotangent space
$T^{*}X$. In  particular we construct an {}``escape function''
which expresses the fact that the trajectories escape towards infinity
in $T^{*}X$ except on a specific subspace called the \textbf{trapped
set} $K$. The vector field is considered as a partial differential
operator of order 1 acting on smooth functions $C^{\infty}\left(X\right)$
and can be extended to the space of distributions. In Section 4 using
a semiclassical approach (with escape function on phase space) we
establish a first result, in Theorem \ref{th:discrete_spectrum},
which shows that the operator $\widehat{H}=-iV$ has a discrete spectrum
of resonances in specific anisotropic Sobolev spaces. This discrete
spectrum is intrinsic to the vector field $V$ in the sense that we
get the same spectrum in an overlap region for two different Sobolev
spaces constructed according to some general principles. This result
has already been obtained by O. Butterley and C. Liverani in \cite[Theorem 1]{liverani_butterley_07}.
The novelty here is to show that this resonance spectrum fits with
the general theory of semiclassical resonances developed by B. Helffer
and J. Sjöstrand \cite{sjostrand_87} and initiated by Aguilar, Baslev,
Combes \cite{aguilar_combes_71,balslev_combes_71}. Our main new result
is Theorem \ref{thm:Upper-bound-for-resonances} page \pageref{thm:Upper-bound-for-resonances}
which provides an upper bound $o\left(\alpha^{n-1/2}\right)$ (with
$n=\dim X$), for the number of resonances in the spectral domain
$\Re\left(\lambda\right)\in\left[\alpha,\alpha+\sqrt{\alpha}\right]$,
$\Im\left(\lambda\right)>-\beta$ (all $\beta$) in the semiclassical
limit $\left|\alpha\right|\rightarrow\infty$. 

The use of escape functions on phase space for resonances has been
introduced by B. Helffer and J. Sjöstrand \cite{sjostrand_87} and
used in many situations \cite{sjostrand_07,sjostrand_90,sjostrand_hyp_02,sjostrand_resonances_02,zworski_resonances_99,zworski_lin_guillope_02,nonnenmacher_07}.
In particular in \cite{zworski_lin_guillope_02}, the authors consider
the geodesic flow associated to Schottky groups and provide an upper
bound for the density of Ruelle resonances (see also \cite{bortwick_book_07}).

In this paper as well as in \cite{fred-roy-sjostrand-07}, one aim
is to make more precise the connection between the spectral study
of Ruelle resonances and the spectral study in quantum chaos \cite{nonnenmacher_08,zelditch-05},
in particular to emphasize the importance of the symplectic properties
of the dynamics in the cotangent space $T^{*}X$ on the spectral properties
of the transfer operator, and long time behavior of the dynamics.

\paragraph{Acknowledgment:}

This work has been supported by {}``Agence Nationale de la Recherche''
under the grants JC05\_52556 and ANR-08-BLAN-0228-01.

\section{Anosov flows}

Let $X$ be a $n$-dimensional smooth compact connected manifold,
with $n\geq3$. Let $\phi_{t}$ be the flow on $X$ generated by a
\textbf{smooth} vector field $V\in C^{\infty}\left(X;TX\right)$:
\begin{equation}
V\left(x\right)=\frac{d}{dt}\left(\phi_{t}\left(x\right)\right)_{/t=0}\in T_{x}X,\qquad x\in X.\label{eq:def_V}\end{equation}

We assume that the flow $\phi_{t}$ is \textbf{Anosov}. We recall
the definition (see \cite{katok_hasselblatt} page 545, or \cite{pesin_04}
page 8)

\selectlanguage{french}%
\vspace{0.cm}\begin{center}{\color{red}\fbox{\color{black}\parbox{16cm}{
\begin{defn}
\label{def:Anosov_flow}\foreignlanguage{english}{On a smooth Riemannian
manifold $\left(X,g\right)$, a vector field $V$ generates an \textbf{Anosov}
\textbf{flow} $\left(\phi_{t}\right)_{t\in\mathbb{R}}$ (or \textbf{uniformly
hyperbolic flow}) if:}
\selectlanguage{english}%
\begin{itemize}
\item For each $x\in X$, there exists a decomposition\begin{equation}
T_{x}X=E_{u}\left(x\right)\oplus E_{s}\left(x\right)\oplus E_{0}\left(x\right),\label{eq:decomposition_of_TX}\end{equation}
where $E_{0}\left(x\right)$ is the one dimensional subspace generated
by $V\left(x\right)$.
\item The decomposition (\ref{eq:decomposition_of_TX}) is invariant by
$\phi_{t}$ for every $t$:\[
\forall x\in X,\quad\left(D_{x}\phi_{t}\right)\left(E_{u}\left(x\right)\right)=E_{u}\left(\phi_{t}\left(x\right)\right)\quad\mbox{and}\quad\left(D_{x}\phi_{t}\right)\left(E_{s}\left(x\right)\right)=E_{s}\left(\phi_{t}\left(x\right)\right).\]

\item There exist constants $c>0$, $\theta>0$ such that for every $x\in X$\begin{eqnarray}
\left|D_{x}\phi_{t}\left(v_{s}\right)\right|_{g} & \leq & ce^{-\theta t}\left|v_{s}\right|_{g},\qquad\forall v_{s}\in E_{s}\left(x\right),\quad t\geq0\label{eq:def_dynamics}\\
\left|D_{x}\phi_{t}\left(v_{u}\right)\right|_{g} & \leq & ce^{-\theta\left|t\right|}\left|v_{u}\right|_{g},\qquad\forall v_{u}\in E_{u}\left(x\right),\quad t\leq0,\nonumber \end{eqnarray}
meaning that $E_{s}$ is the stable distribution and $E_{u}$ the
unstable distribution for positive time.
\end{itemize}
\end{defn}
}}}\end{center}\vspace{0.cm}

\selectlanguage{english}%
\begin{figure}
\begin{centering}
\includegraphics[scale=0.6]{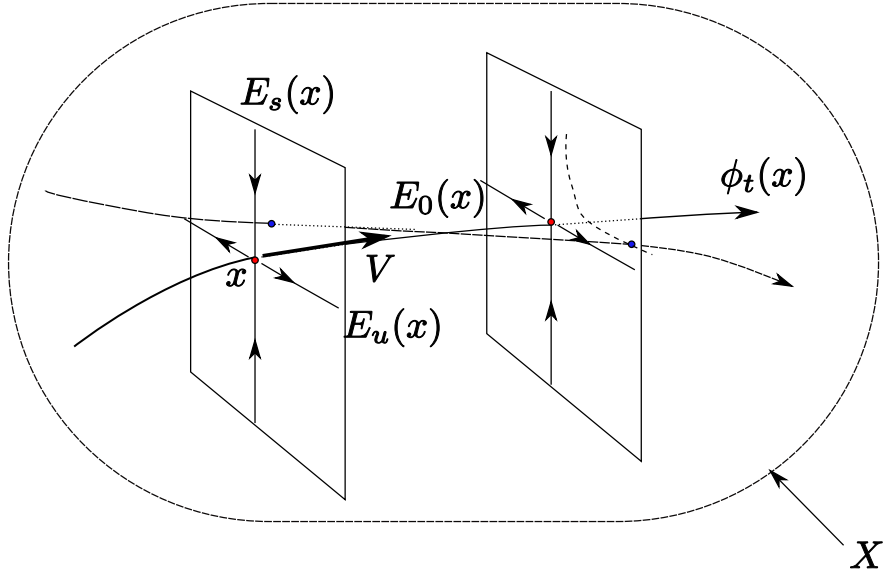}
\par\end{centering}

\caption{\label{fig:hyperbolic_flow}Picture of an Anosov flow in $X$ and
instability of trajectories.}

\end{figure}

\subsection{Remarks:\label{par:Remarks:_smoothness}}

The remarks of this Section give more information on Anosov flows
but are not necessary for the rest of this paper.

\subsubsection{General remarks}
\begin{itemize}
\item Standard examples of Anosov flows are \textbf{suspensions of Anosov
diffeomorphisms} (see \cite{pesin_04} p.8), or \textbf{geodesic flows
on manifolds $M$ with sectional negative curvature} (see \cite{pesin_04}
p.9, or \cite{katok_hasselblatt} p.549, p.551). Notice that in this
case, the geodesic flow is Anosov on the unit cotangent bundle $X=T_{1}^{*}M$.
\item The global hyperbolic structure of Anosov flows or Anosov diffeomorphisms
is a very strong geometric property, so that manifolds carrying such
dynamics satisfy strong topological conditions and the list of known
examples is not so long. See \cite{bonatti_09} for a detailed discussion
and references on that question.
\item Let \[
d_{u}=\dim E_{u}\left(x\right),\qquad d_{s}=\dim E_{s}\left(x\right),\]
(they are independent of $x\in X$). Eq.(\ref{eq:decomposition_of_TX})
implies $d_{u}+d_{s}+1=\dim X=n$. For every $d_{u},d_{s}\geq1$ one
may construct an example of an Anosov flow: one considers a suspension
of a hyperbolic diffeomorphism of $SL_{n-1}\left(\mathbb{Z}\right)$
on $\mathbb{T}^{n-1}$, with $n=d_{u}+d_{s}+1$, such that there are
$d_{s}$ eigenvalues with modulus $\left|\lambda\right|<1$, and $d_{u}$
eigenvalues with modulus $\left|\lambda\right|>1$.
\end{itemize}

\subsubsection{Constructive expressions for $E_{u}$, $E_{s}$}

In the case where $d_{u}=d_{s}=1$, there is a formula which gives
the distributions $E_{u},E_{s}$ \cite{arnold-avez}. Let $\left[v_{0}\right]\in C^{\infty}\left(\mathbb{P}\left(TX\right)\right)$
be a global smooth section of the projective tangent bundle, such
that at every point $x\in X$, $\left[v_{0}\right]_{x}\notin\left(E_{0}\left(x\right)\oplus E_{s}\left(x\right)\right)$
(it is sufficient that the direction $\left[v_{0}\right]_{x}$ is
close enough to the unstable direction $E_{u}\left(x\right)$). Then
for every $x\in X$,\begin{equation}
E_{u}\left(x\right)=\lim_{t\rightarrow+\infty}D\phi_{t}\left(\left[v_{0}\right]_{\phi_{-t}\left(x\right)}\right).\label{eq:construction_Eu}\end{equation}
Similarly if $\left[v_{0}\right]_{x}\notin\left(E_{0}\left(x\right)\oplus E_{u}\left(x\right)\right),\forall x\in X$,
then for every $x\in X$, \[
E_{s}\left(x\right)=\lim_{t\rightarrow-\infty}D\phi_{t}\left(\left[v_{0}\right]_{\phi_{-t}\left(x\right)}\right).\]
See figure \ref{fig:projective_flow}.

\begin{center}
\begin{figure}
\begin{centering}
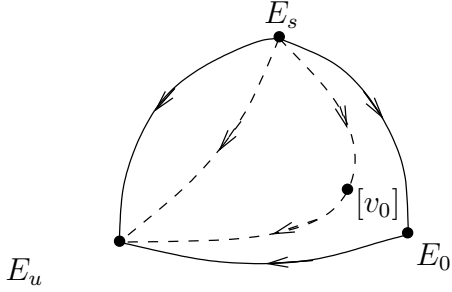
\par\end{centering}

\caption{\label{fig:projective_flow}Picture of the flow on the projective
tangent bundle $\mathbb{P}\left(TX\right)$ induced by the hyperbolic
vector field $V$. A global section $[v_{0}]$ will converge towards
$E_{u}$ or $E_{s}$ for $t\rightarrow\pm\infty$ as explained in
Eq.(\ref{eq:construction_Eu}).}

\end{figure}

\par\end{center}

In the general case, for every $d_{u},d_{s}$, there exists a similar
construction. Let $\left[v_{0}\right]\in C^{\infty}\left(\mbox{Gr}_{d_{u},n}\left(TX\right)\right)$
be a global smooth and non vanishing section of the Grassmanian bundle,
such that at every point $x\in X$, the linear space $\left[v_{0}\right]_{x}$
does not intersect $E_{0}\left(x\right)\oplus E_{s}\left(x\right)$.
Then for every $x\in X$,\[
E_{u}\left(x\right)=\lim_{t\rightarrow+\infty}D\phi_{t}\left(\left[v_{0}\right]_{\phi_{-t}\left(x\right)}\right).\]
Similarly\[
E_{s}\left(x\right)=\lim_{t\rightarrow-\infty}D\phi_{t}\left(\left[v_{0}\right]_{\phi_{-t}\left(x\right)}\right),\]
when $\left[v_{0}\right]\in C^{\infty}\left(\mbox{Gr}_{d_{s},n}\left(TX\right)\right)$
does not intersect $E_{0}\oplus E_{u}$.

\subsubsection{\label{sub:Regularity-of-the-distrib}Anosov one form $\alpha$ and
regularity of the distributions $E_{u}\left(x\right)$, $E_{s}\left(x\right)$}

The distribution $E_{0}\left(x\right)$ is smooth since $V\left(x\right)$
is assumed to be smooth. The distributions $E_{u}\left(x\right)$,
$E_{s}\left(x\right)$ and $E_{u}\left(x\right)\oplus E_{s}\left(x\right)$
are only Hölder continuous in general (see \cite{pesin_04} p.15,
\cite{ghys_92} p.211). Smoothness can be present with additional
hypothesis or with particular models. See the discussion below, section
\ref{sub:Regularity-of-the-distrib} page \pageref{sub:Regularity-of-the-distrib}.

The above hypothesis on the flow implies that there is a particular
continuous one form on $X$, denoted $\alpha\in C^{0}\left(T^{*}X\right)$
called the \textbf{Anosov 1-form }and defined by\begin{equation}
\mbox{ker}\left(\alpha\left(x\right)\right)=E_{u}\left(x\right)\oplus E_{s}\left(x\right),\qquad\left(\alpha\left(x\right)\right)\left(V\left(x\right)\right)=1,\qquad\forall x\in X.\label{eq:def_alpha}\end{equation}
Since $E_{u}$ and $E_{s}$ are invariant by the flow then $\alpha$
is invariant as well, $\phi_{t}^{*}\left(\alpha\right)=\alpha$ for
every $t\in\mathbb{R}$, and therefore (in the sense of distributions)\begin{equation}
\mathcal{L}_{V}\left(\alpha\right)=0,\label{eq:derivee_Lie_alpha}\end{equation}
where $\mathcal{L}_{V}$ denotes the Lie derivative.

We discuss now some known results about the smoothness of the distributions
$E_{u}\left(x\right),E_{s}\left(x\right)$ in some special cases.
\begin{itemize}
\item In the case of a\textbf{ geodesic flow} on a smooth Riemannian negative
curvature manifold $M$ (with $X=T_{1}^{*}M$), $E_{u}\left(x\right)\oplus E_{s}\left(x\right)$
is orthogonal%
\footnote{Proof: since the metric $g$ is preserved by the flow, if $v_{u}\in E_{u}$,
and $V\in E_{0}$, \[
g\left(v_{u},V\right)=g\left(\left(D\phi_{t}\right)v_{u},\left(D\phi_{t}\right)V\right)=g\left(\left(D\phi_{t}\right)v_{u},V\right),\]
goes to zero as $t\rightarrow-\infty$, from Eq.(\ref{eq:def_dynamics}).
Therefore $g\left(E_{u},E_{0}\right)=0$. Similarly $g\left(E_{s},E_{0}\right)=0$.%
} to $E_{0}\left(x\right)$, therefore $E_{u}\left(x\right)\oplus E_{s}\left(x\right)$
is $C^{\infty}$. The distributions $E_{u}\left(x\right)$, $E_{s}\left(x\right)$
are $C^{1}$ individually (see \cite{ghys_87} p.252).
\item More generally, the flow $\phi_{t}$ is \textbf{a contact flow }(or
\textbf{Reeb vector field}, see \cite{mac_duff_98} p.106, \cite[p.55]{da_silva_01})
if the associated one form $\alpha$ defined in Eq.(\ref{eq:def_alpha})
is $C^{\infty}$ and if\begin{equation}
d\alpha_{/\left(E_{u}\oplus E_{s}\right)}\mbox{ is non degenerate (i.e. symplectic)}\label{eq:def_contact}\end{equation}
meaning that $\alpha$ \textbf{is a contact one form}. Equivalently,
$dx:=\alpha\wedge\left(d\alpha\right)^{n}$ is a volume form on $X$
with $n:=\frac{1}{2}\dim\left(E_{u}\oplus E_{s}\right)$. Notice that
(\ref{eq:derivee_Lie_alpha}) implies that the volume form is invariant
by the flow: \begin{equation}
\mathcal{L}_{V}\left(dx\right)=0.\label{eq:volume_form_si_contact}\end{equation}
 In that case, $E_{u}\left(x\right)\oplus E_{s}\left(x\right)=\mbox{ker}\left(\alpha\right)$
is $C^{\infty}$ and $\alpha$ determines $V$ by $d\alpha\left(V\right)=0$
and $\alpha\left(V\right)=1$.
\item (From \cite{ghys_92} p.211) Hurder and Katok in \cite{hurder-90}
showed that if $\phi_{t}$ is an Anosov flow on $X$, $\mbox{dim}X=3$,
and if $\alpha$ is a contact form of class $C^{1}$ then $\alpha$
is $C^{\infty}$ in fact, and $E_{u}\left(x\right)$, $E_{s}\left(x\right)$
are $C^{2-\varepsilon}$ for every $\varepsilon>0$. Moreover if $E_{u}\left(x\right),E_{s}\left(x\right)$
are $C^{2}$ then they are $C^{\infty}$ in fact and $C^{\infty}$
conjugated to an algebraic flow \cite{ghys_87}.
\end{itemize}

\section{Transfer operator and the dynamics lifted on $T^{*}X$}

\subsection{The transfer operator}

The flow $\phi_{t}$ acts in a natural manner on functions by pull
back and this defines the transfer operator:

\vspace{0.cm}\begin{center}{\color{red}\fbox{\color{black}\parbox{16cm}{
\begin{defn}
\label{def:The-Ruelle-transfer}The \textbf{Ruelle transfer operator
$\widehat{M}_{t}:C^{\infty}\left(X\right)\rightarrow C^{\infty}\left(X\right)$,
$t\in\mathbb{R}$ }is defined by:\begin{equation}
\widehat{M}_{t}\varphi=\varphi\circ\phi_{-t},\qquad\varphi\in C^{\infty}\left(X\right),\label{eq:def_M_hat_t}\end{equation}

$\widehat{M}_{t}$ can be expressed in terms of the vector field $V$
as\begin{equation}
\widehat{M}_{t}=\exp\left(-tV\right)=\exp\left(-it\widehat{H}\right)\label{eq:def_M_hat}\end{equation}
with the \textbf{generator}\begin{equation}
\widehat{H}:=-iV.\label{eq:def_H_hat}\end{equation}

\end{defn}
}}}\end{center}\vspace{0.cm}

\paragraph{Remarks:}
\begin{itemize}
\item If $dx$ is a smooth density on $X$, then $\widehat{M}_{t}$ can
be extended to $L^{2}\left(X,dx\right)$. In this space, $\widehat{M}_{t}$
is a bounded operator. The adjoint of $\widehat{H}$ is (\cite{taylor_tome1}
prop.(2.4) p.129) \begin{equation}
\widehat{H}^{*}=-iV-i\,\mbox{div}\left(V\right)=\widehat{H}-i\,\mbox{div}\left(V\right).\label{eq:Adjoint_H}\end{equation}
Hence (\cite{taylor_tome1} def.(2.1) p.125)\begin{eqnarray}
\mbox{div}\left(V\right)=0 & \Leftrightarrow\mathcal{L}_{V}\left(dx\right)=0\Leftrightarrow & \phi_{t}\mbox{ preserves }dx\label{eq:divV0}\\
 & \Leftrightarrow & \widehat{H}\mbox{ is self-adjoint on }L^{2}\left(X,dx\right)\nonumber \\
 & \Leftrightarrow & \widehat{M}_{t}\mbox{ is a unitary operator on }L^{2}\left(X,dx\right).\nonumber \end{eqnarray}
This is the case for the geodesic flow, which preserves the Liouville
measure, or more generally for a contact flow from (\ref{eq:volume_form_si_contact}).
But for a generic Anosov flow there does not exist any smooth invariant
measure.
\item From the probabilistic point of view, it is natural to consider the
\textbf{Perron Frobenius transfer operator} $\widehat{T}_{t}$, $t\in\mathbb{R}$
whose generator is the adjoint $\widehat{H}^{*}$, Eq.(\ref{eq:Adjoint_H}):\begin{equation}
\widehat{T}_{t}:=e^{-i\widehat{H}^{*}t}=\left(\widehat{M}_{\left(-t\right)}\right)^{*}.\label{eq:def_Perron_Frob_Tt}\end{equation}
The reason is that one has the following relation which is interpreted
as a conservation of the total probability measure:\[
\int_{X}\left(\widehat{T}_{t}\psi\right)dx=\int_{X}\psi dx,\qquad\psi\in C^{\infty}\left(X\right).\]

\begin{proof}
$\int_{X}\left(\widehat{T}_{t}\psi\right)dx=\left(1|\widehat{T}_{t}\psi\right)=\left(\widehat{M}_{\left(-t\right)}1|\psi\right)=\int_{X}\psi dx$
since $\widehat{M}_{\left(-t\right)}1=1$ by (\ref{eq:def_M_hat_t}).
\end{proof}
\item We introduce the antilinear operator of complex conjugation $\widehat{C}:C^{\infty}\left(X\right)\rightarrow C^{\infty}\left(X\right)$
by\begin{equation}
\widehat{C}\varphi:=\overline{\varphi},\qquad\varphi\in C^{\infty}\left(X\right),\label{eq:conjugation_operator}\end{equation}
which can be extended to $\mathcal{D}'\left(X\right)$ by duality:
for $\psi\in\mathcal{D}'\left(X\right)$, $\varphi\in C^{\infty}\left(X\right)$,\[
\langle\widehat{C}\psi|\varphi\rangle:=\overline{\langle\psi|\widehat{C}\varphi\rangle}.\]
We have the following relation \begin{equation}
\widehat{H}\widehat{C}+\widehat{C}\widehat{H}=0\label{eq:conjugation_symetry}\end{equation}
(it will imply a symmetry for the Ruelle resonance spectrum, see Proposition
\ref{thm:Symmetry} page \pageref{thm:Symmetry})

\begin{proof}
Since $V$ is real, for every $\varphi\in C^{\infty}\left(X\right)$
one has $\widehat{H}\widehat{C}\varphi=-iV\left(\overline{\varphi}\right)=\overline{iV\left(\varphi\right)}=-\overline{\left(-iV\left(\varphi\right)\right)}=-\widehat{C}\widehat{H}\varphi$.
\end{proof}
\end{itemize}

\subsection{Generators of transfer operators are pseudo-differential operators}

The generator $\widehat{H}$ defined in (\ref{eq:def_H_hat}) is a
differential operator hence a pseudodifferential operator. This allows
us to use the machinery of semiclassical analysis in order to study
the spectral properties \cite[chap. 7]{taylor_tome2}. In particular
it allows us to view $\widehat{M}_{t}=\exp\left(-it\widehat{H}\right)$
as a Fourier integral operator. See proposition \ref{prop_PDO_FIO}
below.

\subsubsection{\label{sub:Symbols-and-pseudodifferential}Symbols and pseudodifferential
operators}

In this Section we recall how pseudodifferential operators are defined
from their symbols on a manifold $X$. We first recall \cite[p.2]{grigis_sjostrand,taylor_tome2}
that:

\vspace{0.cm}\begin{center}{\color{red}\fbox{\color{black}\parbox{16cm}{
\begin{defn}
\label{def:The-symbol-class_S_mu_ro}The \textbf{symbol class} $S^{\mu}$
with order $\mu\in\mathbb{R}$ consist of $C^{\infty}$ functions
$p\left(x,\xi\right)$ on $\mathbb{R}^{n}\times\mathbb{R}^{n}$ such
that\begin{equation}
\forall\alpha,\beta,\quad\left|\partial_{\xi}^{\alpha}\partial_{x}^{\beta}p\right|\leq C_{\alpha,\beta}\left\langle \xi\right\rangle ^{\mu-\left|\alpha\right|},\qquad\mbox{with }\left\langle \xi\right\rangle :=\sqrt{1+\xi^{2}}.\label{eq:def_symbol_p}\end{equation}

\end{defn}
}}}\end{center}\vspace{0.cm}

The value of $\mu$ governs the increase (or decrease) of $p\left(x,\xi\right)$
as $\left|\xi\right|\rightarrow\infty$.

On a manifold $X$ with a given system of coordinates (more precisely
a chart system and a related partition of unity, see \cite[p.30]{taylor_tome2}),
the symbol $p$ determines a \textbf{pseudodifferential operator (PDO
for short)} denoted $\widehat{P}=\mathrm{Op}\left(p\right)$ acting
on $u\in C^{\infty}\left(X\right)$ and defined locally by\begin{equation}
\widehat{P}:u\rightarrow\left(\widehat{P}u\right)\left(x\right)=\int e^{i\xi\left(x-y\right)}p\left(x,\xi\right)u\left(y\right)dyd\xi.\label{eq:left_quantization}\end{equation}
Conversely $p=\sigma\left(\widehat{P}\right)$ is called the \textbf{symbol}
of the PDO $\widehat{P}$. Notation: if $p\in S^{\mu}$ we say that
$\widehat{P}\in\mathrm{Op}\left(S^{\mu}\right)$. 

The value of the order $\mu$ is independent on the choice of coordinates,
but the symbol $p$ of a given PDO $\widehat{P}$ depends on a choice
of a chart and a choice of a partition of unity (\cite[p.30]{taylor_tome2}).
The symbol has not a {}``geometrical meaning''. However it appears
that the change of coordinate systems changes the symbol only at a
subleading order $S^{\mu-1}$. In other words, the \textbf{principal
symbol} $p_{ppal}=p\mbox{ mod }S^{\mu-1}$ is a well defined function
on the manifold $X$ (independently of the charts).

Concerning the operator $\widehat{H}=-iV$ given in Eq.(\ref{eq:def_H_hat}),
one easily checks \cite[p.2]{taylor_tome1} that it is obtained by
$\widehat{H}=\mathrm{Op}\left(H\right)$ with the symbol \[
H\left(x,\xi\right)=V\left(\xi\right)\in S^{1}.\]
Notice that this symbol does not depend on the chart. This is very
particular to differential operators of order 1.

The quantization formula (\ref{eq:left_quantization}) is sometimes
called the \textbf{left-quantization} or ordinary quantization of
differential operators. There are plenty of other quantization formulae
which differ at subleading order $\mathrm{Op}\left(S^{\mu-1}\right)$
so the principal symbol of a PDO is the same for the different quantizations.
Some have interesting properties. For example the \textbf{Weyl quantization
of a symbol $p_{W}$} \cite[(14.5) p.60]{taylor_tome2} denoted by
$\widehat{P}=\mathrm{Op}_{W}\left(p_{W}\right)$ is defined by:\begin{equation}
\widehat{P}:u\rightarrow\left(\widehat{P}u\right)\left(x\right)=\int e^{i\xi\left(x-y\right)}p_{W}\left(\frac{x+y}{2},\xi\right)u\left(y\right)dyd\xi.\label{eq:Weyl_quantization}\end{equation}
In this quantization, a \emph{real} symbol $p_{W}$ is quantized as
a formally self-adjoint operator. 

In our example Eq.(\ref{eq:def_H_hat}), $\widehat{H}=-iV$, the Weyl
symbol is%
\footnote{Indeed from \cite[(14.7) p.60]{taylor_tome2}, in a given chart where
$V=V\left(x\right)\frac{\partial}{\partial x}$, \[
H_{W}\left(x,\xi\right)=\exp\left(\frac{i}{2}\partial_{x}\partial_{\xi}\right)\left(V\left(x\right).\xi\right)=V\left(x\right).\xi+\frac{i}{2}\partial_{x}V=V\left(\xi\right)+\frac{i}{2}\mbox{div}\left(V\right)\]
and $\mbox{div}\left(V\right)$ depends only on the choice of the
volume form, see \cite[p.125]{taylor_tome1}.%
} \begin{equation}
H_{W}\left(x,\xi\right)=V\left(\xi\right)+\frac{i}{2}\mbox{div}\left(V\right).\label{eq:Weyl_symbol_of_H}\end{equation}
Notice that this symbol does not depend on the choice of coordinates
systems provided the volume form is expressed by%
\footnote{On a manifold there always exists charts such that a given volume
form is expressed as $dx_{1}\ldots dx_{n}$%
} $dx=dx_{1}\ldots\dot{dx_{n}}$. The term $\frac{i}{2}\mbox{div}\left(V\right)$
in (\ref{eq:Weyl_symbol_of_H}) belongs to $S^{0}$ and is called
the \textbf{subprincipal symbol}. 

For general symbols and with the Weyl quantization, a change of coordinate
systems preserving the volume form changes the symbol at a subleading
order $S^{\mu-2}$ only. In other words, on a manifold with a fixed
smooth density $dx$, the Weyl symbol $p_{W}$ of a given PDO $\widehat{P}$
is well defined modulo terms in $S^{\mu-2}$. 

In this paper we will also use a \textbf{Toeplitz quantization} (or
F.B.I. quantization) for the proof of Lemma \ref{lem:Chi0} page \pageref{lem:Chi0}.

\subsubsection{Induced dynamics on $T^{*}X$}

Recall that the canonical symplectic two form on $T^{*}X$ is (\cite{mac_duff_98}
p.90)%
\footnote{We take the convention of the {}``semiclassical analysis community''
with $\omega:=d\xi\wedge dx$. The opposite convention $\omega:=dx\wedge d\xi$
is more usual in the {}``symplectic geometry community''.%
} \begin{equation}
\omega:=d\xi\wedge dx.\label{eq:omega_two_form}\end{equation}

The following well known proposition shows that the flow on the cotangent
bundle $T^{*}X$ obtained by lifting the flow $\phi_{t}$ is naturally
associated to the Ruelle transfer operator we are interesting in.

\vspace{0.cm}\begin{center}{\color{blue}\fbox{\color{black}\parbox{16cm}{
\begin{prop}
\label{prop_PDO_FIO}

The symbol of the differential operator $\widehat{H}$ defined in
Eq.(\ref{eq:def_H_hat}) belongs to the symbol class $S^{1}$. Its
principal symbol is equal to: \begin{equation}
H_{0}\left(x,\xi\right)=V\left(\xi\right)\in S^{1}.\label{eq:def_H}\end{equation}

The Ruelle transfer operator $\widehat{M}_{t}$, defined in Eq.(\ref{eq:def_M_hat})
is a semi-classical \textbf{Fourier integral operator} (FIO), whose
associated canonical map denoted by $M_{t}:T^{*}X\rightarrow T^{*}X$
is the canonical lift of the diffeomorphism $\phi_{t}$ on $T^{*}X$
(linear in the fibers). See figure \ref{fig:T*x-1}. More precisely
if $x\in X$, $x'=\phi_{t}\left(x\right)$ then \begin{equation}
M_{t}:\begin{cases}
T^{*}X & \rightarrow T^{*}X\\
x & \longmapsto x'=\phi_{t}\left(x\right)\\
\xi\in T_{x}^{*}X & \longmapsto\xi'=\left(D_{x'}\phi_{-t}\right)^{\mathrm{t}}\xi\in T_{x'}^{*}X\end{cases}\label{eq:def_Mt}\end{equation}
$M_{t}$ is also the Hamiltonian flow generated by the vector field
$\mathbf{X}\in C^{\infty}\left(T^{*}X,T\left(T^{*}X\right)\right)$
defined by \begin{equation}
\omega\left(.,\mathbf{X}\right)=dH_{0}.\label{eq:def_vector_field_X}\end{equation}

The vector field $\mathbf{X}$ is the canonical lift of $V$ on $T^{*}X$.
\end{prop}
}}}\end{center}\vspace{0.cm}
\begin{proof}
For (\ref{eq:def_H}), see \cite{taylor_tome2} p.2. From (\ref{eq:def_M_hat}),
$\widehat{M}_{t}$ is defined by\[
i\frac{d\widehat{M}_{t}}{dt}=\widehat{H}\widehat{M}_{t},\qquad\widehat{M}_{t=0}=Id.\]
For (\ref{eq:def_Mt}) see \cite[ex.2 p.152]{martinez-01} or \cite{taylor_tome1}
Eq.(14.20) p. 77 and Eq.(14.15) p. 76, or \cite{mac_duff_98}, ex.
3.12 p.92.
\end{proof}
\begin{center}
\begin{figure}
\begin{centering}
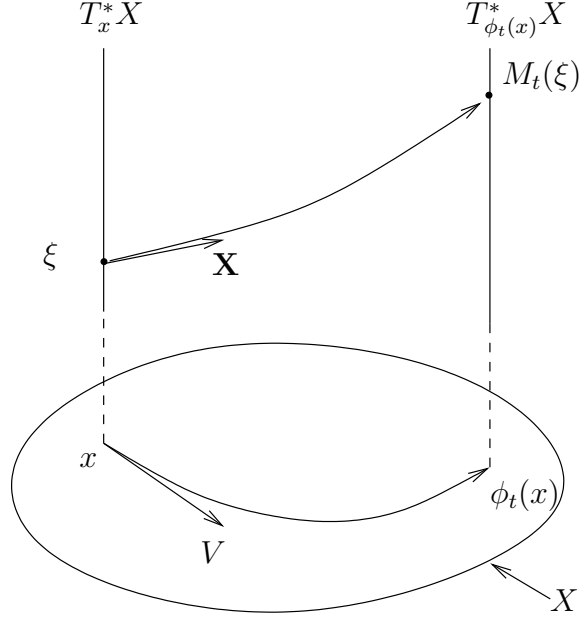
\par\end{centering}

\caption{\label{fig:T*x-1}The flow $\phi_{t}$ generated by the vector field
$V$ on $X$ is lifted on the cotangent bundle $T^{*}X$. This gives
a Hamiltonian flow $M_{t}$ generated by a vector field $\mathbf{X}$.}

\end{figure}

\par\end{center}

\paragraph{Remarks:}
\begin{itemize}
\item The Hamiltonian vector field $\mathbf{X}\in C^{\infty}\left(T^{*}X,T\left(T^{*}X\right)\right)$
can be expressed as usual with respect to a coordinate system by (\cite[p.74]{taylor_tome1})
\[
\mathbf{X}=\frac{\partial H_{0}}{\partial\xi}\frac{\partial}{\partial x}-\frac{\partial H_{0}}{\partial x}\frac{\partial}{\partial\xi}.\]

\item From (\ref{eq:Adjoint_H}) the operators $\widehat{H}$ and $\widehat{H}^{*}$
have the same real principal symbol $H_{0}\left(x,\xi\right)=V\left(\xi\right)$.
Therefore the canonical transform associated to the Perron Frobenius
operator $\widehat{T}_{t}$ , Eq.(\ref{eq:def_Perron_Frob_Tt}), is
also $M_{t}$.
\end{itemize}

\subsubsection{Dual decomposition}

Let

\begin{equation}
\boxed{T_{x}^{*}X=E_{u}^{*}\left(x\right)\oplus E_{s}^{*}\left(x\right)\oplus E_{0}^{*}\left(x\right)}\label{eq:decomp_T*X}\end{equation}
be the decomposition dual to (\ref{eq:decomposition_of_TX}) in the
sense that \begin{align}
\left(E_{0}^{*}\left(x\right)\right)\left(E_{u}\left(x\right)\oplus E_{s}\left(x\right)\right) & =0,\label{eq:dual_decomp}\\
\left(E_{u}^{*}\left(x\right)\right)\left(E_{u}\left(x\right)\oplus E_{0}\left(x\right)\right)=0, & \quad\left(E_{s}^{*}\left(x\right)\right)\left(E_{s}\left(x\right)\oplus E_{0}\left(x\right)\right)=0.\nonumber \end{align}

\paragraph{Remarks:}
\begin{itemize}
\item Let us remark that we have exchanged $E_{u}^{*}$ and $E_{s}^{*}$
with respect to the usual definition of dual spaces. Our choice of
notations will be justified by Eq.(\ref{eq:def_dynamics_T*X}).
\item One has\begin{align*}
\mbox{dim}E_{0}^{*}\left(x\right) & =\mbox{dim}E_{0}\left(x\right)=1,\\
\mbox{dim}E_{u}^{*}\left(x\right)=\mbox{dim}E_{s}\left(x\right)=d_{s}, & \quad\mbox{dim}E_{s}^{*}\left(x\right)=\mbox{dim}E_{u}\left(x\right)=d_{u}.\end{align*}

\item From (\ref{eq:def_alpha}) and (\ref{eq:dual_decomp}), one deduces
that $E_{0}^{*}$ is spanned by the one form $\alpha$: \begin{equation}
E_{0}^{*}\left(x\right)=\mathbb{R}\,\alpha\left(x\right).\label{eq:alpha_E0*}\end{equation}

\item Notice that since $E_{u}\left(x\right)\oplus E_{s}\left(x\right)$
is not smooth in general (see remarks page \pageref{par:Remarks:_smoothness}),
the same holds for $E_{0}^{*}\left(x\right)$. However $\left(E_{u}^{*}\left(x\right)\oplus E_{s}^{*}\left(x\right)\right)$
is smooth since $E_{0}\left(x\right)$ is smooth.
\end{itemize}
\selectlanguage{french}%
\vspace{0.cm}\begin{center}{\color{red}\fbox{\color{black}\parbox{16cm}{
\selectlanguage{english}%
\begin{defn}
For $E\in\mathbb{R}$, let\begin{equation}
\Sigma_{E}:=H_{0}^{-1}\left(E\right)\subset T^{*}X\label{eq:def_Sigma_E}\end{equation}
be the \textbf{{}``energy shell''} (where the function $H_{0}\left(x,\xi\right)=V\left(\xi\right)$
has been defined in (\ref{eq:def_H})).
\end{defn}
\selectlanguage{french}%
}}}\end{center}\vspace{0.cm}

\selectlanguage{english}%
$\Sigma_{E}$ is a smooth hyper-surface in $T^{*}X$. More precisely
for every $x\in X$, \[
\Sigma_{E}\left(x\right):=\Sigma_{E}\cap T_{x}^{*}X\]
is an affine hyperplane in $T_{x}^{*}X$, parallel%
\footnote{Since $E_{0}\left(E_{u}^{*}\oplus E_{s}^{*}\right)=0$ and $V\in E_{0}$
then \[
H_{0}\left(E_{u}^{*}\left(x\right)\oplus E_{s}^{*}\left(x\right)\right)=V\left(E_{u}^{*}\left(x\right)\oplus E_{s}^{*}\left(x\right)\right)=0.\]
} to $E_{u}^{*}\left(x\right)\oplus E_{s}^{*}\left(x\right)$ and therefore
transverse to $E_{0}^{*}\left(x\right)$. See Figure \ref{fig:T*x}.

\begin{center}
\begin{figure}
\begin{centering}
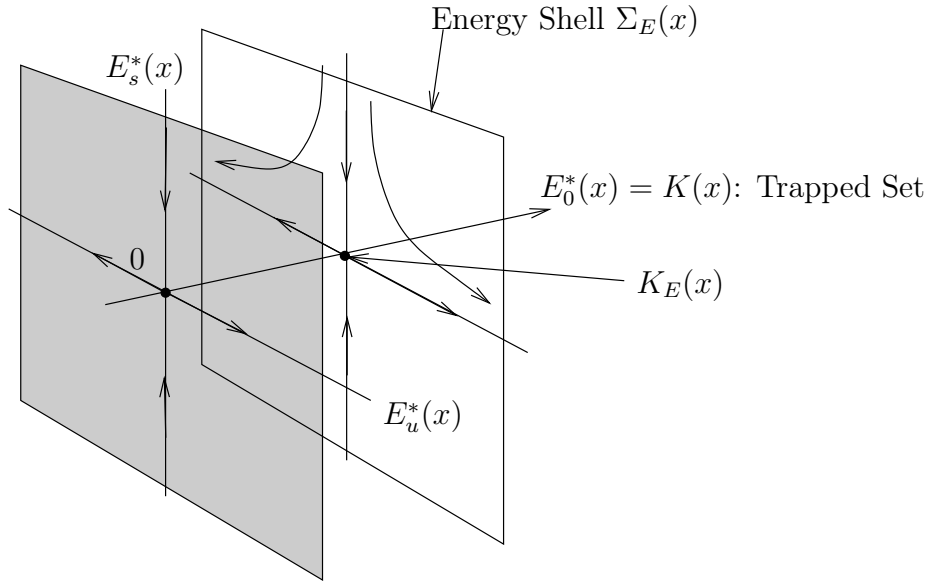
\par\end{centering}

\caption{\label{fig:T*x}Picture of $T_{x}^{*}X$ for a given $x\in X$, and
a given $E\in\mathbb{R}$.}

\end{figure}

\par\end{center}

\vspace{0.cm}\begin{center}{\color{blue}\fbox{\color{black}\parbox{16cm}{
\begin{prop}

The decomposition (\ref{eq:decomp_T*X}) is invariant by the flow
$M_{t}$ and there exists $c>0,\theta>0$ such that \begin{eqnarray}
\left|M_{t}\left(\xi_{s}\right)\right| & \leq & ce^{-\theta t}\left|\xi_{s}\right|,\qquad\forall\xi_{s}\in E_{s}^{*},\quad\forall t\geq0,\label{eq:def_dynamics_T*X}\\
\left|M_{t}\left(\xi_{u}\right)\right| & \leq & ce^{-\theta\left|t\right|}\left|\xi_{u}\right|,\qquad\forall\xi_{u}\in E_{u}^{*},\quad\forall t\leq0.\nonumber \end{eqnarray}

For every $E\in\mathbb{R}$, the energy shell $\Sigma_{E}$ is invariant
by the flow $M_{t}$. In the energy shell $\Sigma_{E}$ the \textbf{trapped
set $K_{E}$} is defined by \begin{equation}
K_{E}:=\left\{ K_{E}\left(x\right)=\left(\Sigma_{E}\left(x\right)\cap E_{0}^{*}\left(x\right)\right),\qquad x\in X\right\} .\label{eq:def_KE}\end{equation}
$K_{E}$ is a global continuous section of the cotangent bundle $T^{*}X$
given in terms of the associated one form (\ref{eq:def_alpha}) by:\[
K_{E}=E\alpha.\]
In general $K_{E}\left(x\right)$ is not smooth but only Hölder continuous
(as $E_{0}^{*}\left(x\right)$). $K_{E}$ is globally invariant under
the flow $M_{t}$.

\end{prop}
}}}\end{center}\vspace{0.cm}
\begin{proof}
By duality with what happens in $TX$ described in (\ref{eq:def_dynamics}).
Since $\alpha\left(V\right)=1$ then $H_{0}\left(\alpha\right)=V\left(\alpha\right)=1$.
Therefore $\alpha\in\Sigma_{E=1}$. Also $\alpha\in E_{0}^{*}$ therefore
$\alpha=K_{E=1}$.
\end{proof}

\paragraph{Remarks:}
\begin{itemize}
\item In general the \textbf{trapped set} $K$ is defined by:\[
K:=\left\{ \left(x,\xi\right)\in T^{*}X,\quad\exists C>0;\forall t\in\mathbb{R},\left|M_{t}\left(x,\xi\right)\right|\leq C\right\} ,\]
i.e. $K$ contains trajectories which do not escape towards infinity
neither in the future nor in the past. We have:\[
K=\bigcup_{E\in\mathbb{R}}K_{E}=E_{0}^{*}.\]

\item Notice that for every $E\in\mathbb{R}$ the trapped set $K_{E}$ is
a sub-manifold of $T^{*}X$ homeomorphic to $X$ hence compact. This
observation is at the origin of the method to prove the existence
of discrete resonance spectrum below (Theorem \ref{th:discrete_spectrum}).
The dynamics of $M_{t}$ restricted to $K_{E}$ is conjugated to the
dynamics of $\phi_{t}$ on $X$ (it is a lift of $\phi_{t}$ on $K_{E}$).
\item For the special case of a\textbf{ contact flow} on $X$ with a $C^{\infty}$
contact 1 form $\alpha$ then $K_{E=1}=\alpha$ and therefore $K_{E}=E\alpha$
is a $C^{\infty}$ section. The restriction of the canonical two form
(\pageref{eq:omega_two_form}) to this section (seen as a submanifold
of $T^{*}X$) is%
\footnote{proof: for a contact flow with contact 1-form $\alpha$ then $K_{E}=E\alpha$,
with $E\in\mathbb{R}$. The restriction of the canonical symplectic
1-form $\eta:=\xi dx$ is then $\eta_{/K_{E}}=\left(\pi^{*}\left(K_{E}\right)\right)_{/K_{E}}=E\left(\pi^{*}\left(\alpha\right)\right)_{/K_{E}}$,
therefore $\omega_{/K_{E}}=d\eta_{/K_{E}}=E\left(\pi^{*}\left(d\alpha\right)\right)_{/K_{E}}$.%
} \begin{equation}
\omega_{/K_{E}}=E\left(\pi^{*}\left(d\alpha\right)\right)_{/K_{E}},\label{eq:omega_contact}\end{equation}
where $\pi:T^{*}X\rightarrow X$ is the bundle projection. We observe
that $K\backslash\left\{ \xi=0\right\} $ is a smooth symplectic submanifold
of $T^{*}X$ (far from being a Lagrangian submanifold of $T^{*}X$),
$\left(K_{E},E\xi dx\right)$ is a contact manifold for $E\neq0$
isomorphic to $\left(X,E\alpha\right)$.
\end{itemize}

\subsection{The escape function}

\begin{figure}
\begin{centering}
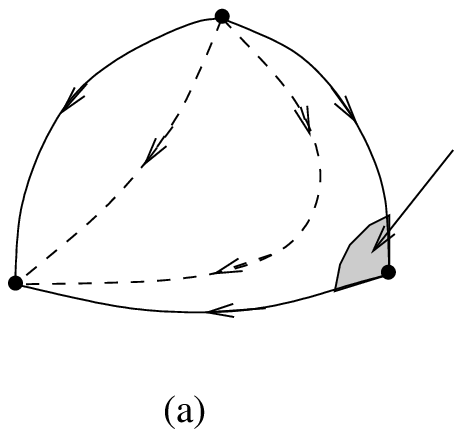$\qquad\qquad$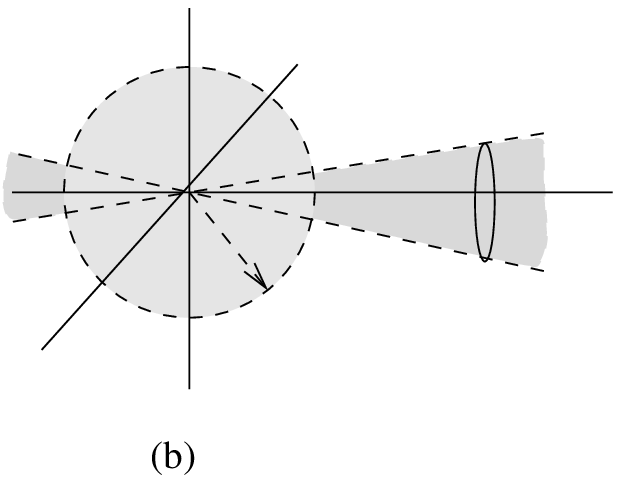
\par\end{centering}

\caption{\label{fig:Set-N0}(a) The induced flow $\widetilde{M}_{t}$ on the
cosphere bundle $S^{*}X:=\left(T^{*}X\backslash\left\{ 0\right\} \right)/\mathbb{R}^{+}$
which is the bundle of directions of cotangent vectors $\xi/\left|\xi\right|$.
(Here the picture is restricted to a fiber $S_{x}^{*}X$).\protect \\
(b) Picture in the cotangent space $T_{x}^{*}X$ which shows in
grey the sets outside of which the escape estimate (\ref{eq:escape_estimate})
holds. }

\end{figure}

In this section we construct a smooth function $G_{m}$ on the cotangent
space $T^{*}X$ called the escape function. We will denote $\frac{\xi}{\left|\xi\right|}$
the direction of a cotangent vector $\xi$ and $S^{*}X:=\left(T^{*}X\backslash\left\{ 0\right\} \right)/\mathbb{R}^{+}$
the \textbf{cosphere bundle} which is the bundle of directions of
cotangent vectors $\xi/\left|\xi\right|$. $S^{*}X$ is a compact
space. The images of $E_{u}^{*},E_{s}^{*},E_{0}^{*}\subset T^{*}X$
by the projection $T^{*}X\backslash\left\{ 0\right\} \rightarrow S^{*}X$
are denoted respectively $\widetilde{E}_{u}^{*},\widetilde{E}_{s}^{*},\widetilde{E}_{0}^{*}\subset S^{*}X$,
see Figure \ref{fig:Set-N0}(a) page \pageref{fig:Set-N0}.

\vspace{0.cm}\begin{center}{\color{blue}\fbox{\color{black}\parbox{16cm}{
\begin{lem}
\label{lem:def_A_s_u}Let $u,n_{0},s\in\mathbb{R}$ with $u<n_{0}<s$.
There exists a smooth function $m\left(x,\xi\right)\in C^{\infty}\left(T^{*}X\right)$
called an \textbf{{}``order function'', }taking values in the interval
$\left[u,s\right]$, and an \textbf{{}``escape function''} on $T^{*}X$
defined by:\begin{equation}
G_{m}\left(x,\xi\right):=m\left(x,\xi\right)\log\sqrt{1+\left(f\left(x,\xi\right)\right)^{2}},\label{eq:def_Gm}\end{equation}
where $f\in C^{\infty}\left(T^{*}X\right)$ and for $\left|\xi\right|\geq1$,
$f>0$ is positively homogeneous of degree 1 in $\xi$. $f\left(x,\xi\right)=\left|\xi\right|$
in a conical neighborhood of $E_{u}^{*}$ and $E_{s}^{*}$. $f\left(x,\xi\right)=H_{0}\left(x,\xi\right)$
in a conical neighborhood of $E_{0}^{*}$, such that:
\begin{enumerate}
\item For $\left|\xi\right|\geq1$, $m\left(x,\xi\right)$ depends only
on the direction $\frac{\xi}{\left|\xi\right|}\in S_{x}^{*}X$ and
takes the value $u$ (respect. $n_{0},s$) in a small neighborhood
of $E_{u}^{*}$ (respect. $E_{0}^{*},E_{s}^{*}$). See figure \ref{fig:Set-N0}(a).
\item $G_{m}$ decreases strictly and uniformly along the trajectories of
the flow $M_{t}$ in the cotangent space, except in a conical vicinity
$\widetilde{N}_{0}$ of the neutral direction $E_{0}^{*}$ and for
small $\left|\xi\right|$: there exists $R>0$ such that\begin{equation}
\forall x\in X,\,\forall\left|\xi\right|\geq R,\,\frac{\xi}{\left|\xi\right|}\notin\widetilde{N}_{0}\qquad\mathbf{X}\left(G_{m}\right)\left(x,\xi\right)<-C_{m}<0,\label{eq:escape_estimate}\end{equation}
with\begin{equation}
C_{m}:=c\min\left(\left|u\right|,s\right)\label{eq:def_C_u_s}\end{equation}
and $c>0$ independent of $u,n_{0},s$.
\item More generally \begin{equation}
\forall x\in X,\,\forall\left|\xi\right|\geq R,\qquad\mathbf{X}\left(G_{m}\right)\left(x,\xi\right)\leq0.\label{eq:XG_negative}\end{equation}
See figure \ref{fig:Set-N0}(b).
\end{enumerate}
\end{lem}
}}}\end{center}\vspace{0.cm}

\paragraph{Remarks\label{rem:C_m}}
\begin{itemize}
\item It is important to notice that we can choose $m$ such that the value
of \textbf{$C_{m}$ is arbitrarily large }(by making $s,\left|u\right|\rightarrow\infty$)
and that the neighborhood \textbf{$\widetilde{N}_{0}$ is arbitrarily
small}.
\item The value of $n_{0}$ could be chosen to be $n_{0}=0$ to simplify.
But it is interesting to observe that letting $n_{0},s\rightarrow+\infty$,
the order function $m\left(x,\xi\right)$ can be made arbitrarily
large for $\left|\xi\right|\geq1$, outside a small vicinity of $E_{u}^{*}$.
We will use this in the proof of Theorem \ref{thm:The-eigenvalues-are_intrinsec}
in order to show that the wavefront of the eigen-distributions are
included in $E_{u}^{*}$.
\item Inspection of the proof shows that with an adapted norm $\left|\xi\right|$
obtained by averaging, $c$ can be chosen arbitrarily close to $\theta$,
defined in (\ref{eq:def_dynamics_T*X}).
\item The constancy of $m$ in the vicinity of the stable/unstable/neutral
directions allows us to have a smooth escape function $G_{m}$ although
the distributions $E_{s}^{*}\left(x\right)$,$E_{u}^{*}\left(x\right)$,$E_{0}^{*}\left(x\right)$
have only Hölder regularity in general.
\end{itemize}

\subsubsection{Proof of Lemma \ref{lem:def_A_s_u}}

We first define a function $m\left(x,\xi\right)$ called the \textbf{order
function }following closely \cite{fred-roy-sjostrand-07} Section
3.1 (and \cite{gerard_88} p.196).

\subparagraph{The function $m$.}

The following Lemma is useful for the construction of escape functions.
Let $M$ be a compact manifold and let $v$ be a smooth vector field
on $M$. We denote $\mathrm{exp}\left(tv\right):M\rightarrow M$ the
flow at time $t$ generated by $v$. Let $\Sigma_{u}$, $\Sigma_{s}$
be compact disjoint subsets of $M$ such that \begin{eqnarray*}
 &  & \mathrm{dist\,}(\exp\left(tv\right)(\rho),\Sigma_{s})\to0,\ t\to+\infty\mbox{ when }\rho\in M\setminus\Sigma_{u}\\
 &  & \mathrm{dist\,}(\exp\left(tv\right)(\rho),\Sigma_{u})\to0,\ t\to-\infty\mbox{ when }\rho\in M\setminus\Sigma_{s}.\end{eqnarray*}

\begin{figure}
\begin{centering}
\includegraphics[scale=0.6]{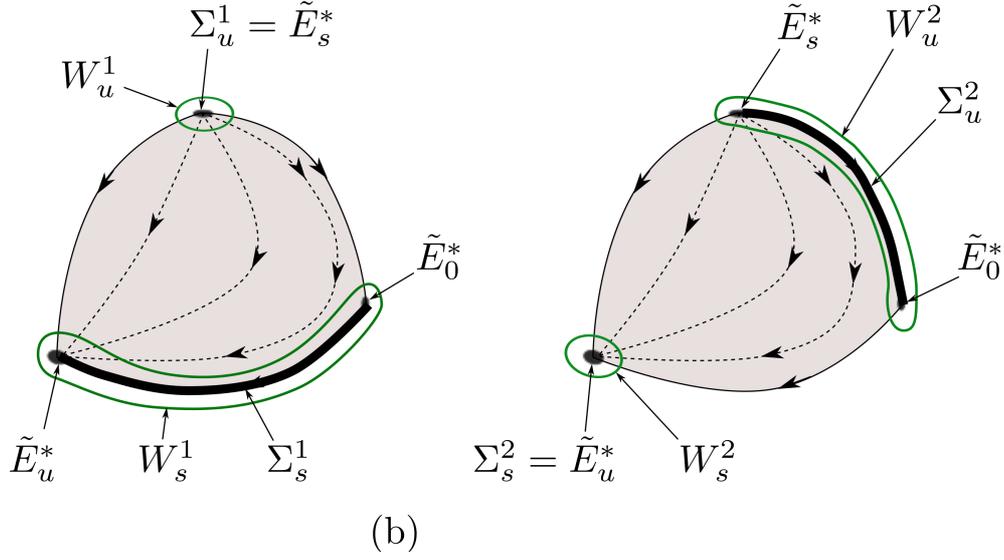}
\par\end{centering}

\caption{\label{fig:schema_m-1}Representation of different sets on $S^{*}X$
used in the proof.}

\end{figure}

\selectlanguage{french}%
\vspace{0.cm}\begin{center}{\color{blue}\fbox{\color{black}\parbox{16cm}{
\selectlanguage{english}%
\begin{lem}
\noindent \label{lem:escape_lemma}Let $V_{u},V_{s}\subset M$ be
open neighborhoods of $\Sigma_{u}$ and $\Sigma_{s}$ respectively
and let $\varepsilon>0$. Then there exist $W_{u}\subset V_{u}$,
$W_{s}\subset V_{s}$, $m\in C^{\infty}(M;[0,1])$, $\eta>0$ such
that $v(m)\ge0$ on $M$, $v(m)>\eta>0$ on $M\setminus\left(W_{u}\cup W_{s}\right)$,
$m\left(\rho\right)>1-\varepsilon$ for $\rho\in W_{s}$ and $m\left(\rho\right)<\varepsilon$
for $\rho\in W_{u}$.
\end{lem}
\selectlanguage{french}%
}}}\end{center}\vspace{0.cm}
\selectlanguage{english}%
\begin{proof}
\noindent After shrinking $V_{u},V_{s}$ we may assume that $V_{u}\cap V_{s}=\emptyset$
and \begin{equation}
t\geq0\Rightarrow\exp\left(tv\right)(V_{s})\subset V_{s},\mbox{ and }t\leq0\Rightarrow\exp\left(tv\right)(V_{u})\subset V_{u}.\label{eq:esc_1}\end{equation}
Let $T>0$ and let $W_{s}:=M\backslash\mathrm{exp}\left(Tv\right)\left(V_{u}\right)=\mathrm{exp}\left(Tv\right)\left(M\backslash V_{u}\right)$
and $W_{u}:=M\backslash\mathrm{exp}\left(-Tv\right)\left(V_{s}\right)=\mathrm{exp}\left(-Tv\right)\left(M\backslash V_{s}\right)$.
See figure \ref{fig:schema_m}.

\begin{figure}
\begin{centering}
\includegraphics[scale=0.6]{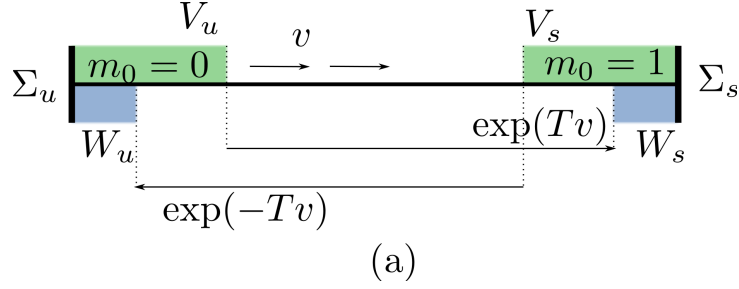}\\

\par\end{centering}

\caption{\label{fig:schema_m}Illustration for the proof of Lemma \ref{lem:escape_lemma}.
The horizontal axis is a schematic picture of $M$ and this shows
the construction and properties of the sets $V_{u},V_{s}$ and $W_{u},W_{s}$.}

\end{figure}

\noindent If $T$ is large enough one has $W_{u}\subset V_{u}$, $W_{s}\subset V_{s}$
and $W_{s}\cap W_{u}=\emptyset$. Let $m_{0}\in C^{\infty}(M;[0,1])$
be equal to 1 on $V_{s}$ and equal to 0 on $V_{u}$. Put \begin{equation}
m=\frac{1}{2T}\int_{-T}^{T}m_{0}\circ\exp\left(tv\right)\, dt.\label{esc.2}\end{equation}
 Then \begin{equation}
v(m)(\rho)=\frac{1}{2T}\left(m_{0}\left(\exp\left(Tv\right)\left(\rho\right)\right)-m_{0}\left(\exp\left(-Tv\right)\left(\rho\right)\right)\right).\label{esc.3}\end{equation}

\begin{itemize}
\item \noindent Let $\rho\in M\setminus\left(W_{u}\cup W_{s}\right)$. From
(\ref{esc.3}) we see that $v\left(m\right)\left(\rho\right)=\frac{1}{2T}(1-0))=\frac{1}{2T}>0$.
\end{itemize}
\noindent For $\rho\in M$ let\[
\mathcal{I}\left(\rho\right):=\left\{ t\in\mathbb{R},\mathrm{exp}\left(tv\right)\left(\rho\right)\in M\backslash\left(V_{u}\cup V_{s}\right)\right\} .\]

\noindent This is a closed connected interval by (\ref{eq:esc_1})
and moreover its length is uniformly bounded: \[
\exists\tau>0,\quad\forall\rho\in M,\quad\left|\mathrm{max}\left(\mathcal{I}\left(\rho\right)\right)-\mathrm{min}\left(\mathcal{I}\left(\rho\right)\right)\right|\leq\tau.\]
In other words, $\tau$ is an upper bound for the travel time in the
domain $M\backslash\left(V_{u}\cup V_{s}\right)$.

\noindent To prove the Lemma, we have to consider two more cases:
\begin{itemize}
\item \noindent Let $\rho\in W_{u}$. If $t\leq T-\tau$ then $m_{0}\left(\mbox{exp}\left(tv\right)\left(\rho\right)\right)=0$
and\[
m\left(\rho\right)=\frac{1}{2T}\left(\int_{-T}^{T-\tau}\underbrace{m_{0}\left(\exp\left(tv\right)\rho\right)}_{=0}dt+\int_{T-\tau}^{T}\underbrace{m_{0}\left(\exp\left(tv\right)\rho\right)}_{\leq1}dt\right)\leq\frac{\tau}{2T}<\varepsilon,\]
where the last inequality holds if one chooses $T$ large enough.
One has $m_{0}\left(\exp\left(-Tv\right)\left(\rho\right)\right)=0$
therefore (\ref{esc.3}) implies that $v(m)(\rho)\ge0$.
\item \noindent Let $\rho\in W_{s}$. One shows similarly that\[
m\left(\rho\right)=\frac{1}{2T}\left(\int_{-T}^{-T+\tau}\underbrace{m_{0}\left(\exp\left(tv\right)\rho\right)}_{\geq0}dt+\int_{-T+\tau}^{T}\underbrace{m_{0}\left(\exp\left(tv\right)\rho\right)}_{=1}dt\right)\geq\frac{2T-\tau}{2T}>1-\varepsilon,\]
for $T$ large enough, and $v(m)(\rho)\ge0$.
\end{itemize}
\end{proof}
We now apply Lemma \ref{lem:escape_lemma} to the case when $M=S^{*}X$
and $v$ is the image $\widetilde{\mathbf{X}}$ on $S^{*}X$ of our
Hamilton field $\mathbf{X}$. See figure \ref{fig:schema_m-1}.
\begin{itemize}
\item We first take $\Sigma_{u}=\Sigma_{u}^{1}=\widetilde{E}_{s}^{*}$ and
let $\Sigma_{s}=\Sigma_{s}^{1}\subset M$ be the set of limit points
$\lim_{j\to+\infty}\exp t_{j}v(\rho)$, where $\rho\in M\setminus\Sigma_{u}^{1}$
and $t_{j}\to+\infty$. $\Sigma_{s}^{1}$ is the union of $\widetilde{E}_{0}^{*}$,
$\widetilde{E}_{u}^{*}$ and all trajectories $\exp(\mathbb{R}v)(\rho)$
where $\rho$ has the property that $\exp tv(\rho)$ converges to
$\widetilde{E}_{0}^{*}$ when $t\to-\infty$ and to $\widetilde{E}_{u}^{*}$
when $t\to+\infty$. Equivalently, $\Sigma_{s}^{1}$ is the image
$\widetilde{E_{u}^{*}\oplus E_{0}^{*}}$ in $S^{*}X$ of $E_{u}^{*}\oplus E_{0}^{*}$.
Applying the Lemma, we get $m_{1}=m\in C^{\infty}(M;[0,1])$ such
that $m_{1}<\varepsilon$ outside an arbitrarily small neighborhood
$W_{u}^{1}$ of $\Sigma_{u}^{1}=\widetilde{E}_{s}^{*}$, $m_{1}>1-\varepsilon$
outside an arbitrarily small neighborhood $W_{s}^{1}$ of $\Sigma_{s}^{1}=\widetilde{E_{u}^{*}\oplus E_{0}^{*}}$
and $\widetilde{\mathbf{X}}(m_{1})\ge0$ everywhere with strict inequality
$\widetilde{\mathbf{X}}(m_{1})>\eta>0$ outside $W_{s}^{1}\cup W_{u}^{1}$.
\item Similarly, we can find $m_{2}=m\in C^{\infty}(M;[0,1])$, such that
$m_{2}<\varepsilon$ outside an arbitrarily small neighborhood $W_{u}^{2}$
of $\Sigma_{u}^{2}=\widetilde{E_{s}\oplus E_{0}^{*}}$, $m_{2}>1-\varepsilon$
outside an arbitrarily small neighborhood $W_{s}^{2}$ of $\Sigma_{s}^{2}=\widetilde{E}_{u}^{*}$
and $\widetilde{\mathbf{X}}(m_{2})\ge0$ everywhere with strict inequality
$\widetilde{\mathbf{X}}(m_{2})>\eta>0$ outside $W_{s}^{2}\cup W_{u}^{2}$.
\end{itemize}

Let $u<n_{0}<s$ and put \[
\widetilde{m}:=s+\left(n_{0}-s\right)m_{1}+\left(u-n_{0}\right)m_{2},\]
\[
\widetilde{N}_{s}:=W_{u}^{1}\cap W_{u}^{2},\quad\widetilde{N}_{0}:=W_{s}^{1}\cap W_{u}^{2},\quad\widetilde{N}_{u}:=W_{s}^{1}\cap W_{s}^{2}.\]

Then
\begin{itemize}
\item on $S^{*}X\backslash\left(\widetilde{N}_{s}\cup\widetilde{N}_{0}\cup\widetilde{N}_{u}\right)=\left(S^{*}X\backslash\left(W_{u}^{1}\cup W_{s}^{1}\right)\right)\cup\left(S^{*}X\backslash\left(W_{u}^{2}\cup W_{s}^{2}\right)\right)$
we have $\widetilde{\mathbf{X}}\left(m_{1}\right)>\eta$ or $\widetilde{\mathbf{X}}\left(m_{2}\right)>\eta$
therefore \begin{align}
\widetilde{\mathbf{X}}\left(\widetilde{m}\right) & =\left(n_{0}-s\right)\widetilde{\mathbf{X}}\left(m_{1}\right)+\left(u-n_{0}\right)\widetilde{\mathbf{X}}\left(m_{2}\right)\label{eq:est_1}\\
 & <-\eta\min\left(\left|n_{0}-s\right|,\left|u-n_{0}\right|\right).\nonumber \end{align}

\item on $\widetilde{N}_{s}=W_{u}^{1}\cap W_{u}^{2}$ we have $m_{1}<\varepsilon$
and $m_{2}<\varepsilon$ therefore\begin{align}
\widetilde{m} & >s+\left(n_{0}-s\right)\varepsilon+\left(u-n_{0}\right)\varepsilon\nonumber \\
 & =s\left(1-\varepsilon\right)+u\varepsilon>\frac{s}{2},\label{eq:est_2}\end{align}
where the last inequality holds if $\varepsilon$ is chosen small
enough.
\item on $\widetilde{N}_{u}=W_{s}^{1}\cap W_{s}^{2}$ we have $m_{1}>1-\varepsilon$
and $m_{2}>1-\varepsilon$ therefore\begin{align}
\widetilde{m} & <s+\left(n_{0}-s\right)\left(1-\varepsilon\right)+\left(u-n_{0}\right)\left(1-\varepsilon\right)\nonumber \\
 & =\varepsilon s+u\left(1-\varepsilon\right)<\frac{u}{2},\label{eq:est_3}\end{align}
where the last inequality holds if $\varepsilon$ is chosen small
enough.
\item on $S^{*}X$ we have \begin{equation}
\widetilde{\mathbf{X}}\left(\widetilde{m}\right)=\left(n_{0}-s\right)\widetilde{\mathbf{X}}\left(m_{1}\right)+\left(u-n_{0}\right)\widetilde{\mathbf{X}}\left(m_{2}\right)\leq0.\label{eq:est_4}\end{equation}

\end{itemize}
We construct a smooth function $m$ on $T^{*}M$ satisfying\begin{eqnarray*}
m\left(x,\xi\right) & = & \widetilde{m}\left(\frac{\xi}{\left|\xi\right|}\right),\qquad\mbox{if }\left|\xi\right|\geq1,\\
 & = & 0\qquad\mbox{if }\left|\xi\right|\leq1/2.\end{eqnarray*}

\subparagraph{The symbol $G_{m}$.}

Let \[
G_{m}\left(x,\xi\right):=m\left(x,\xi\right)\log\sqrt{1+\left(f\left(x,\xi\right)\right)^{2}}\]
with $f\in C^{\infty}\left(T^{*}X\right)$ such that for $\left|\xi\right|\geq1$,
$f>0$ is positively homogeneous of degree 1 in $\xi$, and \[
\frac{\xi}{\left|\xi\right|}\in\widetilde{N}_{u}\cup\widetilde{N}_{s}\Rightarrow f\left(x,\xi\right):=\left|\xi\right|\]
\[
\frac{\xi}{\left|\xi\right|}\in\widetilde{N}_{0}\Rightarrow f\left(x,\xi\right):=H_{0}\left(x,\xi\right).\]

The consequences of these choices are: 
\begin{itemize}
\item Since $\mathbf{X}\left(H_{0}\right)=0$ then $\mathbf{X}\left(\log\sqrt{1+\left(f\left(x,\xi\right)\right)^{2}}\right)=0$
for $\frac{\xi}{\left|\xi\right|}\in\widetilde{N}_{0}$.
\item Since $E_{s}^{*}$ is the stable direction and $E_{u}^{*}$ the unstable
one,\begin{equation}
\exists C>0,\qquad\frac{\xi}{\left|\xi\right|}\in\widetilde{N}_{s}\Rightarrow\mathbf{X}\left(\log\left\langle \xi\right\rangle \right)<-C,\qquad\frac{\xi}{\left|\xi\right|}\in\widetilde{N}_{u}\Rightarrow\mathbf{X}\left(\log\left\langle \xi\right\rangle \right)>C.\label{eq:r-6}\end{equation}
Notice that by averaging, the norm $\left|\xi\right|$ can be chosen
such that for $\left|\xi\right|$ large enough, $C$ is arbitrarily
closed to $\theta$ defined in (\ref{eq:def_dynamics_T*X}).
\item In general $\left|\mathbf{X}\left(\log\sqrt{1+f^{2}}\right)\right|$
is bounded: \[
\exists C_{2}>0,\qquad\forall\xi\in T^{*}X,\quad\left|\mathbf{X}\left(\log\sqrt{1+\left(f\left(x,\xi\right)\right)^{2}}\right)\right|<C_{2}.\]

\end{itemize}
We will show now the uniform escape estimate Eq.(\ref{eq:escape_estimate})
page (\pageref{eq:escape_estimate}). One has \begin{equation}
\mathbf{X}\left(G_{m}\right)=\mathbf{X}\left(m\right)\log\sqrt{1+f^{2}}+m\mathbf{X}\left(\log\sqrt{1+f^{2}}\right).\label{eq:X_G}\end{equation}

We will first consider each term separately assuming $\left|\xi\right|\geq1$.
\begin{itemize}
\item If $\widetilde{\xi}\in S^{*}X\setminus\left(\widetilde{N}_{s}\cup\widetilde{N}_{u}\cup\widetilde{N}_{0}\right)$
then using (\ref{eq:est_1}) and the fact that $\left|\mathbf{X}\left(\log\sqrt{1+f^{2}}\right)\right|$
and $m$ are bounded, one has for $\left|\xi\right|$ large enough
\[
\mathbf{X}\left(G_{m}\right)\left(x,\xi\right)<-c\min\left(s,\left|u\right|\right)\]
with $c>0$ independent of $u,n_{0},s$.
\item If $\widetilde{\xi}\in\widetilde{N}_{u}$ then from (\ref{eq:r-6})
and (\ref{eq:est_3}) there exists $c>0$ such that \[
\mathbf{X}\left(G_{m}\right)=\underbrace{\mathbf{X}\left(m\right)}_{\leq0}\underbrace{\log\left\langle \xi\right\rangle }_{\geq0}+\underbrace{m}_{<\frac{u}{2}}\underbrace{\mathbf{X}\left(\log\left\langle \xi\right\rangle \right)}_{>C}<-c\left|u\right|<0.\]

\item If $\widetilde{\xi}\in\widetilde{N}_{s}$ then from (\ref{eq:r-6})
and (\ref{eq:est_2}) there exists $c>0$ such that \[
\mathbf{X}\left(G_{m}\right)=\underbrace{\mathbf{X}\left(m\right)}_{\leq0}\underbrace{\log\left\langle \xi\right\rangle }_{\geq0}+\underbrace{m}_{>\frac{s}{2}}\underbrace{\mathbf{X}\left(\log\left\langle \xi\right\rangle \right)}_{<-C}<-cs<0.\]

\end{itemize}
We have obtained the uniform escape estimate Eq.(\ref{eq:escape_estimate})
page \pageref{eq:escape_estimate}. Finally for $\widetilde{\xi}\in\widetilde{N}_{0}$,
we have \[
\mathbf{X}\left(G_{m}\right)=\underbrace{\mathbf{X}\left(m\right)}_{\leq0}\underbrace{\log\sqrt{1+f^{2}}}_{\geq0}+m\underbrace{\mathbf{X}\left(\log\sqrt{1+f^{2}}\right)}_{=0}\leq0,\]
and we deduce (\ref{eq:XG_negative}) page \pageref{eq:XG_negative}.
We have finished the proof of Lemma \ref{lem:def_A_s_u} page \pageref{lem:def_A_s_u}.\texttt{ }

\section{Spectrum of resonances}

In this Section we give our main results about the spectrum of the
generator $\widehat{H}$, Eq(\ref{eq:def_H_hat}), in specific Sobolev
spaces. We first define these Sobolev spaces.

\subsection{Anisotropic Sobolev spaces\label{sub:Anisotropic-Sobolev-spaces}}

\subsubsection{\label{sub:Symbol-classes-with}Symbol classes with variable order}

The escape function $G_{m}$ defined in Lemma \ref{lem:def_A_s_u}
has some regularity expressed by the fact that it belongs to some
symbol classes $S^{\mu}$. This will allow us to perform some semiclassical
calculus. In this section, we describe these symbol classes.

\selectlanguage{french}%
\vspace{0.cm}\begin{center}{\color{blue}\fbox{\color{black}\parbox{16cm}{
\selectlanguage{english}%
\begin{lem}
\label{lem:The-order-function_S0}The order function $m\left(x,\xi\right)$
defined in in Lemma \ref{lem:def_A_s_u} belongs to $S^{0}$ ( definition
\ref{def:The-symbol-class_S_mu_ro} page \pageref{def:The-symbol-class_S_mu_ro}).
The escape function $G_{m}$ defined in (\ref{eq:def_Gm}) belongs
to the symbol class $S^{\mu}$ for every $\mu>0$. For short, we will
write $G_{m}\in S^{+0}$.
\end{lem}
\selectlanguage{french}%
}}}\end{center}\vspace{0.cm}

\selectlanguage{english}%
In the paper \cite[Appendix]{fred-roy-sjostrand-07} we have shown
that the order function $m\left(x,\xi\right)\in S^{0}$ can be used
to define the class $S_{\rho}^{m\left(x,\xi\right)}$ of symbols of
variable order $m\left(x,\xi\right)$. We recall the definition:

\selectlanguage{french}%
\vspace{0.cm}\begin{center}{\color{red}\fbox{\color{black}\parbox{16cm}{
\selectlanguage{english}%
\begin{defn}
\label{def_symbol_class_variable_order}Let $m\left(x,\xi\right)\in S^{0}$
and $\frac{1}{2}<\rho\leq1$. A function $p\in C^{\infty}\left(T^{*}X\right)$
belongs to the \textbf{class $S_{\rho}^{m\left(x,\xi\right)}$ of
variable order} if for every trivialization $\left(x,\xi\right):\left.T^{*}X\right|_{U}\rightarrow\mathbb{R}^{2n}$,
for every compact $K\subset U$ and all multi-indices $\alpha,\beta\in\mathbb{N}^{n}$,
there is a constant $C_{K,\alpha,\beta}$ such that \begin{equation}
\left|\partial_{\xi}^{\alpha}\partial_{x}^{\beta}p\left(x,\xi\right)\right|\leq C_{K,\alpha,\beta}\left\langle \xi\right\rangle ^{m\left(x,\xi\right)-\rho\left|\alpha\right|+\left(1-\rho\right)\left|\beta\right|}\label{eq:def_Class_Sm}\end{equation}
for every $\left(x,\xi\right)\in\left.T^{*}X\right|_{U}$.
\end{defn}
\selectlanguage{french}%
}}}\end{center}\vspace{0.cm}

\selectlanguage{english}%
We refer to \cite[Section A.2.2]{fred-roy-sjostrand-07} for a precise
description of semiclassical theorems related to symbols with variable
orders.

\selectlanguage{french}%
\vspace{0.cm}\begin{center}{\color{blue}\fbox{\color{black}\parbox{16cm}{
\selectlanguage{english}%
\begin{prop}
\label{pro:symbol_A_G-1}The operator \begin{equation}
\widehat{A}_{m}:=\mathrm{Op}\left(\exp\left(G_{m}\right)\right)\label{eq:def_Op_Am}\end{equation}
is a PDO whose symbol belongs to the class $S_{\rho}^{m\left(x,\xi\right)}$
for every $\rho<1$ (we write $S_{1-0}^{m\left(x,\xi\right)}$ for
short). Its principal symbol is \[
A_{m}\left(x,\xi\right)=e^{G_{m}\left(x,\xi\right)}=\left\langle \xi\right\rangle ^{m\left(x,\xi\right)}.\]
The symbol $A_{m}$ can be modified at a subleading order (i.e. $S_{\rho}^{m\left(x,\xi\right)-\left(2\rho-1\right)}$)
such that the operator becomes formally \textbf{self-adjoint} and
\textbf{invertible} on $C^{\infty}\left(X\right)$.
\end{prop}
\selectlanguage{french}%
}}}\end{center}\vspace{0.cm}

\selectlanguage{english}%
Remark: one can also show that $A_{m}\in S_{1}^{m\left(x,\xi\right)+0}$
but this is less precise than $A_{m}\in S_{1-0}^{m\left(x,\xi\right)}$.
\begin{proof}
We refer to the appendix in the paper \cite[Lemma 6]{fred-roy-sjostrand-07}.
\end{proof}

\subsubsection{Anisotropic Sobolev spaces}

For every order function $m$ as in Lemma \ref{lem:def_A_s_u}, we
define the \textbf{anisotropic Sobolev space} $H^{m}$ to be the space
of distributions (included in $\mathcal{D}'\left(X\right)$):\begin{equation}
\boxed{H^{m}:=\widehat{A}_{m}^{-1}\left(L^{2}\left(X\right)\right).}\label{eq:def_Hm}\end{equation}
 Some basic properties of the space $H^{m}$, such as embedding properties,
are given in \cite[section 3.2]{fred-roy-sjostrand-07}. 

The generator $\widehat{H}=-iV$ , Eq.(\ref{eq:def_H_hat}), is defined
by duality on the distribution space $\mathcal{D}'\left(X\right)$
and we can therefore consider its restriction to the anisotropic Sobolev
space $H^{m}$.

\subsection{Main results on the spectrum of Ruelle resonances}

The following theorem \ref{th:discrete_spectrum} has been obtained
in \cite[Theorem 1]{liverani_butterley_07} (with the slight difference
that the authors use Banach spaces) . In particular we refer to this
paper for results and discussions concerning the SRB measure. We provide
a new proof below, based on semiclassical analysis in the spirit of
the paper \cite{fred-roy-sjostrand-07}.

\vspace{0.cm}\begin{center}{\color{blue}\fbox{\color{black}\parbox{16cm}{
\begin{thm}
\label{th:discrete_spectrum} \textbf{{}``Discrete spectrum''.}
Let $m$ be a function which satisfies the hypothesis of Lemma \ref{lem:def_A_s_u}
page \pageref{lem:def_A_s_u}. The generator $\widehat{H}=-iV$ ,
Eq.(\ref{eq:def_H_hat}), defines by duality an unbounded operator
on the anisotropic Sobolev space $H^{m}$ , Eq.(\ref{eq:def_Hm}),\[
\widehat{H}:H^{m}\rightarrow H^{m}\]
in the sense of distributions with domain given by \[
\mathcal{D}\left(\widehat{H}\right):=\left\{ \varphi\in H^{m},\quad\widehat{H}\varphi\in H^{m}\right\} .\]

It coincides with the closure of $\left(-iV\right):C^{\infty}\rightarrow C^{\infty}$
in the graph norm for operators. For $z\in\mathbb{C}$ such that $\Im\left(z\right)>-\left(C_{m}-C\right)$
with $C_{m}$ defined in (\ref{eq:def_C_u_s}) and some $C$ independent
of $m$, the operator $\left(\widehat{H}-z\right):\mathcal{D}\left(\widehat{H}\right)\cap H^{m}\rightarrow H^{m}$
is a Fredholm operator with index $0$ depending analytically on $z$.
Recall that $C_{m}$ is arbitrarily large. Consequently the operator\textbf{
$\widehat{H}$ has a discrete spectrum} in the domain $\Im\left(z\right)>-\left(C_{m}-C\right)$,
consisting of eigenvalues $\lambda_{i}$ of finite algebraic multiplicity.
See Figure \ref{fig:Spectrum-of-Ruelle}. Moreover, $\widehat{H}$
has no spectrum in the half plane $\Im\left(z\right)>0$.
\end{thm}
\selectlanguage{french}%
}}}\end{center}\vspace{0.cm}

\selectlanguage{english}%
Concerning Fredholm operators we refer to \cite[p.122]{davies_07}
or \cite[Appendix A p.220]{sjostrand_87}. The proof of Theorem \ref{th:discrete_spectrum}
is given page \pageref{sub:Proof-of-theorem-1}. 

The next Theorem show that the spectrum is intrinsic and describes
the wavefront of the eigenfunctions associated to $\lambda_{i}$.
The wavefront of a distribution has been introduced by Hörmander.
See for instance \cite[p.77]{grigis_sjostrand} of \cite[p.27]{taylor_tome2}
for the definition. The wavefront corresponds to the directions in
$T^{*}X$ where the distribution is not $C^{\infty}$.

\vspace{0.cm}\begin{center}{\color{blue}\fbox{\color{black}\parbox{16cm}{
\begin{thm}
\label{thm:The-eigenvalues-are_intrinsec}\textbf{''The discrete
spectrum is intrinsic to the Anosov vector field''.} More precisely,
let $\widetilde{m},\widetilde{f}$, $\widetilde{G}_{m}=\widetilde{m}\log\sqrt{1+\widetilde{f}^{2}}$
be another set of functions as in Lemma \ref{lem:def_A_s_u} so that
Theorem \ref{th:discrete_spectrum} applies and $\widehat{H}:H^{\widetilde{m}}\rightarrow H^{\widetilde{m}}$
has discrete spectrum in the set $\Im\left(z\right)>-\left(\widetilde{C}_{\widetilde{m}}-\widetilde{C}\right)$.
Then in the set $\Im\left(z\right)>-\min\left(\left(C_{m}-C\right),\left(\widetilde{C}_{\widetilde{m}}-\widetilde{C}\right)\right)$
the eigenvalues of $\widehat{H}:H^{m}\rightarrow H^{m}$ counted with
their multiplicity and their respective eigenspaces coincide with
those of $\widehat{H}:H^{\widetilde{m}}\rightarrow H^{\widetilde{m}}$.

The eigenvalues $\lambda_{i}$ are called the \textbf{Ruelle Resonances}
and we denote the set by $\mathrm{Res}\left(\widehat{H}\right)$.

The \textbf{wavefront} of the associated generalized eigenfunctions
is contained in the unstable direction $E_{u}^{*}$.

The resolvent $\left(z-\widehat{H}\right)^{-1}$ viewed as an operator
$C^{\infty}\left(X\right)\rightarrow\mathcal{D}'\left(X\right)$ has
a meromorphic extension from $\Im\left(z\right)\gg1$ to $\mathbb{C}$.
The poles of this extension are the Ruelle resonances.
\end{thm}
\selectlanguage{french}%
}}}\end{center}\vspace{0.cm}

\selectlanguage{english}%
The proof of Theorem \ref{thm:The-eigenvalues-are_intrinsec} is given
page \pageref{sub:Proof-of-theorem_intrinsic}.

The following proposition is a very simple observation.

\vspace{0.cm}\begin{center}{\color{blue}\fbox{\color{black}\parbox{16cm}{
\begin{prop}
\textbf{\label{thm:Symmetry}''Symmetry''. }The order function $m$
can be chosen such that $m\left(x,-\xi\right)=m\left(x,\xi\right)$.
Then the conjugation operator $\widehat{C}$ defined in (\ref{eq:conjugation_operator})
leaves the space $H^{m}$ invariant. If $\widehat{H}\psi=\lambda\psi$,
$\psi\in H^{m}$ then $\widetilde{\psi}:=\widehat{C}\psi\in H^{m}$
is also an eigenfunction with eigenvalue $\widetilde{\lambda}=-\overline{\lambda}$.
The spectrum of Ruelle resonances is therefore symmetric with respect
to the imaginary axis.
\end{prop}
\selectlanguage{french}%
}}}\end{center}\vspace{0.cm}
\selectlanguage{english}%
\begin{proof}
of Proposition \ref{thm:Symmetry}. We first have to show that the
space $H^{m}\left(X\right)=\widehat{A}^{-1}\left(L^{2}\left(X\right)\right)$
is invariant by $\widehat{C}$, equivalently that $L^{2}\left(X\right)$
is invariant by $\widehat{A}\widehat{C}\widehat{A}^{-1}$. Notice
that $\widehat{C}$ is an {}``anti-linear FIO'' whose associated
transformation is $C:\left(x,\xi\right)\rightarrow\left(x,-\xi\right)$,
which is anti-canonical since $C^{*}\omega=-\omega$. The symbol $A\left(x,\xi\right)$
is invariant under the map $C:\left(x,\xi\right)\rightarrow\left(x,-\xi\right)$.
One can therefore construct $\widehat{A}$ such that $\widehat{C}\widehat{A}\widehat{C}=\widehat{A}$.
Since $\widehat{C}\widehat{A}\widehat{C}=\widehat{A}$$\Leftrightarrow\widehat{A}\widehat{C}\widehat{A}^{-1}=\widehat{C}$
and since the space $L^{2}\left(X\right)$ is invariant under $\widehat{C}$,
we conclude that $L^{2}\left(X\right)$ is invariant by $\widehat{A}\widehat{C}\widehat{A}^{-1}$.
Finally if $\widehat{H}\psi=\lambda\psi$, $\psi\in H^{m}$, let $\widetilde{\psi}=\widehat{C}\psi\in H^{m}$.
Then using (\ref{eq:conjugation_symetry}), $\widehat{H}\widetilde{\psi}=\widehat{H}\widehat{C}\psi=-\widehat{C}\widehat{H}\psi=-\overline{\lambda}\widetilde{\psi}$.
\end{proof}
Here is the new result of this paper:

\vspace{0.cm}\begin{center}{\color{blue}\fbox{\color{black}\parbox{16cm}{ 
\begin{thm}
\textbf{{}``Semiclassical upper bound for the density of resonances''}.\label{thm:Upper-bound-for-resonances}
For every $E\in\mathbb{R}\backslash\left\{ 0\right\} ,$ every $\beta>0$,
in the semiclassical limit $\alpha\rightarrow+\infty$ we have \begin{equation}
\sharp\left\{ \lambda\in\mathrm{Res}\left(\widehat{H}\right),\,\,\left|\Re\left(\lambda\right)-E\alpha\right|\leq\sqrt{\alpha},\,\,\Im\left(\lambda\right)>-\beta\right\} \leq o\left(\alpha^{n-1/2}\right),\label{eq:Upper_bound}\end{equation}
with $n=\dim X$.
\end{thm}
}}}\end{center}\vspace{0.cm}

\begin{center}
\begin{figure}
\begin{centering}
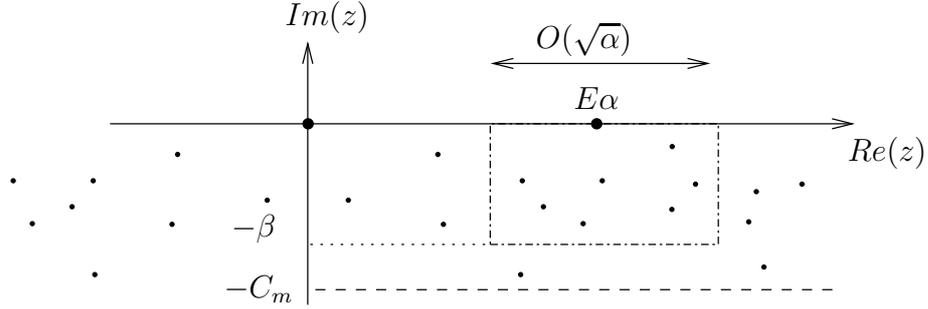
\par\end{centering}

\caption{\label{fig:Spectrum-of-Ruelle}Spectrum of Ruelle resonances of $\widehat{H}=-iV$.
From Theorem \ref{thm:Upper-bound-for-resonances} the number of eigenvalues
in the rectangle is $o\left(\alpha^{n-1/2}\right)$ for $\alpha\rightarrow\infty$. }

\end{figure}

\par\end{center}

\paragraph{Remarks:}
\begin{itemize}
\item Notice that by a simple scaling in $\alpha$ we can reduce the values
of $E$ to $E=\pm1$ in Theorem \ref{thm:Upper-bound-for-resonances}. 
\item The case $E=0$ is excluded in Theorem \ref{thm:Upper-bound-for-resonances}
because the vicinity of the origin $\xi=0$ is excluded in (\ref{eq:XG_negative}).
If one were able to construct an escape function such that in addition
$\mathbf{X}\left(G_{m}\right)\left(x,\xi\right)\leq C$, $\forall\left(x,\xi\right)$,
with some $C$ independent of $m$ then $E=0$ would not be excluded.
\item We recall a simple and well known result (which follows from the property
that $\left\Vert \widehat{M}_{t}\right\Vert _{\infty}=1$), that there
is no eigenvalue in the upper half plane and no Jordan block on the
real axis .
\item The upper bound given in (\ref{eq:Upper_bound}) results from our
method and choice of escape function $A\left(x,\xi\right)$. In the
proof, $o\left(\alpha^{n-1/2}\right)$ comes from a symplectic volume
$\mathcal{V}$ in phase space which contains the trapped set $\Sigma_{E}$
and which is of order $\mathcal{V}\simeq\delta\alpha^{-1/2}$, with
$\delta$ arbitrarily small. Using Weyl inequalities we obtain an
upper bound of order $\alpha^{n}\mathcal{V}\simeq\delta\alpha^{n-1/2}$
in (\ref{eq:Upper_bound}). It is expected that a better choice of
the escape function could improve this upper bound. For specific models,
e.g. geodesic flows on a surface with constant negative curvature,
it is known that the upper bound is $O\left(\alpha^{\frac{n}{2}}\right)$
(see \cite{leboeuf_04}). We reasonably expect this in general.
\item From the upper bound (\ref{eq:Upper_bound}), one can deduce upper
bounds in larger spectral domains. For example: for every $\beta>0$,
in the semiclassical limit $\alpha\rightarrow+\infty$ we have \[
\sharp\left\{ \lambda\in\mathrm{Res}\left(\widehat{H}\right),\,\,\Re\left(\lambda\right)\in\left[-\alpha,\alpha\right],\,\,\Im\left(\lambda\right)>-\beta\right\} \leq o\left(\alpha^{n}\right),\]
with $n=\dim X$.
\end{itemize}

\subsection{\label{sub:Proof-of-theorem-1}Proof of theorem \ref{th:discrete_spectrum}
about the discrete spectrum of resonances}

Here are the different steps that we will follow in the proof.
\begin{enumerate}
\item The operator $\widehat{H}$ on the Sobolev space $H^{m}=\widehat{A}_{m}^{-1}\left(L^{2}\left(X\right)\right)$
is unitarily isomorphic to the operator $\widehat{P}:=\widehat{A}_{m}\widehat{H}\widehat{A}_{m}^{-1}$
on $L^{2}\left(X\right)$. We will show that $\widehat{P}$ is a pseudo-differential
operator. We will compute the symbol $P\left(x,\xi\right)$ of $\widehat{P}$
in Lemma \ref{lem:symbol_K}. The important fact is that the derivative
of the escape function appears in the imaginary part of the symbol
$P\left(x,\xi\right)$.
\item For $\Im\left(z\right)\gg1$, using the Gårding inequality, we will
show that $\left(\widehat{P}-z\right)$ is invertible and therefore
that $\widehat{P}$ has no spectrum in the domain $\Im\left(z\right)\gg1$.
\item Using the Gårding inequality again for a modified operator and analytic
Fredholm theory we will show that $\left(\widehat{P}-z\right)$ is
invertible for $\Im\left(z\right)>-\left(C_{m}-C\right)$ for some
constant $C$ independent of $m$, except for a discrete set of points
$z=\lambda_{i}$ with finite multiplicity.
\end{enumerate}

\subsubsection{Conjugation by the escape function and unique closed extension of
$\widehat{P}$ on $L^{2}\left(X\right)$}

Let us define 

\begin{equation}
\boxed{\widehat{P}:=\widehat{A}_{m}\widehat{H}\widehat{A}_{m}^{-1}.}\label{eq:def_K}\end{equation}
The following commuting diagram shows that the operator $\widehat{P}$
on $L^{2}\left(X\right)$ is unitarily equivalent to $\widehat{H}$
on $H^{m}$.

\[
\begin{array}{ccc}
L^{2}\left(X\right) & \overset{\widehat{P}}{\rightarrow} & L^{2}\left(X\right)\\
\downarrow\widehat{A}_{m}^{-1} & \circlearrowleft & \downarrow\widehat{A}_{m}^{-1}\\
H^{m} & \overset{\widehat{H}}{\rightarrow} & H^{m}\end{array}\]

The definitions of symbol classes $S^{\mu}$ and $S_{\rho}^{m}$ are
given in Sections \ref{sub:Symbols-and-pseudodifferential} and \ref{sub:Symbol-classes-with}.
In the following Lemma, the notation $\mathcal{O}_{m}\left(S^{-1+0}\right)$
means that the term is a symbol in $S^{-1+0}$. We add the index $m$
to emphasize that it depends on the escape function $m$ whereas $\mathcal{O}\left(S^{0}\right)$
means that the term is a symbol in $S^{0}$ which \emph{does not}
depend on $m$.

\selectlanguage{french}%
\vspace{0.cm}\begin{center}{\color{blue}\fbox{\color{black}\parbox{16cm}{
\selectlanguage{english}%
\begin{lem}
\label{lem:symbol_K}The operator $\widehat{P}$ defined in (\ref{eq:def_K})
is a PDO in $\mathrm{Op}\left(S^{1}\right)$. With respect to every
given system of coordinates its symbol is equal to\begin{equation}
P\left(x,\xi\right)=H\left(x,\xi\right)+i\left(\mathbf{X}\left(G_{m}\right)\right)\left(x,\xi\right)+\mathcal{O}_{m}\left(S^{-1+0}\right),\label{eq:K_x_xi}\end{equation}

where $H\left(x,\xi\right)$ is the symbol of $\widehat{H}$:\[
H\left(x,\xi\right)=V\left(\xi\right)+\mathcal{O}\left(S^{0}\right),\]
with principal symbol $V\left(\xi\right)\in S^{1}$, see Eq.(\ref{eq:def_H}),
and $\mathbf{X}\left(G_{m}\right)\in S^{+0}$. $\mathbf{X}$ is the
Hamiltonian vector field of $H$ defined in (\ref{eq:def_vector_field_X}).
\end{lem}
\selectlanguage{french}%
}}}\end{center}\vspace{0.cm}
\selectlanguage{english}%
\begin{proof}
The proof consists in making the following two lines precise and rigorous:\begin{align*}
\widehat{P} & =\widehat{A}\widehat{H}\widehat{A}^{-1}=\mathrm{Op}\left(e^{G_{m}}\right)\widehat{H}\left(\mathrm{Op}\left(e^{G_{m}}\right)\right)^{-1}\simeq\left(1+\mathrm{Op}\left(G_{m}\right)+\ldots\right)\widehat{H}\left(1-\mathrm{Op}\left(G_{m}\right)+\ldots\right)\\
 & =\widehat{H}+\left[\mathrm{Op}\left(G_{m}\right),\widehat{H}\right]+\ldots=\mathrm{Op}\left(H-i\left\{ G_{m},H\right\} +\ldots\right)=\mathrm{Op}\left(H+i\mathbf{X}\left(G_{m}\right)+\ldots\right).\end{align*}

In order to avoid to work with exponentials of operators, let us define\[
\widehat{A}_{m,t}:=\mathrm{Op}\left(e^{tG_{m}}\right)=\mathrm{Op}\left(e^{G_{tm}}\right)=\widehat{A}_{tm},\qquad0\leq t\leq1,\]
and\[
\widehat{H}_{m,t}:=\widehat{A}_{m,t}\widehat{H}\widehat{A}_{m,t}^{-1}\]
which interpolates between $\widehat{H}=\widehat{H}_{m,0}$ and $\widehat{P}=\widehat{H}_{m,1}$.
We have seen in Lemma \ref{lem:The-order-function_S0} that $\widehat{G}_{m}\in\mathrm{Op}\left(S^{+0}\right)$,
in Proposition \ref{pro:symbol_A_G-1} that $\widehat{A}_{m,t}\in\mathrm{Op}\left(S^{tm\left(x,\xi\right)+0}\right)$,
$\widehat{A}_{m,t}^{-1}\in\mathrm{Op}\left(S^{-tm\left(x,\xi\right)+0}\right)$
and in Eq.(\ref{eq:def_H}) that $\widehat{H}\in\mathrm{Op}\left(S^{1}\right)$.
We deduce that%
\footnote{\label{fn:Composition_operators}From the Theorem of composition of
pseudodifferential operators (PDO), see \cite[Prop.(3.3) p.11]{taylor_tome2},
if $A\in S_{\rho}^{m_{1}}$ and $B\in S_{\rho}^{m_{2}}$ then\[
\mathrm{Op}\left(A\right)\mathrm{Op}\left(B\right)=\mathrm{Op}\left(AB\right)+\mathcal{O}\left(\mathrm{Op}(S_{\rho}^{m_{1}+m_{2}-\left(2\rho-1\right)})\right)\]
 i.e. the symbol of $\mathrm{Op}\left(A\right)\mathrm{Op}\left(B\right)$
is the product $AB$ and belongs to $S_{\rho}^{m_{1}+m_{2}}$ modulo
terms in $S_{\rho}^{m_{1}+m_{2}-\left(2\rho-1\right)}$.%
} $\widehat{H}_{m,t}\in\mathrm{Op}\left(S^{1+0}\right)$. Then\[
\frac{d\widehat{A}_{m,t}}{dt}=\mathrm{Op}\left(G_{m}e^{tG_{m}}\right)=\mathrm{Op}\left(G_{m}\right)\mathrm{Op}\left(e^{tG_{m}}\right)+\mathcal{O}_{m}\left(\mathrm{Op}\left(S^{tm-1+0}\right)\right)\]
\[
\left(\frac{d\widehat{A}_{m,t}}{dt}\right)\widehat{A}_{m,t}^{-1}=-\widehat{A}_{m,t}\left(\frac{d\widehat{A}_{m,t}^{-1}}{dt}\right)=\mathrm{Op}\left(G_{m}+r_{m,t}\right)\]
with $r_{m,t}\in S^{-1+0}$ and%
\footnote{From \cite[Eq.(3.24)(3.25) p.13]{taylor_tome2}, if $A\in S_{\rho}^{m_{1}}$
and $B\in S_{\rho}^{m_{2}}$ then the symbol of $\left[\mathrm{Op}\left(A\right),\mathrm{Op}\left(B\right)\right]$
is the Poisson bracket $-i\left\{ A,B\right\} $ and belongs to $S_{\rho}^{m_{1}+m_{2}-\left(2\rho-1\right)}$
modulo $S_{\rho}^{m_{1}+m_{2}-2\left(2\rho-1\right)}$. We also recall
\cite[(10.8) p.43]{taylor_tome1} that $\left\{ A,B\right\} =-\mathbf{X}_{B}\left(A\right)$
where $\mathbf{X}_{B}$ is the Hamiltonian vector field generated
by $B$.%
}\begin{align*}
\frac{d}{dt}\widehat{H}_{m,t} & =\left(\frac{d\widehat{A}_{m,t}}{dt}\widehat{A}_{m,t}^{-1}\right)\widehat{A}_{m,t}\widehat{H}\widehat{A}_{m,t}^{-1}+\widehat{A}_{m,t}\widehat{H}\widehat{A}_{m,t}^{-1}\left(\widehat{A}_{m,t}\frac{d\widehat{A}_{m,t}^{-1}}{dt}\right)\\
 & =\left[\mathrm{Op}\left(G_{m}+r_{m,t}\right),\widehat{H}_{m,t}\right]\in\mathrm{Op}\left(S^{+0}\right).\end{align*}
Therefore $\widehat{H}_{m,t}-\widehat{H}=\left(\int_{0}^{t}\frac{d}{ds}\widehat{H}_{m,s}ds\right)\in\mathrm{Op}\left(S^{+0}\right)$
and\begin{align*}
\frac{d}{dt}\widehat{H}_{m,t} & =\left[\mathrm{Op}\left(G_{m}\right),\widehat{H}\right]+\left[\mathrm{Op}\left(r_{m,t}\right),\widehat{H}\right]+\left[\mathrm{Op}\left(G_{m}+r_{m,t}\right),\widehat{H}_{m,t}-\widehat{H}\right]\\
 & =\left[\mathrm{Op}\left(G_{m}\right),\widehat{H}\right]+\mathcal{O}_{m}\left(\mathrm{Op}\left(S^{-1+0}\right)\right).\end{align*}
We deduce that\[
\widehat{P}=\widehat{H}+\left(\int_{0}^{1}\frac{d}{dt}\widehat{H}_{m,t}dt\right)=\widehat{H}+\left[\mathrm{Op}\left(G_{m}\right),\widehat{H}\right]+\mathcal{O}_{m}\left(\mathrm{Op}\left(S^{-1+0}\right)\right).\]
Since \[
\left[\mathrm{Op}\left(G_{m}\right),\widehat{H}\right]=\mathrm{Op}\left(i\left(\mathbf{X}\left(G_{m}\right)\right)\left(x,\xi\right)+\mathcal{O}_{m}\left(S^{-1+0}\right)\right),\]
we get\[
\widehat{P}=\widehat{H}+\mathrm{Op}\left(i\left(\mathbf{X}\left(G_{m}\right)\right)\left(x,\xi\right)+\mathcal{O}_{m}\left(S^{-1+0}\right)\right).\]

Finally since $\widehat{H}=\mathrm{Op}\left(V\left(\xi\right)+\mathcal{O}\left(S^{0}\right)\right)$
with a remainder in $S^{0}$ which depends on the quantization (see
discussion in Section \ref{sub:Symbols-and-pseudodifferential}) but
which is independent of the escape function $m$, we get (\ref{eq:K_x_xi}).
Notice that with Weyl quantization, Eq. (\ref{eq:Weyl_symbol_of_H}),
the term $\mathcal{O}\left(S^{0}\right)$ is precisely equal to $\frac{i}{2}\mbox{div}\left(V\right)$.
\end{proof}
We have shown that $\widehat{P}$ is a unbounded PDO of order 1 that
we may first equip with the domain $C^{\infty}\left(X\right)$ which
is dense in $L^{2}\left(X\right)$. Lemma \ref{lem:minimal-maximal_extension}
page \pageref{lem:minimal-maximal_extension} shows that $\widehat{P}$
has a unique closed extension as an unbounded operator $\widehat{P}$
on $L^{2}\left(X\right)$. The adjoint $\widehat{P}^{*}$ is also
a PDO of order 1 and it is the unique closed extension from $C^{\infty}\left(X\right)$.

\subsubsection{\label{sub:K-has-empty-spectrum}$\widehat{P}$ has empty spectrum
for $\Im\left(z\right)\gg1$.}

Let us write \[
\widehat{P}=\widehat{P}_{1}+i\widehat{P}_{2}\]
 with $\widehat{P}_{1}:=\frac{1}{2}\left(\widehat{P}+\widehat{P}^{*}\right)$,
$\widehat{P}_{2}:=\frac{i}{2}\left(\widehat{P}^{*}-\widehat{P}\right)$
self-adjoint. From (\ref{eq:K_x_xi}) and (\ref{eq:XG_negative}),
the symbol of the operator $\widehat{P}_{2}$ is \begin{equation}
P_{2}\left(x,\xi\right)=\mathbf{X}\left(G_{m}\right)\left(x,\xi\right)+\mathcal{O}\left(S^{0}\right)+\mathcal{O}_{m}\left(S^{-1+0}\right)\label{eq:symbol_P2}\end{equation}
 belongs to $S^{+0}$ and satisfies \[
\exists C_{0},\forall\left(x,\xi\right),\quad\Re\left(P_{2}\left(x,\xi\right)\right)\leq C_{0}.\]
From the sharp Gårding inequality (\ref{eq:sharp_Garding}) page \pageref{eq:sharp_Garding}
applied here with order $\mu=1$ (since $P_{2}\in S^{+0}\subset S^{1}$)
we deduce that there exists $C>0$ such that $\left(\widehat{P}_{2}u|u\right)\leq\left(C_{0}+C\right)\left\Vert u\right\Vert ^{2}$
which writes:\begin{equation}
\left(\widehat{P}_{2}-\left(C_{0}+C\right)u|u\right)\leq0.\label{eq:ineq_spectre_vide}\end{equation}

\selectlanguage{french}%

\vspace{0.cm}\begin{center}{\color{blue}\fbox{\color{black}\parbox{16cm}{
\selectlanguage{english}%
\begin{lem}
\label{lem:K_negative-1}From the inequality (\ref{eq:ineq_spectre_vide})
we deduce that for every $z\in\mathbb{C}$, $\Im\left(z\right)>C+C_{0}$,
the resolvent $\left(\widehat{P}-z\right)^{-1}$ exists. Therefore
$\widehat{P}$ has \textbf{empty spectrum} for $\Im\left(z\right)>C+C_{0}$.
\end{lem}
\selectlanguage{french}%
}}}\end{center}\vspace{0.cm}
\selectlanguage{english}%
\begin{proof}
Let $\varepsilon=\Im\left(z\right)-\left(C_{0}+C\right)>0$. Then
for $u\in C^{\infty}\left(X\right)$, \[
\Im\left(\left(\widehat{P}-z\right)u|u\right)=\left(\left(\widehat{P}_{2}-\left(C_{0}+C\right)\right)u|u\right)-\left(\Im\left(z\right)-\left(C_{0}+C\right)\right)\left\Vert u\right\Vert ^{2}\leq-\varepsilon\left\Vert u\right\Vert ^{2}.\]
Using Cauchy-Schwarz inequality,\[
\left\Vert \left(\widehat{P}-z\right)u\right\Vert \left\Vert u\right\Vert \geq\left|\left(\left(\widehat{P}-z\right)u|u\right)\right|\geq\left|\Im\left(\left(\widehat{P}-z\right)u|u\right)\right|\geq\varepsilon\left\Vert u\right\Vert ^{2}.\]

Hence for $u\in C^{\infty}\left(X\right)$\begin{equation}
\left\Vert \left(\widehat{P}-z\right)u\right\Vert \geq\varepsilon\left\Vert u\right\Vert .\label{eq:in__1}\end{equation}

By density this extends to all $u\in\mathcal{D}\left(\widehat{P}\right)$
and it follows that $\widehat{P}-z$ is injective with closed range
$\mathcal{R}\left(\widehat{P}-z\right)$.

The same argument for the adjoint $\widehat{P}^{*}=\widehat{P}_{1}-i\widehat{P}_{2}$
gives \begin{equation}
\left\Vert \left(\widehat{P}^{*}-\overline{z}\right)u\right\Vert \geq\varepsilon\left\Vert u\right\Vert ,\qquad\forall u\in\mathcal{D}\left(\widehat{P}^{*}\right),\label{eq:in__2}\end{equation}
so $\widehat{P}^{*}-\overline{z}$ is also injective. If $u\in L^{2}\left(X\right)$
is orthogonal to $\mathcal{R}\left(\widehat{P}-z\right)$ then $u$
belongs to the kernel of $\widehat{P}^{*}-\overline{z}$ which is
$0$. Hence $\mathcal{R}\left(\widehat{P}-z\right)=L^{2}\left(X\right)$
and $\widehat{P}-z:\mathcal{D}\left(\widehat{P}\right)\rightarrow L^{2}\left(X\right)$
is bijective with bounded inverses.
\end{proof}

\subsubsection{\label{sub:The-spectrum-of-P_discrete}The spectrum of $\widehat{P}$
is discrete on $\Im\left(z\right)\geq-\left(C_{m}-C\right)$ with
some $C\geq0$ independent of $m$.}

As usual \cite[p.113]{reed-simon4}, in order to obtain a discrete
spectrum for the operator $\widehat{P}$, we need to construct a relatively
compact perturbation $\widehat{\chi}$ of the operator such that $\left(\widehat{P}-i\widehat{\chi}\right)$
has no spectrum on $\Im\left(z\right)\geq-\left(C_{m}-C\right)$.

Let $\chi_{0}:T^{*}X\rightarrow\mathbb{R}^{+}$ be a smooth non negative
function with $\chi_{0}\left(x,\xi\right)=C_{m}>0$ for $\left(x,\xi\right)\in\widetilde{N}_{0}$
and $\chi_{0}\left(x,\xi\right)=0$ outside a neighborhood of $\widetilde{N}_{0}$
where $R$ and $\widetilde{N}_{0}$ are defined in Eq.(\ref{eq:escape_estimate})
page \pageref{eq:escape_estimate}. See also figure \ref{fig:Set-N0}
(b). We can assume that $\chi_{0}\in S^{0}$.

Let $\widehat{\chi}_{0}:=\mathrm{Op}\left(\chi_{0}\right)$. We can
assume that $\widehat{\chi}_{0}$ is self-adjoint. From Eq.(\ref{eq:escape_estimate}),
for every $\left(x,\xi\right)\in T^{*}X$, $\left|\xi\right|\geq R$,
\[
\left(\mathbf{X}\left(G_{m}\right)\left(x,\xi\right)-\chi_{0}\left(x,\xi\right)\right)\leq-C_{m},\]
hence (\ref{eq:symbol_P2}) gives for every $\left(x,\xi\right)\in T^{*}X$:\[
P_{2}\left(x,\xi\right)-\chi_{0}\left(x,\xi\right)\leq-C_{m}+C+\mathcal{O}_{m}\left(S^{-1+0}\right),\]
with some $C\in\mathbb{R}$ independent of $m$, coming from the $\mathcal{O}\left(S^{0}\right)$
term in (\ref{eq:symbol_P2}). Notice that the remainder term $\mathcal{O}_{m}\left(S^{-1+0}\right)$
could be bounded but by a constant which depends on $m$.

Since $P_{2}\in S^{\mu}$ with every order $0<\mu<1$, the sharp Gårding
inequality (\ref{eq:sharp_Garding}) page \pageref{eq:sharp_Garding}
implies that for every $u\in C^{\infty}\left(X\right)$ there exists
$C_{\mu}>0$ such that\[
\left(\left(\widehat{P}_{2}-\widehat{\chi}_{0}+\left(C_{m}-C\right)\right)u|u\right)\leq C_{\mu}\left\Vert u\right\Vert _{H^{\frac{\mu-1}{2}}}^{2}.\]

The right hand side can be written $C_{\mu}\left\Vert u\right\Vert _{H^{\frac{\mu-1}{2}}}^{2}=C_{\mu}\left(\left\langle \widehat{\xi}\right\rangle ^{\mu-1}u|u\right)=\Im\left(i\widehat{\chi}_{1}u|u\right)$
with $\widehat{\chi}_{1}=\mathrm{Op}\left(\chi_{1}\right)$, $\chi_{1}=C_{\mu}\left\langle \xi\right\rangle ^{\mu-1}\in S^{\mu-1}$
and can be absorbed on the left by defining\[
\chi:=\chi_{0}+\chi_{1},\qquad\widehat{\chi}:=\mathrm{Op}\left(\chi\right).\]
We can assume that $\widehat{\chi}$ is self-adjoint. We obtain:\[
\left(\left(\widehat{P}_{2}-\widehat{\chi}+\left(C_{m}-C\right)\right)u|u\right)\leq0.\]

As in the proof of Lemma \ref{lem:K_negative-1} page \pageref{lem:K_negative-1}
we obtain that the resolvent $\left(\widehat{P}-i\widehat{\chi}-z\right)^{-1}$
exists for $\Im\left(z\right)>-\left(C_{m}-C\right)$. The following
lemma is the central observation for the proof of Theorem \ref{th:discrete_spectrum}.

\selectlanguage{french}%
\vspace{0.cm}\begin{center}{\color{blue}\fbox{\color{black}\parbox{16cm}{
\selectlanguage{english}%
\begin{lem}
\label{lem:Compact_operator_result}For every $z\in\mathbb{C}$ such
that $\Im\left(z\right)>-\left(C_{m}-C\right)$, the operator $\widehat{\chi}\left(\widehat{P}-i\widehat{\chi}-z\right)^{-1}$
is compact.
\end{lem}
\selectlanguage{french}%
}}}\end{center}\vspace{0.cm}
\selectlanguage{english}%
\begin{proof}
On the cone $\widetilde{N}_{0}$, the operator $\left(\widehat{P}-i\widehat{\chi}-z\right)$
is elliptic of order 1. We can therefore invert it micro-locally on
$\widetilde{N}_{0}$, namely construct $E\in S^{-1}$ and $R_{1},R_{2}\in S^{0}$
such that \begin{equation}
\left(\widehat{P}-i\widehat{\chi}-z\right)\widehat{E}=1+\widehat{R}_{1},\qquad\widehat{E}\left(\widehat{P}-i\widehat{\chi}-z\right)=1+\widehat{R}_{2},\label{eq:E2}\end{equation}

\[
\forall j=1,2,\qquad\mathrm{WF}\left(\widehat{R}_{j}\right)\cap\widetilde{N}_{0}=\emptyset.\]
In particular $\mathrm{WF}\left(\chi\widehat{R}_{j}\right)=\emptyset$
therefore $\widehat{\chi}\widehat{R}_{2}$ is a compact operator.
Also $\widehat{E}$ is a compact operator (since $E\in S^{-1}$).
Then from (\ref{eq:E2}), we write:\[
\left(\widehat{P}-i\chi-z\right)^{-1}=\widehat{E}-\widehat{R}_{2}\left(\widehat{P}-i\widehat{\chi}-z\right)^{-1},\]
\[
\widehat{\chi}\left(\widehat{P}-i\widehat{\chi}-z\right)^{-1}=\underbrace{\widehat{\chi}}_{bounded}\underbrace{\widehat{E}}_{compact}-\underbrace{\widehat{\chi}\widehat{R}_{2}}_{compact}\underbrace{\left(\widehat{P}-i\chi-z\right)^{-1}}_{bounded},\]
and deduce that $\widehat{\chi}\left(\widehat{P}-i\widehat{\chi}-z\right)^{-1}$
is a compact operator.
\end{proof}
With the following Lemma we finish the proof of Theorem \ref{th:discrete_spectrum}.

\selectlanguage{french}%
\vspace{0.cm}\begin{center}{\color{blue}\fbox{\color{black}\parbox{16cm}{
\selectlanguage{english}%
\begin{lem}
\label{lem:discrete_spectrum-1}From the facts that for every $z\in\mathbb{C},\Im\left(z\right)>-\left(C_{m}-C\right)$,
$\left(\widehat{P}-i\widehat{\chi}-z\right)$ is invertible, $\widehat{\chi}\left(\widehat{P}-i\widehat{\chi}-z\right)^{-1}$
is compact and that $\left(\widehat{P}-z_{0}\right)$ is invertible
for at least one point $z_{0}\in D$, we deduce that $\widehat{P}$
has \textbf{discrete spectrum with locally finite multiplicity} on
$\Im\left(z\right)>-\left(C_{m}-C\right)$.
\end{lem}
\selectlanguage{french}%
}}}\end{center}\vspace{0.cm}
\selectlanguage{english}%
\begin{proof}
Write for $\Im\left(z\right)>-\left(C_{m}-C\right)$:\[
\widehat{P}-z=\left(1+i\widehat{\chi}\left(\widehat{P}-i\widehat{\chi}-z\right)^{-1}\right)\left(\widehat{P}-i\widehat{\chi}-z\right).\]
Here $\left(\widehat{P}-i\widehat{\chi}-z\right):\mathcal{D}\left(\widehat{P}\right)\rightarrow L^{2}\left(X\right)$
is bijective with bounded inverse and hence Fredholm of index $0$.
Similarly $\left(1+i\widehat{\chi}\left(\widehat{P}-i\widehat{\chi}-z\right)^{-1}\right):L^{2}\left(X\right)\rightarrow L^{2}\left(X\right)$
is Fredholm of index $0$ by Lemma \ref{lem:Compact_operator_result}.
Thus \[
\widehat{P}-z:\mathcal{D}\left(\widehat{P}\right)\rightarrow L^{2}\left(X\right),\qquad\Im\left(z\right)>C_{m}-C,\]
is a holomorphic family of Fredholm operators (of index $0$) invertible
for $\Im\left(z\right)\gg0$. It then suffices to apply the analytic
Fredholm theorem (\cite[p.201, case (b)]{reed-simon1}, see also \cite[p.220 Appendix A]{sjostrand_87}).
\end{proof}

\subsection{\label{sub:Proof-of-theorem_intrinsic}Proof of theorem \ref{thm:The-eigenvalues-are_intrinsec}
that the eigenvalues are intrinsic to the Anosov vector field $V$}

Let $m$ and $G_{m}$ be as in Lemma \ref{lem:def_A_s_u}. Let $\widehat{m}=f\left(m\right)$
where $f\in C^{\infty}\left(\mathbb{R}\right)$, $f\left(t\right)\geq\max\left(0,t\right)$,
$f'\geq0$, $f\left(t\right)=0$ for $t\leq u/2$ and $f\left(t\right)=t$
for $t\geq s/2$. $\widehat{H}$ viewed as a closed unbounded operator
in $H^{\widehat{m}}$ has no spectrum in the half plane $\Im\left(z\right)\geq C_{1}$
for $C_{1}\gg0$. The same holds for $\widehat{H}:L^{2}\rightarrow L^{2}$.
Since $\widehat{m}\geq0$ we have $H^{\widehat{m}}\subset L^{2}$
so if $v\in H^{\widehat{m}}$ then $R_{L^{2}}\left(z\right)v=R_{H^{\widehat{m}}}\left(z\right)v$
for $\Im\left(z\right)\geq C_{1}$ where $R_{L^{2}}$ denotes the
resolvent of $\widehat{H}:L^{2}\rightarrow L^{2}$ and similarly for
$R_{H^{\widehat{m}}}$.

Since $\widehat{m}\geq m$ we also have $H^{\widehat{m}}\subset H^{m}$
and hence $R_{H^{\widehat{m}}}\left(z\right)v=R_{H^{m}}\left(z\right)v$
for $\Im\left(z\right)\geq C_{1}$, $v\in H^{\widehat{m}}$. Especially
when $v\in C^{\infty}$, we get $R_{L^{2}}\left(z\right)v=R_{H^{m}}\left(z\right)v$,
$\Im\left(z\right)\geq C_{1}$. Applying Theorem \ref{th:discrete_spectrum},
we conclude that $R_{L^{2}}\left(z\right)$, viewed as an operator
$C^{\infty}\rightarrow\mathcal{D}'$ has a meromorphic extension $R\left(z\right)$
from the half plane $\Im\left(z\right)\geq C_{1}$ to the half plane
$\Im\left(z\right)>-\left(C_{m}-C\right)$ which coincide with $R_{H^{m}}$
restricted to $C^{\infty}$.

If $\gamma$ is a simple positively oriented closed curve in the half
plane $\Im\left(z\right)>-\left(C_{m}-C\right)$ which avoids the
eigenvalues of $\widehat{H}:H^{m}\rightarrow H^{m}$ then the spectral
projection, associated to the spectrum of $\widehat{H}$ inside $\gamma$,
is given by\[
\pi_{\gamma}^{H^{m}}=\frac{1}{2\pi i}\int_{\gamma}R_{H^{m}}\left(z\right)dz.\]
For $v\in C^{\infty}$, we have\[
\pi_{\gamma}^{H^{m}}v=\pi_{\gamma}v:=\frac{1}{2\pi i}\left(\int_{\gamma}R_{H^{m}}\left(z\right)dz\right)v.\]
Now $C^{\infty}$ is dense in $H^{m}$ and $\pi_{\gamma}^{H^{m}}$
is of finite rank, hence its range $\pi_{\gamma}^{H^{m}}\left(H^{m}\right)$
is equal to the image $\pi_{\gamma}\left(C^{\infty}\right)$ of $C^{\infty}$.
The latter space is independent of the choice of $H^{m}$. More precisely
if $\widetilde{m},\widetilde{f}$ are as in Theorem \ref{th:discrete_spectrum}
and we choose $\gamma$ as above, now in the half plane $\Im\left(z\right)>-\min\left(\left(C_{m}-C\right),\left(C_{\widetilde{m}}-C\right)\right)$
and avoiding the spectrum of $\widehat{H}:H^{m}\rightarrow H^{m}$
and $\widehat{H}:H^{\widetilde{m}}\rightarrow H^{\widetilde{m}}$,
then the spectral projections $\pi_{\gamma}^{H^{m}}$ and $\pi_{\gamma}^{H^{\widetilde{m}}}$
have the same range. 

Since one can find order functions $m$ which are arbitrarily large
in every direction except $E_{u}^{*}$, see remark \ref{rem:C_m}
page \pageref{rem:C_m}, we deduce that the eigen-distributions are
smooth in every direction except $E_{u}^{*}$. The Theorem follows.

\subsection{Proof of theorem \ref{thm:Upper-bound-for-resonances} for the upper
bound on the density of resonances}

The asymptotic regime $\Re\left(z\right)\gg1$ which is considered
in Theorem \ref{thm:Upper-bound-for-resonances} is a semiclassical
regime in the sense that it involves large values of $H\left(\xi\right)=V\left(\xi\right)\gg1$,
hence large values of $\left|\xi\right|$ in the cotangent space $T^{*}X$.

For convenience, we will switch to $h$-semiclassical analysis. Let
$0<h\ll1$ be a small parameter (we will set $\alpha=1/h$ in Theorem
\ref{thm:Upper-bound-for-resonances}). In $h$-semiclassical analysis
the symbol $\xi$ will be quantized into the operator $\mathrm{Op}_{h}\left(\xi\right):=hD_{x}=-ih\partial/\partial x$
whereas for ordinary PDO, $\xi$ is quantized into $\mathrm{Op}\left(\xi\right):=D_{x}=-i\partial/\partial x$.
This is simply a rescaling of the cotangent space $T^{*}X$ by a factor
$h$, i.e. \begin{equation}
h\,\mathrm{Op}\left(\xi\right)=\mathrm{Op}_{h}\left(\xi\right).\label{eq:Opxi_Op_h_xi}\end{equation}

In this Section we first recall the definition of symbols in $h$-semiclassical
analysis. In Lemma \ref{lem:We-define-P_h} we derive again the expression
of the symbol of $\widehat{P}$. In Section \ref{sub:Main-idea-of}
we give the main idea of the proof and the next Sections give the
details of this proof.

\subsubsection{$h$-semiclassical class of symbols}

We first define the symbol classes we will need in $h$-semiclassical
analysis.

\selectlanguage{french}%
\vspace{0.cm}\begin{center}{\color{red}\fbox{\color{black}\parbox{16cm}{
\selectlanguage{english}%
\begin{defn}
\label{def:The-symbol-class_S_mu_ro-1-1-1}The \textbf{symbol class}
$\left(h^{-k}S_{\rho}^{\mu}\right)$ with $1/2<\rho\leq1$, order
$\mu\in\mathbb{R}$ and $k\in\mathbb{R}$ consists of $C^{\infty}$
functions $p\left(x,\xi;h\right)$ on $T^{*}X$, indexed by $0<h\ll1$
such that in every trivialization $\left(x,\xi\right):\left.T^{*}X\right|_{U}\rightarrow\mathbb{R}^{2n}$,
for every compact $K\subset U$ \begin{equation}
\forall\alpha,\beta,\quad\left|\partial_{\xi}^{\alpha}\partial_{x}^{\beta}p\right|\leq C_{K,\alpha,\beta}h^{-k+\left(\rho-1\right)\left(\left|\alpha\right|+\left|\beta\right|\right)}\left\langle \xi\right\rangle ^{\mu-\rho\left|\alpha\right|+\left(1-\rho\right)\left|\beta\right|}.\label{eq:def_h_class_symbol_p}\end{equation}

\end{defn}
\selectlanguage{french}%
}}}\end{center}\vspace{0.cm}

\selectlanguage{english}%
For short we will write $S_{\rho}^{\mu}$ instead of $\left(h^{-k}S_{\rho}^{\mu}\right)$
when $k=0$, and write $S^{\mu}$ instead of $S_{\rho}^{\mu}$ when
$\rho=1$.

For symbols of variable orders we have:

\selectlanguage{french}%
\vspace{0.cm}\begin{center}{\color{red}\fbox{\color{black}\parbox{16cm}{
\selectlanguage{english}%
\begin{defn}
\label{def_symbol_class_variable_order-1-1}Let $m\left(x,\xi\right)\in S^{0}$,
$\frac{1}{2}<\rho\leq1$ and $k\in\mathbb{R}$. A family of functions
$p\left(x,\xi;h\right)\in C^{\infty}\left(T^{*}X\right)$ indexed
by $0<h\ll1$, belongs to the \textbf{class $\left(h^{-k}S_{\rho}^{m\left(x,\xi\right)}\right)$
of variable order} if in every trivialization $\left(x,\xi\right):\left.T^{*}X\right|_{U}\rightarrow\mathbb{R}^{2n}$,
for every compact $K\subset U$ and all multi-indices $\alpha,\beta\in\mathbb{N}^{n}$,
there is a constant $C_{K,\alpha,\beta}$ such that \begin{equation}
\left|\partial_{\xi}^{\alpha}\partial_{x}^{\beta}p\left(x,\xi\right)\right|\leq C_{K,\alpha,\beta}h^{-k-\left(1-\rho\right)\left(\left|\alpha\right|+\left|\beta\right|\right)}\left\langle \xi\right\rangle ^{m\left(x,\xi\right)-\rho\left|\alpha\right|+\left(1-\rho\right)\left|\beta\right|},\label{eq:def_Class_h_Sm}\end{equation}
for every $\left(x,\xi\right)\in\left.T^{*}X\right|_{U}$.
\end{defn}
\selectlanguage{french}%
}}}\end{center}\vspace{0.cm}

\selectlanguage{english}%

\subsubsection{The symbol of the conjugated operator}

Since the symbol $H=V\left(\xi\right)\mbox{ mod }S^{0}$ of $\widehat{H}$,
given in (\ref{eq:def_H}), \emph{is linear} in $\xi$, it follows
from (\ref{eq:Opxi_Op_h_xi}) that\[
h\widehat{H}=h\mathrm{Op}\left(H\right)=\mathrm{Op}_{h}\left(H\right).\]
Therefore we also rescale the spectral domain $z\in\mathbb{C}$ by
defining:\begin{equation}
z_{h}:=hz,\qquad\widehat{H}_{h}:=h\widehat{H}.\label{eq:def_scaling}\end{equation}
and get \[
\widehat{H}_{h}=\mathrm{Op}_{h}\left(V\left(\xi\right)+\mathcal{O}\left(hS^{0}\right)\right)\quad\in\mathrm{Op}_{h}\left(S^{1}\right).\]
From now on we will work with these new variables and \textbf{we will
often drop the indices $h$ }for short.

We will take again the escape function to be $G_{m}\left(x,\xi\right):=m\left(x,\xi\right)\log\left\langle \xi\right\rangle $
as in (\ref{eq:def_Gm}) but with the rescaled variable $\xi$, i.e.
quantized by \[
\widehat{G}_{m}:=\mathrm{Op}_{h}\left(G_{m}\right).\]
Since the vector field $\mathbf{X}$ is linear in the fibers of the
bundle $T^{*}X$ we get the same estimates (\ref{eq:escape_estimate})
and (\ref{eq:XG_negative}). We can now proceed as in Section \ref{sub:Anisotropic-Sobolev-spaces}:
$G_{m}$ is a $h$-semiclassical symbol , $G_{m}\in S^{+0}$ and quantization
gives \[
\widehat{A}_{m}:=\mathrm{Op}_{h}\left(\exp\left(G_{m}\right)\right),\]
 which is a $h$-PDO with symbol $A_{m}\in S_{1-0}^{m\left(x,\xi\right)}$
(the invertibility of $\widehat{A}_{m}$ is automatic if $h$ is small
enough). Notice that the Sobolev space defined now by $H^{m}:=\widehat{A}_{m}^{-1}\left(L^{2}\left(X\right)\right)=\left(\mathrm{Op}_{h}\left(A_{m}\right)\right)^{-1}\left(L^{2}\left(X\right)\right)$
is identical to (\ref{eq:def_Hm}) as a linear space. However the
norm in $H^{m}$ depends on $h$.

In the following Lemma, we will use again the notation $\mathcal{O}_{m}\left(h^{2}S^{-1+0}\right)$
which means that the term is a symbol in $h^{2}S^{-1+0}$. We add
the index $m$ to emphasize that it depends on the escape function
$m$ whereas $\mathcal{O}\left(hS^{0}\right)$ means that the term
is a symbol in $hS^{0}$ which \emph{does not} depend on $m$.

\selectlanguage{french}%
\vspace{0.cm}\begin{center}{\color{blue}\fbox{\color{black}\parbox{16cm}{
\selectlanguage{english}%
\begin{lem}
\label{lem:We-define-P_h}We define \[
\widehat{P}:=\widehat{A}_{m}\widehat{H}\widehat{A}_{m}^{-1},\]
 as in (\ref{eq:def_K}). Its symbol $P\in S^{1}$ is \begin{equation}
P\left(x,\xi\right)=V\left(\xi\right)+ih\mathbf{X}\left(G_{m}\right)\left(x,\xi\right)+\mathcal{O}\left(hS^{0}\right)+\mathcal{O}_{m}\left(h^{2}S^{-1+0}\right).\label{eq:h_symbol_P}\end{equation}

\end{lem}
\selectlanguage{french}%
}}}\end{center}\vspace{0.cm}
\selectlanguage{english}%
\begin{proof}
Eq. (\ref{eq:h_symbol_P}) follows from Lemma \ref{lem:symbol_K}.
But for clarity we re-derive it. Let us define\[
\widehat{A}_{m,t}:=\mathrm{Op}_{h}\left(e^{tG_{m}}\right)=\mathrm{Op}_{h}\left(e^{G_{tm}}\right)=\widehat{A}_{tm},\qquad0\leq t\leq1\]
and\[
\widehat{H}_{m,t}:=\widehat{A}_{m,t}\widehat{H}\widehat{A}_{m,t}^{-1}\]
which interpolates between $\widehat{H}=\widehat{H}_{m,0}$ and $\widehat{P}=\widehat{H}_{m,1}$.
We have%
\footnote{\label{fn:h-Composiiton_operators}The Theorem of composition of $h$-semiclassical
PDO says that if $A\in S_{\rho}^{m_{1}}$ and $B\in S_{\rho}^{m_{2}}$
then the symbol of $\mathrm{Op}_{h}\left(A\right)\mathrm{Op}_{h}\left(B\right)$
is the product $AB$ and belongs to $S_{\rho}^{m_{1}+m_{2}}$ modulo
$hS_{\rho}^{m_{1}+m_{2}-\left(2\rho-1\right)}$.%
} $\widehat{A}_{m,t}\in\mathrm{Op}_{h}\left(S^{tm+0}\right)$,$\widehat{A}_{m,t}^{-1}\in\mathrm{Op}_{h}\left(S^{-tm+0}\right)$,
$\widehat{H}\in\mathrm{Op}_{h}\left(S^{1}\right)$ therefore $\widehat{H}_{m,t}\in\mathrm{Op}_{h}\left(S^{1+0}\right)$.
Then\[
\left(\frac{d}{dt}\widehat{A}_{m,t}\right)\widehat{A}_{m,t}^{-1}=-\widehat{A}_{m,t}\left(\frac{d}{dt}\widehat{A}_{m,t}^{-1}\right)=\mathrm{Op}\left(G_{m}+r_{m,t}\right)\]
with $r_{m,t}\in hS^{-1+0}$ and\[
\frac{d}{dt}\widehat{H}_{m,t}=\left[\mathrm{Op}_{h}\left(G_{m}+r_{m,t}\right),\widehat{H}_{m,t}\right].\]
We deduce that $\frac{d}{dt}\widehat{H}_{m,t}\in\mathrm{Op}_{h}\left(hS^{+0}\right)$
therefore $\widehat{H}_{m,t}-\widehat{H}=\left(\int_{0}^{t}\frac{d}{ds}\widehat{H}_{m,s}ds\right)\in\mathrm{Op}_{h}\left(hS^{+0}\right)$
also and\begin{align*}
\frac{d}{dt}\widehat{H}_{m,t} & =\left[\mathrm{Op}_{h}\left(G_{m}\right),\widehat{H}\right]+\left[\mathrm{Op}_{h}\left(r_{m,t}\right),\widehat{H}\right]+\left[\mathrm{Op}_{h}\left(G_{m}+r_{m,t}\right),\widehat{H}_{m,t}-\widehat{H}\right]\\
 & =\left[\mathrm{Op}_{h}\left(G_{m}\right),\widehat{H}\right]+\mathcal{O}_{m}\left(\mathrm{Op}_{h}\left(h^{2}S^{-1+0}\right)\right).\end{align*}
We deduce that\[
\widehat{P}=\widehat{H}+\left(\int_{0}^{1}\frac{d}{dt}\widehat{H}_{m,t}dt\right)=\widehat{H}+\left[\mathrm{Op}_{h}\left(G_{m}\right),\widehat{H}\right]+\mathcal{O}_{m}\left(\mathrm{Op}_{h}\left(h^{2}S^{-1+0}\right)\right).\]
Since%
\footnote{If $A\in S_{\rho}^{m_{1}}$ and $B\in S_{\rho}^{m_{2}}$ then the
symbol of $\left[\mathrm{Op}_{h}\left(A\right),\mathrm{Op}_{h}\left(B\right)\right]$
is the Poisson bracket $-ih\left\{ A,B\right\} =ih\mathbf{X}_{B}\left(A\right)$
and belongs to $hS_{\rho}^{m_{1}+m_{2}-\left(2\rho-1\right)}$ modulo
$h^{2}S_{\rho}^{m_{1}+m_{2}-2\left(2\rho-1\right)}$. Here $\mathbf{X}_{B}$
is the Hamiltonian vector field generated by $B$.%
} \[
\left[\mathrm{Op}\left(G_{m}\right),\widehat{H}\right]=\mathrm{Op}_{h}\left(ih\left(\mathbf{X}\left(G_{m}\right)\right)\left(x,\xi\right)+\mathcal{O}_{m}\left(h^{2}S^{-1+0}\right)\right),\]
we get\[
\widehat{P}=\widehat{H}+\mathrm{Op}_{h}\left(ih\left(\mathbf{X}\left(G_{m}\right)\right)\left(x,\xi\right)+\mathcal{O}_{m}\left(h^{2}S^{-1+0}\right)\right).\]

Finally, since $\widehat{H}=\mathrm{Op}_{h}\left(V\left(\xi\right)+\mathcal{O}\left(hS^{0}\right)\right)$
with a remainder in $hS^{0}$ which depends on the quantization (see
discussion in Section \ref{sub:Symbols-and-pseudodifferential}) but
which is independent of the escape function $m$, we get (\ref{eq:K_x_xi}).
\end{proof}
We recall the main properties of the different terms in (\ref{eq:h_symbol_P}).
First $V\left(\xi\right)\in S^{1}$ is real. In each fiber $T_{x}^{*}X$,
$V\left(\xi\right)$ is linear in $\xi$ and for every $E\in\mathbb{R}$
the characteristic set $\Sigma_{E}:=\left\{ \left(x,\xi\right),\, V\left(\xi\right)-E=0\right\} $
is the energy shell defined in (\ref{eq:def_Sigma_E}) and transverse
to $E_{0}^{*}$.

The second term $ih\mathbf{X}\left(G_{m}\right)\in hS^{+0}$ is purely
imaginary and

\begin{equation}
\mathbf{X}\left(G_{m}\right)\left(x,\xi\right)\mbox{ is }\begin{cases}
\leq0 & \mbox{for }\left|\xi\right|\geq R\\
\leq\mathcal{O}_{m}\left(1\right) & \mbox{for }\left|\xi\right|<R\\
\leq-C_{m}, & \quad C_{m}>0,\mbox{ for }\left(x,\xi\right)\notin\left(D_{R}\cup\widetilde{N}_{0}\right)\end{cases}\label{eq:X_G-1}\end{equation}
where $D_{R}=\left\{ \xi,\left|\xi\right|\leq R\right\} $ and $\widetilde{N}_{0}$
is the cone defined in Lemma \ref{lem:def_A_s_u}. With a convenient
choice of the order function $m\left(x,\xi\right)$ we have independently:\begin{equation}
\begin{cases}
\widetilde{N}_{0} & \mbox{ with arbitrarily small aperture}\\
R & \mbox{ arbitrarily small}\\
C_{m}>0 & \mbox{ arbitrarily large}\end{cases}\label{eq:arbitray_small_large}\end{equation}

\begin{figure}
\begin{centering}
\includegraphics[scale=0.7]{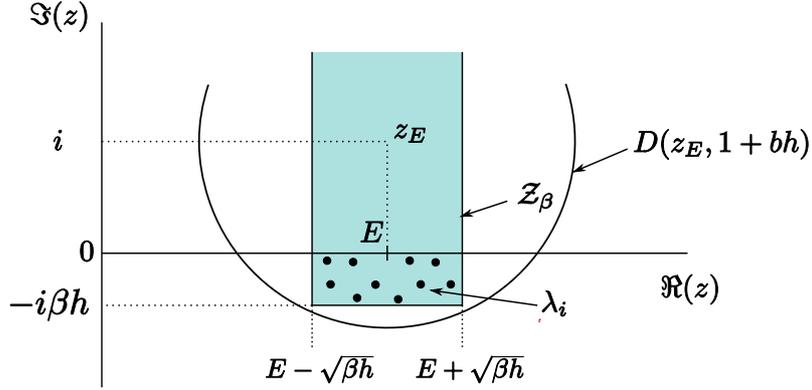}
\par\end{centering}

\caption{\label{fig:spectral_domain}The objective is to bound from above the
number of eigenvalues $\lambda_{i}$ in the domain $\mathcal{Z}_{\beta}$.
For that purpose, we will bound the number of resonances in the disk
of radius $1+bh$ and center $z_{E}=E+i$. }

\end{figure}

\subsubsection{\label{sub:Main-idea-of}Main idea of the proof}

Before giving the details of the proof we give here the main arguments
that we will use in order to prove (\ref{eq:Upper_bound}).

\begin{figure}
\begin{centering}
\includegraphics{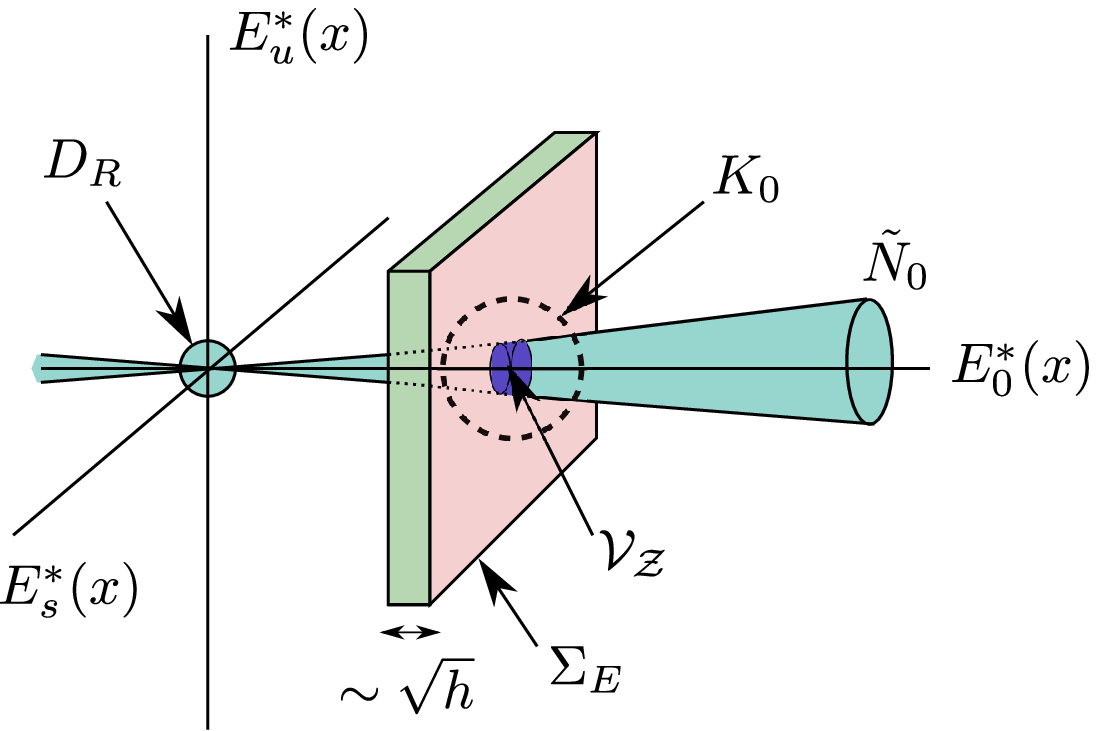}
\par\end{centering}

\caption{\label{fig:volume_VZ}Picture in $T_{x}^{*}X$ with $x\in X$, of
the volume $\mathcal{V}_{Z}$ which supports micro-locally the eigenvalues
$\lambda_{i}\in\mathcal{Z}$ of figure \ref{fig:spectral_domain}.}

\end{figure}

Let us consider the following complex valued function $\widetilde{p}\left(x,\xi\right)\in C^{\infty}\left(T^{*}X\right)$
made from the first two leading terms of the symbol (\ref{eq:h_symbol_P}):\begin{equation}
\widetilde{p}\left(x,\xi\right):=V\left(\xi\right)+ih\mathbf{X}\left(G_{m}\right).\label{eq:def_p_tilde}\end{equation}
Let $E\in\mathbb{R}\backslash\left\{ 0\right\} $ and $h\ll1$. We
define the spectral domain $\mathcal{Z}\subset\mathbb{C}$:\[
\mathcal{Z}:=\left\{ z\in\mathbb{C},\quad\left|\Re\left(z\right)-E\right|\leq\sqrt{C_{m}h},\quad\Im\left(z\right)\geq-C_{m}h\right\} .\]
See Figure \ref{fig:spectral_domain}. Let\begin{equation}
\mathcal{V}_{\mathcal{Z}}:=\left\{ \left(x,\xi\right)\in T^{*}X,\quad\widetilde{p}\left(x,\xi\right)\in\mathcal{Z}\right\} .\label{eq:def_VZ}\end{equation}
We have from (\ref{eq:X_G-1}) \begin{equation}
\left(x,\xi\right)\in\mathcal{V}_{\mathcal{Z}}\Leftrightarrow\begin{cases}
\left|V\left(\xi\right)-E\right|^{2}\leq C_{m}h\\
h\mathbf{X}\left(G_{m}\right)\left(x,\xi\right)\geq-C_{m}h\end{cases}\Rightarrow\begin{cases}
\left(x,\xi\right) & \in\mathbf{\Sigma}_{E\pm\sqrt{C_{m}h}}\\
\left(x,\xi\right) & \in\left(D_{R}\cup\widetilde{N}_{0}\right)\end{cases},\label{eq:V_Z}\end{equation}
where $\mathbf{\Sigma}_{E\pm\sqrt{C_{m}h}}:=\left(\bigcup_{\left|E'-E\right|\leq\sqrt{C_{m}h}}\Sigma_{E'}\right)$
is a union a energy shells (\ref{eq:def_Sigma_E}). We deduce that
the symplectic volume of $\mathcal{V}_{\mathcal{Z}}$ is\begin{equation}
\mathrm{Vol}\left(\mathcal{V}_{\mathcal{Z}}\right)\leq C\,\mathrm{Vol}\left(X\right)\mathrm{Vol}\left(\widetilde{N}_{0}\right)\sqrt{h},\label{eq:Vol_VZ}\end{equation}
with some constant $C>0$. See Figure \ref{fig:volume_VZ}.

Using the {}``max-min formula'' and {}``Weyl inequalities'' we
will obtain an upper bound for the number of eigenvalues (in a smaller
domain $\mathcal{Z}_{\beta}\subset\mathcal{Z}$) in terms of this
upper bound:\[
\sharp\left\{ \lambda_{i}\in\mathcal{Z}_{\beta}\right\} \leq\frac{C_{m}\mathrm{Vol}\left(\mathcal{V}_{\mathcal{Z}}\right)}{h^{n}}=C_{m}C\,\mathrm{Vol}\left(X\right)\mathrm{Vol}\left(\widetilde{N}_{0}\right)h^{1/2-n},\qquad\beta<\frac{1}{4}C_{m},\]
with\begin{equation}
\mathcal{Z}_{\beta}:=\left\{ \lambda\in\mathbb{C},\quad\left|\Re\left(\lambda\right)-E\right|\leq\sqrt{\beta h},\quad\Im\left(\lambda\right)\geq-\beta h\right\} .\label{eq:def_Z_beta}\end{equation}
Using $C_{m}$ arbitrarily large and that $\mathrm{Vol}\left(\widetilde{N}_{0}\right)$
is independently arbitrarily small, from Eq.(\ref{eq:arbitray_small_large}),
we deduce that \[
\sharp\left\{ \lambda_{i}\in\mathcal{Z}_{\beta}\right\} \leq o\left(h^{1/2-n}\right),\]
which is precisely (\ref{eq:Upper_bound}) with $\alpha=1/h$.

The proof below follows these ideas but is not so simple because $P\left(x,\xi\right)$
in Eq.(\ref{eq:h_symbol_P}) is a symbol and not simply a function
(symbols belongs to a non commutative algebra of star product) and
because the term $h\mathbf{X}\left(G_{m}\right)$ is subprincipal.
We will have to decompose the phase space $T^{*}X$ in different parts
in order to separate the different contributions as in (\ref{eq:V_Z}).
Another technical difficulty is that the width of the volume $\mathcal{V}_{\mathcal{Z}}$
is of order $\sqrt{h}$. We will use FBI quantization which is convenient
for a sharp control on phase space at the scale $\sqrt{h}$.

\subsubsection{Proof of Theorem \ref{thm:Upper-bound-for-resonances}}

We present in reverse order the main steps we will follow in the proof.

\paragraph{Steps of the proof:}
\begin{itemize}
\item Our purpose is to bound the cardinal of the spectrum $\sigma\left(\widehat{P}\right)$
of the operator $\widehat{P}$ in the rectangular domain $\mathcal{Z}_{\beta}$
given by (\ref{eq:def_Z_beta}). But as suggested by figure \ref{fig:spectral_domain}
and confirmed by Lemma \ref{lem:Z_disk} below, it suffices to bound
the number of eigenvalues of $\widehat{P}$ in the disk\[
D\left(z_{E},1+bh\right):=\left\{ z\in\mathbb{C},\quad\left|z-z_{E}\right|\leq\left(1+bh\right)\right\} ,\qquad b>0,\]
with radius $\left(1+bh\right)$ and center: \[
z_{E}:=E+i\in\mathbb{C}.\]

\selectlanguage{french}%
\vspace{0.cm}\begin{center}{\color{blue}\fbox{\color{black}\parbox{14cm}{
\selectlanguage{english}%
\begin{lem}
\label{lem:Z_disk}If $b>2\beta$ and $h$ small enough then\[
\left(\sigma\left(\widehat{P}\right)\bigcap\mathcal{Z}_{\beta}\right)\subset D\left(z_{E},1+bh\right).\]

\end{lem}
\selectlanguage{french}%
}}}\end{center}\vspace{0.cm}
\selectlanguage{english}%
\begin{proof}
We know from a remark after Theorem \ref{th:discrete_spectrum} that
$z\in\sigma\left(\widehat{P}\right)\Rightarrow\Im\left(z\right)\leq0$.
Also, Pythagora's Theorem in the corner of $\mathcal{Z}$ gives the
condition $\left(1+bh\right)^{2}>\left(1+\beta h\right)^{2}+\left(\sqrt{\beta h}\right)^{2}$
which is fulfilled if $b>2\beta$ and $h$ small enough.
\end{proof}
\item In order to bound the number of eigenvalues of $\widehat{P}$ in the
disk $D\left(z_{E},1+bh\right)$, we will use Weyl inequalities in
Corollary \ref{cor:Using-Weyl-inequalities} page \pageref{cor:Using-Weyl-inequalities}
and a bound for the number of small singular values of the operator
$\left(\widehat{P}-z_{E}\right)$ (i.e. eigenvalues of $\left(\widehat{P}-z_{E}\right)^{*}\left(\widehat{P}-z_{E}\right)$)
obtained in Lemma \ref{lem:sngular-values}.
\item In order to get this bound on singular values, we will bound from
below the expressions $\left\Vert \left(\widehat{P}-z_{E}\right)u\right\Vert ^{2}=\left(\left(\widehat{P}-z_{E}\right)^{*}\left(\widehat{P}-z_{E}\right)u|u\right)$.
From symbolic calculus (see footnote \ref{fn:h-Composiiton_operators}
page \pageref{fn:h-Composiiton_operators}) we can compute the symbol
of this operator and get:\begin{align}
\left(\widehat{P}-z_{E}\right)^{*}\left(\widehat{P}-z_{E}\right) & =\mathrm{Op}\left(\left|V\left(\xi\right)-E\right|^{2}+\left|1-h\mathbf{X}\left(G_{m}\right)\right|^{2}+\mathcal{O}\left(hS^{1}\right)+\mathcal{O}_{m}\left(h^{2}S^{+0}\right)\right)\label{eq:(P-E)2}\\
 & =\mathrm{Op}\left(\left|V\left(\xi\right)-E\right|^{2}+1-2h\mathbf{X}\left(G_{m}\right)+\mathcal{O}\left(hS^{1}\right)+\mathcal{O}_{m}\left(h^{2}S^{+0}\right)\right).\nonumber \end{align}
However it is not possible to deduce directly estimates from this
symbol because for large $\left|\xi\right|$ the remainders $\mathcal{O}\left(hS^{1}\right)$
and $\mathcal{O}_{m}\left(h^{2}S^{+0}\right)$ may dominate the important
term $2h\mathbf{X}\left(G_{m}\right)\in hS^{+0}$. Therefore we first
have to perform a partition of unity on phase space. 
\end{itemize}

\paragraph{Partition of unity on phase space:}

Let $K_{0}\subset T^{*}X$ be a compact subset (independent of $h$)
such that $\mathcal{V}_{\mathcal{Z}}\subset K_{0}$ with $\mathcal{V}_{\mathcal{Z}}$
defined in (\ref{eq:def_VZ}). See figure \ref{fig:volume_VZ}. Lemma
\ref{lem:Let-partition_Ki0-Ki1} page \pageref{lem:Let-partition_Ki0-Ki1}
associates a {}``quadratic partition of unity of PDO'' to the compact
set $K_{0}$:\begin{equation}
\boxed{\widehat{\chi}_{0}^{2}+\widehat{\chi}_{1}^{2}=1+\mathrm{Op}\left(h^{\infty}S^{-\infty}\right)}\label{eq:Quad_Partition}\end{equation}
with self-adjoint operators $\widehat{\chi}_{0},\widehat{\chi}_{1}$
with symbols $\chi_{0}\in S^{-\infty}$, $\chi_{1}\in S^{0}$. On
the compact set $K_{0}$, $\chi_{0}=1+\mathcal{O}\left(h^{\infty}\right)$,
$\chi_{1}=\mathcal{O}\left(h^{\infty}\right)$.

Then from Lemma \ref{lem:IMS-localization-formula} page \pageref{lem:IMS-localization-formula}
called {}``IMS localization formula'' we have: for every $u\in L^{2}\left(X\right)$,\begin{equation}
\boxed{\left\Vert \left(\widehat{P}-z_{E}\right)u\right\Vert ^{2}=\left\Vert \left(\widehat{P}-z_{E}\right)\widehat{\chi}_{0}u\right\Vert ^{2}+\left\Vert \left(\widehat{P}-z_{E}\right)\widehat{\chi}_{1}u\right\Vert ^{2}+\mathcal{O}\left(h^{2}\right)\left\Vert u\right\Vert ^{2}.}\label{eq:IMS_result}\end{equation}

We will now study the different terms of (\ref{eq:IMS_result}) separately.

\paragraph{Informal remark:}

In order to show that the Lemma \ref{lem:Chi1} below is expected,
let us give an informal remark (non necessary for the proof). Using
the function $\widetilde{p}\left(x,\xi\right):=V\left(\xi\right)+ih\mathbf{X}\left(G_{m}\right)$,
as in (\ref{eq:def_p_tilde}), which is the dominant term of the symbol
$P\left(x,\xi\right)$, we write:\begin{align}
\left|\widetilde{p}\left(x,\xi\right)-z_{E}\right|^{2} & =\left|V\left(\xi\right)-E\right|^{2}+\left|1-h\mathbf{X}\left(G_{m}\right)\right|^{2}\label{eq:discuss_1-1}\\
 & =\left|V\left(\xi\right)-E\right|^{2}+1-2h\mathbf{X}\left(G_{m}\right)+\mathcal{O}\left(h^{2}S^{+0}\right).\end{align}
If $\left(x,\xi\right)\notin K_{0}$ there are two cases, according
to (\ref{eq:X_G-1}):
\begin{enumerate}
\item Either $\mathbf{X}\left(G_{m}\right)\left(x,\xi\right)\leq-C_{m}$
, therefore:\[
\left|\widetilde{p}\left(x,\xi\right)-z_{E}\right|^{2}\geq1+2hC_{m}.\]

\item or $\left|V\left(\xi\right)-E\right|^{2}\geq C_{0}>0$ and $\mathbf{X}\left(G_{m}\right)\leq\mathcal{O}\left(1\right)$
from (\ref{eq:X_G-1}). Therefore\[
\left|\widetilde{p}\left(x,\xi\right)-z_{E}\right|^{2}\geq1+C_{0}+\mathcal{O}\left(h\right).\]

\end{enumerate}
In both cases we have \begin{equation}
\left(x,\xi\right)\notin K_{0}\Rightarrow\left|\widetilde{p}\left(x,\xi\right)-z_{E}\right|^{2}\geq1+2hC_{m}.\label{eq:estimate_1}\end{equation}

Since $\chi_{1}$ is negligible on $K_{0}$, the following Lemma \ref{lem:Chi1}
is not surprising in the light of property (\ref{eq:estimate_1}).
It gives a lower bound for the second term in the right side of (\ref{eq:IMS_result}).

\selectlanguage{french}%
\vspace{0.cm}\begin{center}{\color{blue}\fbox{\color{black}\parbox{16cm}{
\selectlanguage{english}%
\begin{lem}
\label{lem:Chi1}For every $u\in L^{2}\left(X\right)$,\begin{equation}
\left\Vert \left(\widehat{P}-z_{E}\right)\widehat{\chi}_{1}u\right\Vert ^{2}\geq\left(1+2h\left(C_{m}-C\right)\right)\left\Vert \widehat{\chi}_{1}u\right\Vert ^{2}-\mathcal{O}\left(h^{\infty}\right)\left\Vert u\right\Vert ^{2}.\label{eq:result_Ki1}\end{equation}

\end{lem}
\selectlanguage{french}%
}}}\end{center}\vspace{0.cm}
\selectlanguage{english}%
\begin{proof}
In order to prove (\ref{eq:result_Ki1}) we have to consider a partition
of unity in order to take into account two contributions as in the
discussion after (\ref{eq:discuss_1-1}). Let $\Psi_{0}\in S^{0}$
which has its support inside the region where $\chi_{1}=1$ and we
set $\Psi_{0}=1$ away from a conical neighborhood of the energy shell
$\Sigma_{E}$, Eq.(\ref{eq:def_Sigma_E}), which is the characteristic
set $V\left(\xi\right)-E=0$. See figure \ref{fig:Decoupe_TX}.

\begin{figure}
\begin{centering}
\includegraphics{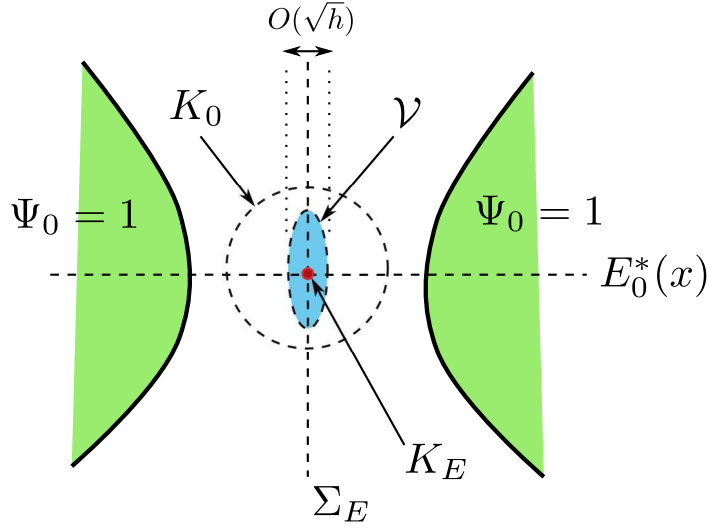}
\par\end{centering}

\caption{\label{fig:Decoupe_TX}Picture in $T_{x}^{*}X$ with $x\in X$, which
shows the partition of unity of  phase space used in the proof of
Lemma \ref{lem:Chi1}. The support of $\chi_{1}$ is outside the set
$K_{0}$.}

\end{figure}

Since $\left(V\left(\xi\right)-E\right)$ is the principal symbol
of $\left(P\left(x,\xi\right)-E\right)\in S^{1}$ and is non vanishing
on the support of $\Psi_{0}$, there exists $\widehat{Q}\in\mathrm{Op}\left(S^{-1}\right)$
such that\[
\widehat{Q}\left(\widehat{P}-E\right)=\widehat{\Psi}_{0}+\widehat{R},\qquad\widehat{R}\in\mathrm{Op}\left(h^{\infty}S^{-\infty}\right).\]
Since $\widehat{Q}$ is continuous in $L^{2}\left(X\right)$, there
exists $C_{0}>0$ such that for every $v\in L^{2}\left(X\right)$,
$\left\Vert v\right\Vert ^{2}\geq\frac{1}{C_{0}}\left\Vert \widehat{Q}v\right\Vert ^{2}$,
hence for every $u\in L^{2}\left(X\right)$\begin{align}
\left\Vert \left(\widehat{P}-E\right)\widehat{\chi}_{1}u\right\Vert ^{2} & \geq\frac{1}{C_{0}}\left\Vert \widehat{Q}\left(\widehat{P}-E\right)\widehat{\chi}_{1}u\right\Vert ^{2}=\frac{1}{C_{0}}\left\Vert \left(\widehat{\Psi}_{0}+\widehat{R}\right)\widehat{\chi}_{1}u\right\Vert ^{2}\label{eq:ineg_1}\\
 & \geq\frac{1}{2C_{0}}\left\Vert \widehat{\Psi}_{0}\widehat{\chi}_{1}u\right\Vert ^{2}-\mathcal{O}\left(h^{\infty}\right)\left\Vert u\right\Vert ^{2}.\nonumber \end{align}

Writing $\widehat{P}=\widehat{P}_{1}+i\widehat{P}_{2}$ with $\widehat{P}_{i}$
self-adjoint, we have \begin{align}
\left\Vert \left(\widehat{P}-z_{E}\right)\widehat{\chi}_{1}u\right\Vert ^{2} & =\left(\left(\left(\widehat{P}-E\right)^{*}+i\right)\left(\left(\widehat{P}-E\right)-i\right)\widehat{\chi}_{1}u|\widehat{\chi}_{1}u\right)\label{eq:eq_3}\\
 & =\left\Vert \left(\widehat{P}-E\right)\widehat{\chi}_{1}u\right\Vert ^{2}+\left\Vert \widehat{\chi}_{1}u\right\Vert ^{2}-\left(2\widehat{P}_{2}\widehat{\chi}_{1}u|\widehat{\chi}_{1}u\right).\nonumber \end{align}
Using (\ref{eq:ineg_1}) in (\ref{eq:eq_3}) we get for every $a>0$:\begin{align}
\left\Vert \left(\widehat{P}-z_{E}\right)\widehat{\chi}_{1}u\right\Vert ^{2} & =\left(1-ah\right)\left\Vert \left(\widehat{P}-E\right)\widehat{\chi}_{1}u\right\Vert ^{2}+ah\left\Vert \left(\widehat{P}-E\right)\widehat{\chi}_{1}u\right\Vert ^{2}+\left\Vert \widehat{\chi}_{1}u\right\Vert ^{2}\label{eq:eq_4}\\
 & \qquad-\left(2\widehat{P}_{2}\widehat{\chi}_{1}u|\widehat{\chi}_{1}u\right)\nonumber \\
 & \geq\left(1-ah\right)\left\Vert \left(\widehat{P}-E\right)\widehat{\chi}_{1}u\right\Vert ^{2}+\left\Vert \widehat{\chi}_{1}u\right\Vert ^{2}-\left(2\widehat{P}_{2}\widehat{\chi}_{1}u|\widehat{\chi}_{1}u\right)\nonumber \\
 & \qquad+\frac{ah}{2C_{0}}\left\Vert \widehat{\Psi}_{0}\widehat{\chi}_{1}u\right\Vert ^{2}-\mathcal{O}\left(h^{\infty}\right)\left\Vert u\right\Vert ^{2}\nonumber \\
 & =\left(1-ah\right)\left\Vert \left(\widehat{P}-E\right)\widehat{\chi}_{1}u\right\Vert ^{2}+\left\Vert \widehat{\chi}_{1}u\right\Vert ^{2}\nonumber \\
 & \qquad+\left(\left(-2\widehat{P}_{2}+\frac{ah}{2C_{0}}\widehat{\Psi}_{0}^{*}\widehat{\Psi}_{0}\right)\widehat{\chi}_{1}u|\widehat{\chi}_{1}u\right)-\mathcal{O}\left(h^{\infty}\right)\left\Vert u\right\Vert ^{2}.\nonumber \end{align}
Recall from (\ref{eq:h_symbol_P}) that \[
\widehat{P}_{2}=\mathrm{Op}\left(h\mathbf{X}\left(G_{m}\right)+\mathcal{O}\left(hS^{0}\right)\right)\in\mathrm{Op}\left(hS^{+0}\right).\]
Therefore \[
\left(-2\widehat{P}_{2}+\frac{ah}{2C_{0}}\widehat{\Psi}_{0}^{*}\widehat{\Psi}_{0}\right)\in\mathrm{Op}\left(hS^{+0}\right).\]
Assume $a\geq4C_{0}\left(C_{m}-C\right)$. Then from (\ref{eq:X_G-1})
and the hypothesis on $\Psi_{0}$, for every $\left(x,\xi\right)\in\mathrm{supp}\left(\chi_{1}\right)$
we have \[
\left(-2P_{2}+\frac{ah}{2C_{0}}\Psi_{0}^{*}\Psi_{0}\right)\left(x,\xi\right)\geq\min\left(2h\left(C_{m}-C\right),\frac{ah}{2C_{0}}\right)\geq2h\left(C_{m}-C\right).\]

We can add a symbol $\Psi_{1}\in S^{0}$ positive, which vanishes
on $\mathrm{supp}\left(\chi_{1}\right)$ so that $\widehat{\Psi}_{1}\widehat{\chi}_{1}\in\mathrm{Op}\left(h^{\infty}S^{-\infty}\right)$
and such that for \emph{every} $\left(x,\xi\right)\in T^{*}X$ we
have\[
\left(-2P_{2}+\frac{ah}{2C_{0}}\Psi_{0}^{*}\Psi_{0}+\Psi_{1}\right)\left(x,\xi\right)\geq\min\left(2h\left(C_{m}-C\right),\frac{ah}{2C_{0}}\right)\geq2h\left(C_{m}-C\right).\]
The semiclassical sharp Gårding inequality implies that:\begin{align*}
\forall u\in L^{2}\left(X\right),\quad\left(\left(-2\widehat{P}_{2}+\frac{ah}{2C_{0}}\widehat{\Psi}_{0}^{*}\widehat{\Psi}_{0}\right)\widehat{\chi}_{1}u|\widehat{\chi}_{1}u\right)\geq & \left(2h\left(C_{m}-C\right)-\mathcal{O}\left(h^{2}\right)\right)\left\Vert \widehat{\chi}_{1}u\right\Vert ^{2}\\
 & -\mathcal{O}\left(h^{\infty}\right)\left\Vert u\right\Vert ^{2},\end{align*}
where the remainder term $\mathcal{O}\left(h^{\infty}\right)\left\Vert u\right\Vert ^{2}$
comes from $\left(\widehat{\Psi}_{1}\widehat{\chi}_{1}u|\widehat{\chi}_{1}u\right)$.
With (\ref{eq:eq_4}) we get:\begin{align*}
\left\Vert \left(\widehat{P}-z_{E}\right)\widehat{\chi}_{1}u\right\Vert ^{2} & \geq\left(1-ah\right)\left\Vert \left(\widehat{P}-E\right)\widehat{\chi}_{1}u\right\Vert ^{2}+\left\Vert \widehat{\chi}_{1}u\right\Vert ^{2}\\
 & \qquad+\left(2h\left(C_{m}-C\right)-\mathcal{O}\left(h^{2}\right)\right)\left\Vert \widehat{\chi}_{1}u\right\Vert ^{2}-\mathcal{O}\left(h^{\infty}\right)\left\Vert u\right\Vert ^{2}\\
 & \geq\left(1+2h\left(C_{m}-C\right)-\mathcal{O}\left(h^{2}\right)\right)\left\Vert \widehat{\chi}_{1}u\right\Vert ^{2}-\mathcal{O}\left(h^{\infty}\right)\left\Vert u\right\Vert ^{2}.\end{align*}
The term $\mathcal{O}\left(h^{2}\right)\left\Vert \widehat{\chi}_{1}u\right\Vert ^{2}$
can be absorbed in the constant $C$.
\end{proof}
\selectlanguage{french}%
\vspace{0.cm}\begin{center}{\color{blue}\fbox{\color{black}\parbox{16cm}{
\selectlanguage{english}%
\begin{lem}
\label{lem:Chi0}There exists a family of trace class operators $\widehat{B}_{h}$
(depending on $h$) such that\begin{equation}
\left\Vert \widehat{B}_{h}\right\Vert _{\mathrm{Tr}}\leq\mathcal{O}\left(1\right)C_{m}\mathrm{Vol}\left(\widetilde{N}_{0}\right)h^{1/2-n},\qquad\widehat{B}_{h}\geq0,\label{eq:result_trace_Bh}\end{equation}
(where the constant $\mathcal{O}\left(1\right)$ does not depend on
the escape function $m$) and for every $u\in L^{2}\left(X\right)$,\begin{equation}
\left\Vert \left(\widehat{P}-z_{E}\right)\widehat{\chi}_{0}u\right\Vert ^{2}+\left(h\widehat{B}_{h}u|u\right)\geq\left(1+2h\left(C_{m}-\mathcal{O}\left(1\right)\right)\right)\left\Vert \widehat{\chi}_{0}u\right\Vert ^{2}-\mathcal{O}\left(h^{\infty}\right)\left\Vert u\right\Vert ^{2},\label{eq:result_Ki0}\end{equation}
and\begin{equation}
\left\Vert \left(\widehat{P}-z_{E}\right)\widehat{\chi}_{0}u\right\Vert ^{2}\geq\left(1-\mathcal{O}\left(h\right)\right)\left\Vert \widehat{\chi}_{0}u\right\Vert ^{2}-\mathcal{O}\left(h^{\infty}\right)\left\Vert u\right\Vert ^{2},\label{eq:result_Ki0_general}\end{equation}
where the term $\mathcal{O}\left(h\right)$ does not depend on $m$.
\end{lem}
\selectlanguage{french}%
}}}\end{center}\vspace{0.cm}

\selectlanguage{english}%

\paragraph{Remarks}

Lemma \ref{lem:Chi0} concerns the first term of the right hand side
of (\ref{eq:IMS_result}). In order to obtain (\ref{eq:result_Ki0}),
which is similar to (\ref{eq:result_Ki1}), it has been necessary
to add a new term which involves a trace class operator $\widehat{B}_{h}$.
Its role is to {}``hide'' the domain $\mathcal{V}_{\mathcal{Z}}$
(\ref{eq:V_Z}). Eq.(\ref{eq:result_Ki0_general}) shows that without
this term the lower bound is smaller.
\begin{proof}
The construction is based on ideas around Anti-Wick quantization,
Berezin quantization, FBI transforms, Bargmann-Segal transforms, Gabor
frames and Toeplitz operators, see e.g. \cite{sjostrand_hitrick_07}.
We review some definitions in Appendix \ref{sub:FBI-transform-and}
page \pageref{sub:FBI-transform-and}. We will use the following two
properties for an operator obtained by Toeplitz quantization of a
symbol $A\left(x,\xi;h\right)$. Let \[
\widehat{A}:=\mathrm{Op}_{T}\left(A\right):=\int A\left(x,\xi;h\right)\widehat{\pi}_{\alpha}d\alpha.\]
Gårding's inequality writes\begin{equation}
A\left(x,\xi\right)\geq0\Rightarrow\left(\widehat{A}u|u\right)\geq0+\mathcal{O}\left(h^{\infty}\right)\left\Vert u\right\Vert ^{2},\label{eq:garding_teplitz}\end{equation}
also\begin{equation}
\mathrm{Tr}\left(\widehat{A}\right)=\frac{\mathcal{O}\left(1\right)}{h^{n}}\int A\left(x,\xi\right)dxd\xi+\mathcal{O}\left(h^{\infty}\right),\label{eq:trace_Toeplitz}\end{equation}
and \begin{equation}
\left(\forall\left(x,\xi\right),\quad A\left(x,\xi\right)\geq0\right)\quad\Rightarrow\quad\left\Vert \widehat{A}\right\Vert _{\mathrm{Tr}}=\mathrm{Tr}\left(\widehat{A}\right)+\mathcal{O}\left(h^{\infty}\right).\label{eq:trace_norm_Toeplitz}\end{equation}

From (\ref{eq:(P-E)2}) we have\begin{equation}
\widehat{\chi}_{0}\left(\widehat{P}-z_{E}\right)^{*}\left(\widehat{P}-z_{E}\right)\widehat{\chi}_{0}=\widehat{\chi}_{0}\widehat{S}\widehat{\chi}_{0}+\widehat{R},\label{eq:chi0_P-E}\end{equation}
with $\widehat{R}\in\mathrm{Op}\left(h^{\infty}S^{-\infty}\right)$
and\[
\widehat{S}=\mathrm{Op}_{T}\left(S\right),\]
with the Toeplitz symbol\[
S\left(x,\xi;h\right)=\left|V\left(\xi\right)-E\right|^{2}+1-2h\mathbf{X}\left(G_{m}\right)\left(x,\xi\right)+\mathcal{O}\left(hS^{-\infty}\right)+\mathcal{O}_{m}\left(h^{2}S^{-\infty}\right)\]
(the remainders are in $S^{-\infty}$ since $\chi_{0}$ has compact
support in (\ref{eq:chi0_P-E}). Since $\mathbf{X}\left(G_{m}\right)\leq0$
from (\ref{eq:X_G-1}), we deduce (\ref{eq:result_Ki0_general}) using
Gårding's inequality (\ref{eq:garding_teplitz}). 

In order to improve this lower bound and get (\ref{eq:result_Ki0}),
let $B_{h}\in C_{0}^{\infty}\left(T^{*}X\right)$ such that $\forall\left(x,\xi\right),\, B_{h}\left(x,\xi\right)\geq0$
and \begin{equation}
\left(x,\xi\right)\in\mathcal{V}_{\mathcal{Z}}\Rightarrow B_{h}\left(x,\xi\right)\geq2C_{m}.\label{eq:eee2}\end{equation}
Notice that from (\ref{eq:Vol_VZ}) $B_{h}$ can be chosen such that
\begin{equation}
\int_{T^{*}X}B_{h}\left(x,\xi\right)dxd\xi\leq\mathcal{O}\left(1\right)C_{m}\mathrm{Vol}\left(\mathcal{V}_{\mathcal{Z}}\right)=\mathcal{O}\left(1\right)C_{m}\mathrm{Vol}\left(X\right)\mathrm{Vol}\left(\widetilde{N}_{0}\right)\sqrt{h}.\label{eq:int_Bh}\end{equation}
From (\ref{eq:trace_norm_Toeplitz}), (\ref{eq:trace_Toeplitz}) and
(\ref{eq:int_Bh}) we deduce (\ref{eq:result_trace_Bh}). Recall that
from (\ref{eq:V_Z}) we have\[
\left(x,\xi\right)\notin\mathcal{V}_{\mathcal{Z}}\Rightarrow\left|V\left(\xi\right)-E\right|^{2}\geq hC_{m}\mbox{ or }-h\mathbf{X}\left(G_{m}\right)\geq hC_{m}.\]
Therefore in view of (\ref{eq:eee2}) for every $\left(x,\xi\right)\in T^{*}X$
we have\begin{align*}
S\left(x,\xi;h\right)+hB_{h}\left(x,\xi\right) & =\left|V\left(\xi\right)-E\right|^{2}+1-2h\mathbf{X}\left(G_{m}\right)+hB_{h}\left(x,\xi\right)\\
 & \qquad+\mathcal{O}\left(hS^{-\infty}\right)+\mathcal{O}_{m}\left(h^{2}S^{-\infty}\right)\\
 & \geq1+2hC_{m}+\mathcal{O}\left(hS^{-\infty}\right)+\mathcal{O}_{m}\left(h^{2}S^{-\infty}\right).\end{align*}
Let $\widehat{B}_{h}:=\mathrm{Op}_{T}\left(B_{h}\right)$. After multiplying
both sides by $\widehat{\chi}_{0}$, using $\widehat{\chi}_{0}\widehat{B}_{h}\widehat{\chi}_{0}=\widehat{B}_{h}+\mathrm{Op}\left(h^{\infty}S^{-\infty}\right)$
and Gårding's inequality we deduce that\[
\forall u\in L^{2}\left(X\right),\quad\left(\widehat{\chi}_{0}\widehat{S}\widehat{\chi}_{0}u|u\right)+\left(h\widehat{B}_{h}u|u\right)\geq\left(\left(\widehat{\chi}_{0}\left(1+2h\left(C_{m}-\mathcal{O}\left(1\right)\right)\right)\widehat{\chi}_{0}\right)u|u\right)+\mathcal{O}\left(h^{\infty}\right)\left\Vert u\right\Vert ^{2}.\]
Replacing the first term by (\ref{eq:chi0_P-E}) this gives (\ref{eq:result_Ki0}).
\end{proof}
\selectlanguage{french}%
\vspace{0.cm}\begin{center}{\color{blue}\fbox{\color{black}\parbox{16cm}{
\selectlanguage{english}%
\begin{cor}
Eq. (\ref{eq:IMS_result}) with (\ref{eq:result_Ki1}), (\ref{eq:result_Ki0}),
(\ref{eq:Quad_Partition}) gives:\begin{equation}
\forall u\in L^{2}\left(X\right),\qquad\left\Vert \left(\widehat{P}-z_{E}\right)u\right\Vert ^{2}+\left(h\widehat{B}_{h}u|u\right)\geq\left(1+2\left(C_{m}-\mathcal{O}\left(1\right)\right)h\right)\left\Vert u\right\Vert ^{2},\label{eq:e}\end{equation}
where $\mathcal{O}\left(1\right)$ does not depend on $m$. Using
(\ref{eq:result_Ki0_general}) instead we get \begin{equation}
\forall u\in L^{2}\left(X\right),\qquad\left\Vert \left(\widehat{P}-z_{E}\right)u\right\Vert ^{2}\geq\left(1-\mathcal{O}\left(h\right)\right)\left\Vert u\right\Vert ^{2}.\label{eq:e-3}\end{equation}

\end{cor}
\selectlanguage{french}%
}}}\end{center}\vspace{0.cm}

\selectlanguage{english}%
Let us show that these last relations imply an upper bound for the
number of eigenvalues of the operator $\left(\widehat{P}-z_{E}\right)^{*}\left(\widehat{P}-z_{E}\right)$
smaller than $\left(1+2\left(C_{m}-\mathcal{O}\left(1\right)\right)h\right)$.

\selectlanguage{french}%
\vspace{0.cm}\begin{center}{\color{blue}\fbox{\color{black}\parbox{16cm}{
\selectlanguage{english}%
\begin{lem}
\label{lem:sngular-values}Let $s_{1}\leq s_{2}\leq\ldots$ be the
singular values $\left(\widehat{P}-z_{E}\right)$ sorted from below.
More precisely, $s_{1}^{2}\leq s_{2}^{2}\leq\ldots$ are the eigenvalues
of the positive self-adjoint operator $\widehat{A}:=\left(\widehat{P}-z_{E}\right)^{*}\left(\widehat{P}-z_{E}\right)$
below the infimum of the essential spectrum of $\widehat{A}$, possibly
completed with an infinite repetition of that infimum if there are
only finitely many such eigenvalues. Then the first eigenvalue is\begin{equation}
s_{1}\geq1-\mathcal{O}\left(h\right)\label{eq:e-2}\end{equation}
and \begin{equation}
\mbox{if }j>\mathcal{O}\left(1\right)C_{m}\mathrm{Vol}\left(\widetilde{N}_{0}\right)h^{\frac{1}{2}-n}\mbox{ then }s_{j}\geq1+\left(C_{m}-\mathcal{O}\left(1\right)\right)h.\label{eq:e-1}\end{equation}
In other words the number of singular values of $\left(\widehat{P}-z_{E}\right)$
below $1+\left(C_{m}-\mathcal{O}\left(1\right)\right)h$ is $\mathcal{O}\left(1\right)C_{m}\mathrm{Vol}\left(\widetilde{N}_{0}\right)h^{\frac{1}{2}-n}$.
\end{lem}
\selectlanguage{french}%
}}}\end{center}\vspace{0.cm}
\selectlanguage{english}%
\begin{proof}
Eq.(\ref{eq:e-2}) is a direct consequence of (\ref{eq:e-3}). We
use the {}``max-min formula'' for self-adjoint operators \cite[p.78]{reed-simon4}
and Eq.(\ref{eq:e}). Put $\lambda_{m}:=C_{m}-\mathcal{O}\left(1\right)$.
We have for every $j$\begin{align*}
s_{j}^{2} & =\max_{U\subseteq L^{2}\left(X\right),\,\mathrm{codim}\left(U\right)\leq j-1}\quad\min_{u\in U,\left\Vert u\right\Vert =1}\left(u,\widehat{A}u\right)\\
 & \geq1+2\lambda_{m}h+\max_{U\subseteq L^{2}\left(X\right),\,\mathrm{codim}\left(U\right)\leq j-1}\quad\min_{u\in U,\left\Vert u\right\Vert =1}\left(-\left(h\widehat{B}_{h}u|u\right)\right)\\
 & =1+2\lambda_{m}h-h\min_{U\subseteq L^{2}\left(X\right),\,\mathrm{codim}\left(U\right)\leq j-1}\quad\max_{u\in U,\left\Vert u\right\Vert =1}\left(\left(\widehat{B}_{h}u|u\right)\right)\\
 & =1+2\lambda_{m}h-hb_{j},\end{align*}
where $U$ varies in the set of closed subspaces of $L^{2}\left(X\right)$
and $b_{1}\geq b_{2}\geq\ldots$ denote the eigenvalues of $\widehat{B}_{h}$
(possibly completed with an infinite repetition of $0$ if there are
only finitely many such eigenvalues). We have \[
\left\Vert \widehat{B}_{h}\right\Vert _{\mathrm{Tr}}=\mathrm{Tr}\left(\widehat{B}_{h}\right)=b_{1}+b_{2}+\ldots\]
Eq.(\ref{eq:result_trace_Bh}) implies that for every $\varepsilon_{0}>0$,
if $b_{j}\geq\varepsilon_{0}$ then $j\varepsilon_{0}\leq\mathrm{Tr}\left(\widehat{B}_{h}\right)\leq\mathcal{O}\left(1\right)C_{m}\mathrm{Vol}\left(\widetilde{N}_{0}\right)h^{\frac{1}{2}-n}$
then $j\leq\frac{1}{\varepsilon_{0}}\mathcal{O}\left(1\right)C_{m}\mathrm{Vol}\left(\widetilde{N}_{0}\right)h^{\frac{1}{2}-n}$.
Equivalently if $j>\frac{1}{\varepsilon_{0}}\mathcal{O}\left(1\right)C_{m}\mathrm{Vol}\left(\widetilde{N}_{0}\right)$
then $b_{j}<\varepsilon_{0}$ and $s_{j}^{2}\geq1+2\lambda_{m}h-hb_{j}\geq1+2\left(\lambda_{m}-\varepsilon_{0}\right)h$.
Taking the square root we get (\ref{eq:e-1}).
\end{proof}
We deduce now an upper bound for the number of eigenvalues of $\widehat{P}$.

\selectlanguage{french}%
\vspace{0.cm}\begin{center}{\color{blue}\fbox{\color{black}\parbox{16cm}{
\selectlanguage{english}%
\begin{cor}
\label{cor:Using-Weyl-inequalities}We have\begin{equation}
\sharp\left\{ \sigma\left(\widehat{P}\right)\cap D\left(z_{E},1+\frac{C_{m}}{2}h\right)\right\} \leq\mathcal{O}\left(1\right)C_{m}\mathrm{Vol}\left(\widetilde{N}_{0}\right)h^{\frac{1}{2}-n}.\label{eq:upper_bound}\end{equation}

\end{cor}
\selectlanguage{french}%
}}}\end{center}\vspace{0.cm}
\selectlanguage{english}%
\begin{proof}
Let $\lambda_{1},\lambda_{2},\lambda_{3}\ldots$ denote the eigenvalues
of $\widehat{P}$ sorted such that $j\rightarrow\left|\lambda_{j}-z_{E}\right|$
is increasing. The Weyl inequalities (see \cite[(a.8) p.38]{sjostrand_90}
for a proof) give\begin{equation}
\prod_{j=1}^{N}s_{j}\leq\prod_{j=1}^{N}\left|\lambda_{j}-z_{E}\right|,\qquad\forall N,\label{eq:Weyl_inequalities_original}\end{equation}
where $\left(s_{j}\right)_{j}$ are the singular values defined in
Lemma \ref{lem:sngular-values} above. Let\[
\widetilde{N}:=\sharp\left\{ \lambda_{j}:\quad\left|\lambda_{j}-z_{E}\right|\leq1+\frac{C_{m}}{2}h\right\} =\sharp\left\{ \sigma\left(P\right)\cap D\left(z_{E},1+\frac{C_{m}}{2}h\right)\right\} \]
and let\[
\widetilde{M}:=\mathcal{O}\left(1\right)C_{m}\mathrm{Vol}\left(\widetilde{N}_{0}\right)h^{\frac{1}{2}-n}\]
be the factor which appears in (\ref{eq:e-1}). We want to show the
bound: \begin{equation}
\widetilde{N}\leq\left(2+\mathcal{O}\left(\frac{1}{C_{m}}\right)\right)\widetilde{M}\label{eq:bound}\end{equation}
for $C_{m}\gg1$.

If $\widetilde{N}\leq\widetilde{M}$ then (\ref{eq:bound}) is true.
Conversely let us suppose that $\widetilde{N}\geq\widetilde{M}$.
Using (\ref{eq:Weyl_inequalities_original}) we have\[
\left(\prod_{j=1}^{\widetilde{M}}s_{j}\right)\left(\prod_{j=\widetilde{M}}^{\widetilde{N}}s_{j}\right)\leq\left(1+\frac{C_{m}}{2}h\right)^{\widetilde{N}}.\]
Then using (\ref{eq:e-2}) and (\ref{eq:e-1}) we have\[
\left(1-\mathcal{O}\left(h\right)\right)^{\widetilde{M}}\left(1+\left(C_{m}-\mathcal{O}\left(1\right)\right)h\right)^{\widetilde{N}-\widetilde{M}}\leq\left(1+\frac{C_{m}}{2}h\right)^{\widetilde{N}}.\]
We take the logarithm and since $h\ll1$ we get:\[
-\widetilde{M}\mathcal{O}\left(h\right)+\left(\widetilde{N}-\widetilde{M}\right)\left(C_{m}-\mathcal{O}\left(1\right)\right)h\leq\widetilde{N}\frac{C_{m}}{2}h\]
\[
\Leftrightarrow\widetilde{N}\left(\frac{C_{m}}{2}-\mathcal{O}\left(1\right)\right)\leq\widetilde{M}\left(C_{m}+\mathcal{O}\left(1\right)\right).\]
Now since $C_{m}\gg1$, \[
\Leftrightarrow\widetilde{N}\leq\widetilde{M}\frac{\left(1+\mathcal{O}\left(\frac{1}{C_{m}}\right)\right)}{\left(\frac{1}{2}+\mathcal{O}\left(\frac{1}{C_{m}}\right)\right)}=\widetilde{M}\left(2+\mathcal{O}\left(\frac{1}{C_{m}}\right)\right),\]
so we have obtained (\ref{eq:bound}). This implies (\ref{eq:upper_bound}).
\end{proof}
From Lemma \ref{lem:Z_disk} with $b=\frac{C_{m}}{2}$ and $\beta=\frac{C_{m}}{4}$
we deduce that the upper bound (\ref{eq:upper_bound}) implies an
upper bound:\[
\sharp\left\{ \lambda_{i}\in\sigma\left(\widehat{P}\right),\quad\left|\Re\left(\lambda_{i}-E\right)\right|\leq\sqrt{\frac{C_{m}}{4}h},\quad\Im\left(\lambda_{i}\right)\geq-\frac{C_{m}}{4}h\right\} =\mathcal{O}\left(1\right)C_{m}\mathrm{Vol}\left(\widetilde{N}_{0}\right)h^{\frac{1}{2}-n}.\]
 We take $\alpha=\frac{1}{h}\gg1$ and return to the original spectral
variable $z=\frac{z_{h}}{h}=\alpha z_{h}$ after the scaling (\ref{eq:def_scaling}).
From (\ref{eq:arbitray_small_large}) we can choose the escape function
$m$ such that $C_{m}\gg1$ is arbitrarily large and $\mathrm{Vol}\left(\widetilde{N}_{0}\right)<o\left(\frac{1}{C_{m}}\right)$
is arbitrarily small so that $\mathcal{O}\left(1\right)C_{m}\mathrm{Vol}\left(\widetilde{N}_{0}\right)h^{\frac{1}{2}-n}=o\left(h^{\frac{1}{2}-n}\right)$.
Since the spectrum does not depend on the escape function $m$, we
get (\ref{eq:Upper_bound}). We have finished the proof of Theorem
\ref{thm:Upper-bound-for-resonances}.

\appendix

\section{Some results in operator theory}

\subsection{On minimal and maximal extensions}

We show here that the pseudodifferential operator $\widehat{P}$ defined
in Eq.(\ref{eq:def_K}), has a unique closed extension on $L^{2}\left(X\right)$.
This a well known procedure for the case of elliptic PDO, we refer
to \cite[chap.13 p.125]{wong_99}, and in general this is not true
for PDO of order 2. The fact that $P$ has order 1 (since it is defined
from a vector field on $X$) is therefore important.

The domain of the minimal closed extension $\widehat{P}_{min}$ of
the operator $\widehat{P}$ with domain $C^{\infty}\left(X\right)$
is\begin{equation}
\mathcal{D}_{min}:=\left\{ u\in L^{2}\left(X\right),\quad u_{j}\in C^{\infty}\left(X\right)\rightarrow u\mbox{ in }L^{2}\left(X\right)\mbox{ and }\widehat{P}u_{j}\rightarrow v\in L^{2}\left(X\right)\right\} .\label{eq:def_Dmin}\end{equation}
The maximal closed extension $\widehat{P}_{max}$ has domain\[
\mathcal{D}_{max}:=\left\{ u\in L^{2}\left(X\right),\quad\widehat{P}u\in L^{2}\left(X\right)\right\} .\]
(Recall that $\widehat{P}$ is defined a priori on $C^{\infty}\left(X\right)$
and $\mathcal{D}'\left(X\right)$).

\selectlanguage{french}%
\vspace{0.cm}\begin{center}{\color{blue}\fbox{\color{black}\parbox{16cm}{
\selectlanguage{english}%
\begin{lem}
\label{lem:minimal-maximal_extension}For a PDO $\widehat{P}$ of
order $1$ (i.e $\widehat{P}\in\mathrm{Op}\left(S^{1}\right)$), the
minimal and maximal extensions coincide: $\mathcal{D}\left(\widehat{P}\right):=\mathcal{D}_{min}=\mathcal{D}_{max}$,
i.e. there is a unique closed extension of the operator $\widehat{P}$
in $L^{2}\left(X\right)$.
\end{lem}
\selectlanguage{french}%
}}}\end{center}\vspace{0.cm}
\selectlanguage{english}%
\begin{proof}
$\mathcal{D}_{min}\subset\mathcal{D}_{max}$ is clear. Let us check
that $\mathcal{D}_{max}\subset\mathcal{D}_{min}$. Let $u\in\mathcal{D}_{max}$,
i.e. $u\in L^{2}\left(X\right)$, $v:=\widehat{P}u\in L^{2}\left(X\right)$.
We will construct a sequence $u_{h}\in C^{\infty}\left(X\right)$
with $h\rightarrow0$, such that $u_{h}\rightarrow u$ in $L^{2}\left(X\right)$
and show that $\widehat{P}u_{h}\rightarrow v$ in $L^{2}$.

Let $\chi:T^{*}X\rightarrow\mathbb{R}^{+}$ be a $C^{\infty}$ function
such that $\chi(x,\xi)=1$ for $\left|\xi\right|\leq1$, and $\chi\left(x,\xi\right)=0$
for $\left|\xi\right|\geq2$. For $h>0$, let the function $\chi_{h}$
on $T^{*}X$ be defined by $\chi_{h}\left(x,\xi\right)=\chi\left(x,h\xi\right)$.
Let the \textbf{truncation operator} be:\[
\widehat{\chi}_{h}:=\mathrm{Op}\left(\chi_{h}\right).\]
Notice that $\widehat{\chi}_{h}$ is a smoothing operator which truncates
large components in $\xi$ (larger than $1/h$), $\widehat{\chi}_{h}$
is similar to a convolution in $x$ coordinates.

Let\[
u_{h}:=\widehat{\chi}_{h}u.\]
It is clear that $u_{h}\rightarrow u$ in $L^{2}\left(X\right)$ as
$h\rightarrow0$. We have\[
\widehat{P}u_{h}=\widehat{P}\widehat{\chi}_{h}u=\widehat{\chi}_{h}\widehat{P}u+\left[\widehat{P},\widehat{\chi}_{h}\right]u.\]
The first term converges $\widehat{\chi}_{h}\widehat{P}u\rightarrow v=\widehat{P}u$
as $h\rightarrow0$. The principal symbol of the PDO $\left[\widehat{P},\widehat{\chi}_{h}\right]$
is \[
\frac{1}{i}\left\{ P,\chi_{h}\right\} =\frac{1}{i}\left(\partial_{\xi}P\partial_{x}\chi_{h}-\partial_{x}P\partial_{\xi}\chi_{h}\right).\]
Now we use the fact that $P\in S^{1}$ has order 1. In the first term,
$\partial_{\xi}P\in S^{0}$ is bounded (from (\ref{def:The-symbol-class_S_mu_ro}))
and $\partial_{x}\chi_{h}$ is non zero only on a large ring $\frac{1}{h}\leq\left|\xi\right|\leq\frac{2}{h}$.
In the second term $\partial_{x}P\in S^{1}$ has order 1 but $\partial_{\xi}\chi_{h}=h\partial_{\xi}\chi\left(x,h\xi\right)$
is non zero on the same large ring and therefore of order $\left(-1\right)$
(since $h\simeq\left|\xi\right|^{-1}$ on the ring). Therefore the
PDO $\left[\widehat{P},\widehat{\chi}_{h}\right]$ converges strongly
to zero in $L^{2}\left(X\right)$ as $h\rightarrow0$. Hence $\left[\widehat{P},\widehat{\chi}_{h}\right]u\rightarrow0$
as $h\rightarrow0$. We deduce that $\widehat{P}u_{h}\rightarrow v=\widehat{P}u$,
and that $u\in\mathcal{D}_{min}$.
\end{proof}

\subsection{The sharp Gårding inequality}

References: \cite[p.52]{grigis_sjostrand} or (\ref{eq:garding}),
\cite[p.99]{martinez-01}, \cite[p.1157]{wunsh-zworski_01} for a
short proof using Toeplitz quantization.

\selectlanguage{french}%
\vspace{0.cm}\begin{center}{\color{blue}\fbox{\color{black}\parbox{16cm}{
\selectlanguage{english}%
\begin{prop}
If $\widehat{P}$ is a PDO with symbol $P\in S^{\mu}$, $\mu\in\mathbb{R}$,
$\Re\left(P\right)\geq0$ then there exists $C\in\mathbb{R}$ such
that\begin{equation}
\forall u\in C^{\infty}\left(X\right),\quad\Re\left(\widehat{P}u|u\right)\geq-C\left\Vert u\right\Vert _{H^{\frac{\mu-1}{2}}}^{2}\label{eq:sharp_Garding}\end{equation}
where $\left\Vert u\right\Vert _{H^{\mu}}^{2}:=\left(\left\langle \widehat{\xi}\right\rangle ^{\mu}u|\left\langle \widehat{\xi}\right\rangle ^{\mu}u\right)_{L^{2}\left(X\right)}$
denotes the norm in the Sobolev space $H^{\mu}$.
\end{prop}
\selectlanguage{french}%
}}}\end{center}\vspace{0.cm}

\selectlanguage{english}%

\subsection{Quadratic partition of unity on phase space}

As usual in this paper, we denote $\widehat{A}:=\mathrm{Op}_{h}\left(A\right)$
for a symbol $A$.

\selectlanguage{french}%
\vspace{0.cm}\begin{center}{\color{blue}\fbox{\color{black}\parbox{16cm}{
\selectlanguage{english}%
\begin{lem}
\label{lem:Let-partition_Ki0-Ki1}Let $K_{0}\subset T^{*}X$ compact.
There exists symbols $\chi_{0}\in S^{-\infty}$ and $\chi_{1}\in S^{0}$
of self-adjoint operators $\widehat{\chi}_{0},\widehat{\chi}_{1}$
such that\[
\widehat{\chi}_{0}^{2}+\widehat{\chi}_{1}^{2}=1+\widehat{R}.\]
The symbol $R\in\left(h^{\infty}S^{-\infty}\right)$ is negligible,
$\mathrm{supp}\left(\chi_{0}\right)$ is compact and on $K_{0}$,
$\chi_{1}\left(x,\xi\right)=\mathcal{O}\left(h^{\infty}\right)$,
$\chi_{0}\left(x,\xi\right)=1+\mathcal{O}\left(h^{\infty}\right)$.
\end{lem}
\selectlanguage{french}%
}}}\end{center}\vspace{0.cm}
\selectlanguage{english}%
\begin{proof}
Let $K_{0}\subset T^{*}X$ be compact. We can find symbols $0\leq\chi_{0}\in C_{0}^{\infty}\left(T^{*}X\right)$
(with compact support) and $0\leq\chi_{1}\in C^{\infty}\left(T^{*}X\right)$
such that\[
\chi_{1}=\begin{cases}
0 & \mbox{ on }K_{0}\\
1 & \mbox{ for }\left|\xi\right|\gg1\end{cases}\]
and\[
A:=\chi_{0}^{2}+\chi_{1}^{2}\mbox{ is }\begin{cases}
>0 & \mbox{ everywhere}\\
=1 & \mbox{ for }\left|\xi\right|\gg1.\end{cases}\]
We replace $\chi_{0},\chi_{1}$ respectively by $\chi_{0}A^{-1/2}$,
$\chi_{1}A^{-1/2}$. We obtain $1=\chi_{0}^{2}+\chi_{1}^{2}$.

Let $\widehat{R}:=\widehat{\chi}_{0}^{2}+\widehat{\chi}_{1}^{2}-1$.
Then $\widehat{R}\in\mathrm{Op}\left(hS^{-\infty}\right)$. We write
$R=hr_{0}\left(x,\xi\right)+h^{2}\ldots$.

We replace $\widehat{\chi}_{j}$, $j=0,1$ by\[
\widehat{\chi}_{j}^{'}:=\left(1+h\widehat{r}_{0}\right)^{-1/4}\widehat{\chi}_{j}\left(1+h\widehat{r}_{0}\right)^{-1/4}.\]

Which is also self-adjoint. We obtain\begin{align*}
\widehat{\chi}_{0}^{'2}+\widehat{\chi}_{1}^{'2} & =\left(1-h\widehat{r}_{0}\right)\widehat{\chi}_{0}^{2}+\left(1-h\widehat{r}_{1}\right)\widehat{\chi}_{1}^{2}+\mathcal{O}\left(\mathrm{Op}\left(h^{2}S^{-\infty}\right)\right)\\
 & =1+\mathcal{O}\left(\mathrm{Op}\left(h^{2}S^{-\infty}\right)\right).\end{align*}
If we iterate this algorithm, we obtain the Lemma.
\end{proof}

\subsubsection{I.M.S. localization formula}

The following Lemma is similar to the {}``I.M.S localization formula''
given in \cite[p.27]{simon_87}. It uses the quadratic partition of
phase space obtained in Lemma \ref{lem:Let-partition_Ki0-Ki1} above.

\selectlanguage{french}%
\vspace{0.cm}\begin{center}{\color{blue}\fbox{\color{black}\parbox{16cm}{
\selectlanguage{english}%
\begin{lem}
\textbf{\label{lem:IMS-localization-formula}} Suppose that $\widehat{P}\in\mathrm{Op}\left(S^{\mu}\right)$
for some $\mu\in\mathbb{R}$ and that $\left(P-P^{*}\right)\in\mathrm{Op}\left(hS^{\mu}\right)$.
Then for every $u\in L^{2}\left(X\right)$, $z\in\mathbb{C}$,\begin{equation}
\left\Vert \left(\widehat{P}-z\right)u\right\Vert ^{2}=\left\Vert \left(\widehat{P}-z\right)\widehat{\chi}_{0}u\right\Vert ^{2}+\left\Vert \left(\widehat{P}-z\right)\widehat{\chi}_{1}u\right\Vert ^{2}+\mathcal{O}\left(h^{2}\right)\left\Vert u\right\Vert ^{2}.\label{eq:result_IMS}\end{equation}

\end{lem}
\selectlanguage{french}%
}}}\end{center}\vspace{0.cm}
\selectlanguage{english}%
\begin{proof}
For simplicity, we suppose $z=i\beta$ with $\beta\in\mathbb{R}$,
i.e. $\Re\left(z\right)=0$ (this is equivalent to replacing $\widehat{P}-\Re\left(z\right)$
by some operator $\widehat{P}'$). We use (\ref{eq:Quad_Partition})
and write\begin{align}
\left\Vert \left(\widehat{P}-i\beta\right)u\right\Vert ^{2}= & \left(\left(\widehat{P}-i\beta\right)^{*}\left(\widehat{P}-i\beta\right)u|u\right)=\sum_{k=0,1}\left(\left(\widehat{P}-i\beta\right)^{*}\widehat{\chi}_{k}^{2}\left(\widehat{P}-i\beta\right)u|u\right)\label{eq:IMS1}\\
 & +\mathcal{O}\left(h^{\infty}\right)\left\Vert u\right\Vert ^{2}.\end{align}
The aim is to move the operators $\widehat{\chi}_{k}$ outside. One
has for $k=0,1$:\begin{align}
\left(\widehat{P}-i\beta\right)^{*} & \widehat{\chi}_{k}^{2}\left(\widehat{P}-i\beta\right)-\widehat{\chi}_{k}\left(\widehat{P}-i\beta\right)^{*}\left(\widehat{P}-i\beta\right)\widehat{\chi}_{k}=\label{eq:IMS}\\
= & \widehat{\chi}_{k}\left(\widehat{P}-i\beta\right)^{*}\widehat{\chi}_{k}\left(\widehat{P}-i\beta\right)+\left[\widehat{P}^{*},\widehat{\chi}_{k}\right]\widehat{\chi}_{k}\left(\widehat{P}-i\beta\right)-\widehat{\chi}_{k}\left(\widehat{P}-i\beta\right)^{*}\left(\widehat{P}-i\beta\right)\widehat{\chi}_{k}\nonumber \\
= & \left[\widehat{P}^{*},\widehat{\chi}_{k}\right]\widehat{\chi}_{k}\left(\widehat{P}-i\beta\right)-\widehat{\chi}_{k}\left(\widehat{P}-i\beta\right)^{*}\left[\widehat{P},\widehat{\chi}_{k}\right]\nonumber \\
= & \left(\underbrace{\left[\widehat{P}^{*},\widehat{\chi}_{k}\right]\widehat{\chi}_{k}\widehat{P}-\widehat{\chi}_{k}\widehat{P}^{*}\left[\widehat{P},\widehat{\chi}_{k}\right]}_{\mathrm{I}_{k}}\right)-i\beta\left(\underbrace{\left[\widehat{P}^{*},\widehat{\chi}_{k}\right]\widehat{\chi}_{k}+\widehat{\chi}_{k}\left[\widehat{P},\widehat{\chi}_{k}\right]}_{\mathrm{II_{k}}}\right).\nonumber \end{align}
First remark that for every PDO $\widehat{A}\in\mathrm{Op}\left(S^{\mu}\right)$
with some $\mu\in\mathbb{R}$, then \[
\left[\widehat{A},\widehat{\chi}_{k}\right]\in\mathrm{Op}\left(hS^{-\infty}\right).\]
This is obvious for $k=0$ since $\widehat{\chi}_{0}\in\mathrm{Op}\left(S^{-\infty}\right)$
and for $k=1$ this is because $\left(\widehat{\chi}_{1}-1\right)\in\mathrm{Op}\left(S^{-\infty}\right)$
and $\left[\widehat{A},1\right]=0$. We have assumed that\[
\left(\widehat{P}^{*}-\widehat{P}\right)\in\mathrm{Op}\left(hS^{\mu}\right),\]
therefore \[
\left[\widehat{P}^{*}-\widehat{P},\widehat{\chi}_{k}\right]\in\mathrm{Op}\left(h^{2}S^{-\infty}\right).\]
Also\[
\left[\widehat{P},\widehat{\chi}_{k}\right]\in\mathrm{Op}\left(hS^{-\infty}\right).\]
The first term of (\ref{eq:IMS}) is\begin{align*}
\mathrm{I_{k}} & =\left[\widehat{P}^{*},\widehat{\chi}_{k}\right]\widehat{\chi}_{k}\widehat{P}-\widehat{\chi}_{k}\widehat{P}^{*}\left[\widehat{P},\widehat{\chi}_{k}\right]\\
 & =\left[\widehat{P},\widehat{\chi}_{k}\right]\widehat{\chi}_{k}\widehat{P}-\widehat{\chi}_{k}\widehat{P}\left[\widehat{P},\widehat{\chi}_{k}\right]+\mathcal{O}\left(\mathrm{Op}\left(h^{2}S^{-\infty}\right)\right)\\
 & =\left[\left[\widehat{P},\widehat{\chi}_{k}\right],\widehat{\chi}_{k}\widehat{P}\right]+\mathcal{O}\left(\mathrm{Op}\left(h^{2}S^{-\infty}\right)\right)\\
 & =\mathcal{O}\left(\mathrm{Op}\left(h^{2}S^{-\infty}\right)\right).\end{align*}
The second term of (\ref{eq:IMS}) is\begin{align*}
\mathrm{II_{k}} & =\left[\widehat{P}^{*},\widehat{\chi}_{k}\right]\widehat{\chi}_{k}+\widehat{\chi}_{k}\left[\widehat{P},\widehat{\chi}_{k}\right]\\
 & =\left[\widehat{P},\widehat{\chi}_{k}\right]\widehat{\chi}_{k}+\widehat{\chi}_{k}\left[\widehat{P},\widehat{\chi}_{k}\right]+\mathcal{O}\left(\mathrm{Op}\left(h^{2}S^{-\infty}\right)\right)\\
 & =\left[\widehat{P},\widehat{\chi}_{k}^{2}\right]+\mathcal{O}\left(\mathrm{Op}\left(h^{2}S^{-\infty}\right)\right).\end{align*}
Therefore using (\ref{eq:Quad_Partition}) \begin{align*}
\mathrm{II_{0}+II_{1}} & =\left[\widehat{P},\widehat{\chi}_{0}^{2}+\widehat{\chi}_{1}^{2}\right]+\mathcal{O}\left(\mathrm{Op}\left(h^{2}S^{-\infty}\right)\right)\\
 & =\mathcal{O}\left(\mathrm{Op}\left(h^{2}S^{-\infty}\right)\right).\end{align*}
We have shown that\[
\sum_{k=0,1}\left(\widehat{P}-i\beta\right)^{*}\widehat{\chi}_{k}^{2}\left(\widehat{P}-i\beta\right)=\sum_{k=0,1}\widehat{\chi}_{k}\left(\widehat{P}-i\beta\right)^{*}\left(\widehat{P}-i\beta\right)\widehat{\chi}_{k}+\mathcal{O}\left(\mathrm{Op}\left(h^{2}S^{-\infty}\right)\right).\]
Coming back to (\ref{eq:IMS1}) we get (\ref{eq:result_IMS}).
\end{proof}

\subsection{\label{sub:FBI-transform-and}FBI transform and Toeplitz operators}

References: \cite{sjostrand_hitrick_07},\cite{martinez-01},\cite{wunsh-zworski_01},
\cite{sjostrand_density_resonances_96}.

The manifold $X$ is equipped with a smooth Riemannian metric so that
we have a well-defined exponential map $\exp_{x}:T_{x}X\rightarrow X$
which is a diffeomorphism from a neighborhood of $0\in T_{x}X$ onto
a neighborhood of $x\in X$. Define the \textbf{coherent state }at
point $\alpha=\left(\alpha_{x},\alpha_{\xi}\right)\in T^{*}X$ to
be the function of $y\in X$:\[
e_{\alpha}\left(y\right):=\chi\left(\alpha_{x},y\right)\exp\left(\frac{i}{h}\alpha_{\xi}\left(\exp_{\alpha_{x}}^{-1}\left(y\right)\right)-\frac{1}{2h}\left\langle \alpha_{\xi}\right\rangle \mathrm{dist}\left(\alpha_{x},y\right)^{2}\right),\qquad\left\langle \alpha_{\xi}\right\rangle :=\left(1+\alpha_{\xi}^{2}\right)^{1/2},\]
where $\chi\in C^{\infty}\left(X\times X\right)$ is a standard cutoff
to a small neighborhood of the diagonal. In the Euclidean case $X=\mathbb{R}^{n}$,
the cutoff is often superfluous and we get the complex Gaussian {}``wave
packet''\[
e_{\alpha}\left(y\right)=\exp\left(\frac{i}{h}\alpha_{\xi}\left(y-\alpha_{x}\right)-\frac{1}{2h}\left\langle \alpha_{\xi}\right\rangle \left|y-\alpha_{x}\right|^{2}\right).\]
We can define the \textbf{FBI-transform} of $u\in C^{\infty}\left(X\right)$
by\[
\left(Tu\right)\left(\alpha;h\right):=h^{-\frac{3n}{4}}\left(e_{\alpha}|u\right)=h^{-\frac{3n}{4}}\int_{X}\overline{e_{\alpha}\left(y\right)}u\left(y\right)dy,\]
which can be made asymptotically isometric after multiplication to
the left by an elliptic symbol of order $0$ and we can keep this
point of view in mind. We have the following known facts \cite{sjostrand_density_resonances_96,sjostrand_hitrick_07}:
\begin{itemize}
\item There exists $a_{0}\left(\alpha;h\right)\in h^{-\frac{3n}{2}}S^{n/2}$
elliptic and $a_{0}>0$ such that\[
u=\int_{T^{*}X}\left(\widehat{\pi}_{\alpha}u\right)d\alpha+\widehat{R}u,\quad\forall u\in L^{2}\left(X\right)\]
with\[
\widehat{\pi}_{\alpha}:=a_{0}\left(\alpha;h\right)e_{\alpha}\left(e_{\alpha}|.\right)\]
and $\widehat{R}\in\mathrm{Op}_{h}\left(h^{\infty}S^{-\infty}\right)$
negligible.
\item $\widehat{\pi}_{\alpha}\geq0$ and\[
\left\Vert \widehat{\pi}_{\alpha}\right\Vert _{tr}=\mathrm{Tr}\left(\widehat{\pi}_{\alpha}\right)=a_{0}\left(\alpha;h\right)\left\Vert e_{\alpha}\right\Vert ^{2}=\mathcal{O}\left(1\right)h^{-n}.\]

\item If $\widehat{B}\in\mathrm{Op}_{h}\left(S^{m}\right)$ has the principal
symbol $b_{0}$ (modulo $hS^{m-1}$), then\[
\widehat{B}=\int_{T^{*}X}b\left(\alpha;h\right)\widehat{\pi}_{\alpha}d\alpha+\widehat{R}\]
where $\widehat{R}$ is negligible as above, $b\in S^{m}$ and $b=b_{0}\mbox{ mod}\left(hS^{m-1}\right)$.
\item For a function $A\left(x,\xi;h\right)$, we define the \textbf{Toeplitz
quantization of $A$} by \[
\mathrm{Op}_{T}\left(A\right):=\int A\left(\alpha;h\right)\widehat{\pi}_{\alpha}d\alpha,\]
then the previous results imply a {}``Gårding's inequality'':\begin{equation}
A\left(x,\xi\right)\geq0\Rightarrow\left(\mathrm{Op}_{T}\left(A\right)u|u\right)\geq0+\mathcal{O}\left(h^{\infty}\right)\left\Vert u\right\Vert ^{2}\label{eq:garding}\end{equation}
and\[
\mathrm{Tr}\left(\widehat{A}\right)=\frac{\mathcal{O}\left(1\right)}{h^{n}}\int A\left(x,\xi\right)dxd\xi+\mathcal{O}\left(h^{\infty}\right).\]

\end{itemize}
\bibliographystyle{plain}
\bibliography{/home/faure/articles/articles}

\end{document}

%% file: projective_flow.pstex_t
\begin{picture}(0,0)%
\includegraphics{projective_flow}%
\end{picture}%
\setlength{\unitlength}{4144sp}%
\begingroup\makeatletter\ifx\SetFigFont\undefined%
\gdef\SetFigFont#1#2#3#4#5{%
  \reset@font\fontsize{#1}{#2pt}%
  \fontfamily{#3}\fontseries{#4}\fontshape{#5}%
  \selectfont}%
\fi\endgroup%
\begin{picture}(2460,1791)(2191,-1264)
\put(4276,-781){\makebox(0,0)[lb]{\smash{{\SetFigFont{12}{14.4}{\familydefault}{\mddefault}{\updefault}{\color[rgb]{0,0,0}$[v_0]$}%
}}}}
\put(2206,-1186){\makebox(0,0)[lb]{\smash{{\SetFigFont{12}{14.4}{\familydefault}{\mddefault}{\updefault}{\color[rgb]{0,0,0}$E_u$}%
}}}}
\put(3736,344){\makebox(0,0)[lb]{\smash{{\SetFigFont{12}{14.4}{\familydefault}{\mddefault}{\updefault}{\color[rgb]{0,0,0}$E_s$}%
}}}}
\put(4636,-1096){\makebox(0,0)[lb]{\smash{{\SetFigFont{12}{14.4}{\familydefault}{\mddefault}{\updefault}{\color[rgb]{0,0,0}$E_0$}%
}}}}
\end{picture}%

%% file: lifted_Mt.pstex_t
\begin{picture}(0,0)%
\includegraphics{lifted_Mt}%
\end{picture}%
\setlength{\unitlength}{3947sp}%
\begingroup\makeatletter\ifx\SetFigFont\undefined%
\gdef\SetFigFont#1#2#3#4#5{%
  \reset@font\fontsize{#1}{#2pt}%
  \fontfamily{#3}\fontseries{#4}\fontshape{#5}%
  \selectfont}%
\fi\endgroup%
\begin{picture}(3455,3927)(-1483,-2755)
\put(-1049,-1786){\makebox(0,0)[lb]{\smash{{\SetFigFont{12}{14.4}{\rmdefault}{\mddefault}{\updefault}{\color[rgb]{0,0,0}$x$}%
}}}}
\put(1501,-2011){\makebox(0,0)[lb]{\smash{{\SetFigFont{12}{14.4}{\rmdefault}{\mddefault}{\updefault}{\color[rgb]{0,0,0}$\phi_t(x)$}%
}}}}
\put(-1049,989){\makebox(0,0)[lb]{\smash{{\SetFigFont{12}{14.4}{\rmdefault}{\mddefault}{\updefault}{\color[rgb]{0,0,0}$T^*_xX$}%
}}}}
\put(1351,989){\makebox(0,0)[lb]{\smash{{\SetFigFont{12}{14.4}{\rmdefault}{\mddefault}{\updefault}{\color[rgb]{0,0,0}$T^*_{\phi_t(x)}X$}%
}}}}
\put(1876,-2686){\makebox(0,0)[lb]{\smash{{\SetFigFont{12}{14.4}{\rmdefault}{\mddefault}{\updefault}{\color[rgb]{0,0,0}$X$}%
}}}}
\put(1576,614){\makebox(0,0)[lb]{\smash{{\SetFigFont{12}{14.4}{\rmdefault}{\mddefault}{\updefault}{\color[rgb]{0,0,0}$M_t(\xi)$}%
}}}}
\put(-1274,-511){\makebox(0,0)[lb]{\smash{{\SetFigFont{12}{14.4}{\rmdefault}{\mddefault}{\updefault}{\color[rgb]{0,0,0}$\xi$}%
}}}}
\put(-299,-2386){\makebox(0,0)[lb]{\smash{{\SetFigFont{12}{14.4}{\rmdefault}{\mddefault}{\updefault}{\color[rgb]{0,0,0}$V$}%
}}}}
\put(-224,-586){\makebox(0,0)[lb]{\smash{{\SetFigFont{12}{14.4}{\rmdefault}{\mddefault}{\updefault}{\color[rgb]{0,0,0}$\mathbf{X}$}%
}}}}
\end{picture}%

%% file: Cotangent.pstex_t
\begin{picture}(0,0)%
\includegraphics{Cotangent}%
\end{picture}%
\setlength{\unitlength}{3947sp}%
\begingroup\makeatletter\ifx\SetFigFont\undefined%
\gdef\SetFigFont#1#2#3#4#5{%
  \reset@font\fontsize{#1}{#2pt}%
  \fontfamily{#3}\fontseries{#4}\fontshape{#5}%
  \selectfont}%
\fi\endgroup%
\begin{picture}(3927,3633)(814,-4498)
\put(1576,-2536){\makebox(0,0)[lb]{\smash{{\SetFigFont{12}{14.4}{\rmdefault}{\mddefault}{\updefault}{\color[rgb]{0,0,0}$0$}%
}}}}
\put(1426,-1336){\makebox(0,0)[lb]{\smash{{\SetFigFont{12}{14.4}{\rmdefault}{\mddefault}{\updefault}{\color[rgb]{0,0,0}$E^*_s(x)$}%
}}}}
\put(4126,-2086){\makebox(0,0)[lb]{\smash{{\SetFigFont{12}{14.4}{\rmdefault}{\mddefault}{\updefault}{\color[rgb]{0,0,0}$E^*_0(x)=K(x)$: Trapped Set}%
}}}}
\put(3151,-3511){\makebox(0,0)[lb]{\smash{{\SetFigFont{12}{14.4}{\rmdefault}{\mddefault}{\updefault}{\color[rgb]{0,0,0}$E^*_u(x)$}%
}}}}
\put(4726,-2686){\makebox(0,0)[lb]{\smash{{\SetFigFont{12}{14.4}{\rmdefault}{\mddefault}{\updefault}{\color[rgb]{0,0,0}$K_E(x)$}%
}}}}
\put(3451,-1036){\makebox(0,0)[lb]{\smash{{\SetFigFont{12}{14.4}{\rmdefault}{\mddefault}{\updefault}{\color[rgb]{0,0,0}Energy Shell $\Sigma_E(x)$}%
}}}}
\end{picture}%

%% file: cosphere.pstex_t
\begin{picture}(0,0)%
\includegraphics{cosphere}%
\end{picture}%
\setlength{\unitlength}{4144sp}%
\begingroup\makeatletter\ifx\SetFigFont\undefined%
\gdef\SetFigFont#1#2#3#4#5{%
  \reset@font\fontsize{#1}{#2pt}%
  \fontfamily{#3}\fontseries{#4}\fontshape{#5}%
  \selectfont}%
\fi\endgroup%
\begin{picture}(2356,2158)(2551,-1655)
\put(3736,344){\makebox(0,0)[lb]{\smash{{\SetFigFont{12}{14.4}{\familydefault}{\mddefault}{\updefault}{\color[rgb]{0,0,0}$\tilde{E}^*_s$}%
}}}}
\put(2566,-1051){\makebox(0,0)[lb]{\smash{{\SetFigFont{12}{14.4}{\familydefault}{\mddefault}{\updefault}{\color[rgb]{0,0,0}$\tilde{E}^*_u$}%
}}}}
\put(4636,-916){\makebox(0,0)[lb]{\smash{{\SetFigFont{12}{14.4}{\familydefault}{\mddefault}{\updefault}{\color[rgb]{0,0,0}$\tilde{E}^*_0$}%
}}}}
\put(4892,-348){\makebox(0,0)[lb]{\smash{{\SetFigFont{12}{14.4}{\rmdefault}{\mddefault}{\updefault}{\color[rgb]{0,0,0}$\tilde{N}_0$}%
}}}}
\end{picture}%

%% file: Set_N0.pstex_t
\begin{picture}(0,0)%
\includegraphics{Set_N0}%
\end{picture}%
\setlength{\unitlength}{4144sp}%
\begingroup\makeatletter\ifx\SetFigFont\undefined%
\gdef\SetFigFont#1#2#3#4#5{%
  \reset@font\fontsize{#1}{#2pt}%
  \fontfamily{#3}\fontseries{#4}\fontshape{#5}%
  \selectfont}%
\fi\endgroup%
\begin{picture}(2816,2403)(485,-3185)
\put(3286,-2041){\makebox(0,0)[lb]{\smash{{\SetFigFont{12}{14.4}{\familydefault}{\mddefault}{\updefault}{\color[rgb]{0,0,0}$E^*_0$}%
}}}}
\put(2971,-1771){\makebox(0,0)[lb]{\smash{{\SetFigFont{12}{14.4}{\familydefault}{\mddefault}{\updefault}{\color[rgb]{0,0,0}$\tilde{N}_0$}%
}}}}
\put(541,-2806){\makebox(0,0)[lb]{\smash{{\SetFigFont{12}{14.4}{\familydefault}{\mddefault}{\updefault}{\color[rgb]{0,0,0}$E^*_u$}%
}}}}
\put(1350,-941){\makebox(0,0)[lb]{\smash{{\SetFigFont{12}{14.4}{\familydefault}{\mddefault}{\updefault}{\color[rgb]{0,0,0}$E^*_s$}%
}}}}
\put(1756,-2401){\makebox(0,0)[lb]{\smash{{\SetFigFont{12}{14.4}{\rmdefault}{\mddefault}{\updefault}{\color[rgb]{0,0,0}$R$}%
}}}}
\end{picture}%

%% file: spectrum.pstex_t
\begin{picture}(0,0)%
\includegraphics{spectrum}%
\end{picture}%
\setlength{\unitlength}{4144sp}%
\begingroup\makeatletter\ifx\SetFigFont\undefined%
\gdef\SetFigFont#1#2#3#4#5{%
  \reset@font\fontsize{#1}{#2pt}%
  \fontfamily{#3}\fontseries{#4}\fontshape{#5}%
  \selectfont}%
\fi\endgroup%
\begin{picture}(5015,1881)(-1222,-3424)
\put(406,-1726){\makebox(0,0)[lb]{\smash{{\SetFigFont{12}{14.4}{\familydefault}{\mddefault}{\updefault}{\color[rgb]{0,0,0}$Im(z)$}%
}}}}
\put( 46,-3346){\makebox(0,0)[lb]{\smash{{\SetFigFont{12}{14.4}{\rmdefault}{\mddefault}{\updefault}{\color[rgb]{0,0,0}$-C_m$}%
}}}}
\put( 91,-2986){\makebox(0,0)[lb]{\smash{{\SetFigFont{12}{14.4}{\rmdefault}{\mddefault}{\updefault}{\color[rgb]{0,0,0}$-\beta$}%
}}}}
\put(3736,-2536){\makebox(0,0)[lb]{\smash{{\SetFigFont{12}{14.4}{\familydefault}{\mddefault}{\updefault}{\color[rgb]{0,0,0}$Re(z)$}%
}}}}
\put(2116,-2221){\makebox(0,0)[lb]{\smash{{\SetFigFont{12}{14.4}{\rmdefault}{\mddefault}{\updefault}{\color[rgb]{0,0,0}$E\alpha$}%
}}}}
\put(1891,-1861){\makebox(0,0)[lb]{\smash{{\SetFigFont{12}{14.4}{\rmdefault}{\mddefault}{\updefault}{\color[rgb]{0,0,0}$O(\sqrt{\alpha})$}%
}}}}
\end{picture}%